%% file: main.tex
\makeatletter
\def\cl@chapter{\@elt {theorem}}
\makeatother

\documentclass[smallextended,envcountsame,nospthms]{svjour3}
\journalname{}
\smartqed  
\usepackage{bbm}

\usepackage{amsmath, amssymb, amsthm}
\usepackage{thmtools}
\usepackage{thm-restate}
\PassOptionsToPackage{hyphens}{url}
\usepackage[bookmarksopen,bookmarksdepth=3]{hyperref}
\usepackage{lineno}
\usepackage{tikz}
\usetikzlibrary{shapes,calc,arrows,automata,decorations.pathmorphing}
\usepackage{xifthen}
\usepackage{xspace}
\usepackage[noline,noend]{algorithm2e}
\usepackage[capitalize,nameinlink]{cleveref}
\theoremstyle{plain}
\newtheorem{theorem}{Theorem}
\numberwithin{theorem}{section}

\numberwithin{notation}{section}

\numberwithin{conjecture}{section}
\newtheorem{lemma}[theorem]{Lemma}
\numberwithin{lemma}{section}
\newtheorem{corollary}[theorem]{Corollary}
\numberwithin{corollary}{section}
\theoremstyle{definition}

\numberwithin{remark}{section}
\newtheorem{definition}[theorem]{Definition}
\numberwithin{definition}{section}
\theoremstyle{example}
\newtheoremstyle{example} 
        {\topsep}                    
        {\topsep}                    
        {\slshape\fontfamily{ptm}\selectfont}                   
        {}                           
        {\fontfamily{ptm}\selectfont\scshape\color{blue}}                   
        {:}                          
        {.5em}                       
        {}  

\newtheorem{example}[theorem]{Example}
\numberwithin{example}{section}
\usepackage{stackengine}
\usepackage[utf8]{inputenc}
\usepackage[inline]{enumitem}
\setlist[enumerate,1]{label=(\roman*), wide=0pt, leftmargin=*}

\renewcommand{\emptyset}{\varnothing}

\usepackage{mathtools}

\usepackage{stmaryrd}
\usepackage{etoolbox}
\usepackage{mymatrix}
\usepackage{multirow}
\usepackage{bm}
\usepackage[english]{babel}
\usepackage[location=appendix,prependtoappendix=true]{moveproofs}
\usepackage{todonotes}
\usepackage{accents}
\usepackage{caption}
\usepackage{subcaption}

\makeatletter
\let\xx@thm\@thm
\AtBeginDocument{\let\@thm\xx@thm}
\makeatother

\usepackage[nottoc]{tocbibind}
\usepackage{cite}
\usepackage[numbers]{natbib}
\usepackage{doi}

\allowdisplaybreaks

\hypersetup{
    pdftitle={Termination of Triangular Polynomial Loops},
    colorlinks=true,
    linkcolor=blue,
    citecolor=olive,
    filecolor=magenta,
    urlcolor=cyan
}

\input{macros}

\begin{document}
\title{Termination of Triangular Polynomial Loops\thanks{funded
  by the Deutsche
     Forschungsgemeinschaft (DFG, German Research Foundation) -
     235950644 (Project GI 274/6-2),
 by the Deutsche
Forschungsgemeinschaft (DFG, 
     German Research Foundation) - 389792660
as part of \href{https://perspicuous-computing.science}{TRR~248},     
     and 
     by the DFG Research Training Group 2236 UnRAVeL.
  }}

\author{Marcel Hark \and Florian Frohn \and J\"urgen Giesl}
    
\institute{
  Marcel Hark \and  J\"urgen Giesl \at LuFG Informatik 2, RWTH Aachen University, Germany \and
  Florian Frohn \at Max Planck Institute for Informatics,
  Saarland Informatics Campus, Saarbr\"ucken, Germany,
LuFG Informatik 2, RWTH Aachen University, Germany
}

\date{}

\maketitle

\vspace*{-1cm}

\begin{abstract}
 \input{abstract}

\end{abstract}


\input{introduction}

\input{preliminaries}

\input{closed}

\input{deciding}

\input{transformations}

\input{automorphisms}

\input{adaptions}

\input{linearizing}

\input{complexity}

\input{ptime}

\input{integer}

\input{conclusion}
\input{related_work}

\bibliographystyle{spbasic}      
\input{main.bbl}

\clearpage
\section*{\huge\appendixname}
\appendix
\end{document}

%% file: macros.tex
\newcommand{\RA}{\mathbb{R}_{\mathbb{A}}}
\newcommand{\ZZ}{\mathbb{Z}}
\newcommand{\RR}{\mathbb{R}}
\newcommand{\NN}{\mathbb{N}}

\newcommand{\LL}{\mathbf{L}}
\newcommand{\Lan}{\mathcal{L}}
\newcommand{\QQ}{\mathbb{Q}}
\newcommand{\Q}{\mathcal{Q}}
\newcommand{\VV}{\mathcal{V}}
\newcommand{\setcomp}[2]{\left\{ {#1} \relmiddle{|} {#2} \right\}}

\newcommand{\linpoly}[1]{\ensuremath{\ring[#1]_{\mathrm{lin}}}}
\newcommand{\qlinpoly}[1]{\ensuremath{\Q_\ring[#1]_{\mathrm{lin}}}}

\newcommand{\PE}[1]{\mathbb{PE}_{#1}}
\newcommand{\PEN}[1]{\mathbb{NPE}_{#1}}
\newcommand{\OO}{\mathcal{O}}
\newcommand{\assign}{\ensuremath{\leftarrow}}

\newcommand{\relmiddle}[1]{\mathrel{}\middle#1\mathrel{}}
\newcommand{\name}[1]{\textit{#1}\xspace}
\newcommand{\tnn}{\name{tnn}}
\newcommand{\twn}{\name{twn}}
\newcommand{\coeffs}[1]{\coeffsop\left(#1\right)}
\newcommand{\sign}[1]{\signop\left(#1\right)}
\newcommand{\base}[1]{\baseop\left(#1\right)}
\newcommand{\unmark}[1]{\unmarkop\left(#1\right)}
\newcommand{\succc}{\succ_{\mathit{coef}}}
\newcommand{\preccc}{\prec_{\mathit{coef}}}
\newcommand{\maxsucc}[1]{\max\nolimits_{\succc}\left(#1\right)}
\newcommand{\EFO}[1]{\efoop(#1)}
\newcommand{\QFFO}[1]{\qffoop(#1)}
\newcommand{\TRUE}{\mathrm{true}}
\newcommand{\FALSE}{\mathrm{false}}
\renewcommand{\ring}{\mathcal{S}}

\DeclareMathOperator{\TrOp}{\mathit{T\!r}}
\newcommand{\Tr}[3]{\TrOp_{#1}(#2,#3)}
\newcommand{\Trnv}[3]{\TrOp_{#1}(#2,#3)}

\newcommand{\Aut}[1]{\mathrm{Aut}_{#1}\left(#1[\vec{x}]\right)}
\newcommand{\End}[1]{\mathrm{End}_{#1}\left(#1[\vec{x}]\right)}

\newcommand{\cc}[1]{\textsf{#1}}
\newcommand{\card}[1]{\left|#1\right|}

\newcommand{\update}{\vec{u}}
\newcommand{\updateElem}{u}

\newcommand{\cond}{\varphi}
\newcommand{\witness}{\vec{e}}
\newcommand{\closed}{\vec{q}}
\newcommand{\closednorm}{\closed_{\mathit{norm}}}

\newcommand{\maybe}{\ensuremath{\vDash^{\scriptscriptstyle ?}}}

\newcommand{\minialg}[3]{
	\ifstrequal{#2}{}{
		\begin{minipage}{#1}
			\algorithmstyle{plain}
			\begin{algorithm}[H]
				\SetAlgoShortEnd
				\SetStartEndCondition{ (}{)}{)}
				\SetAlgoBlockMarkers{}{\}}%
				\SetKwFor{While}{while}{\ do\ }{}%
				#3
			\end{algorithm}
		\end{minipage}
	}
	{
		\begin{minipage}{#1}
			\algorithmstyle{#2}
			\begin{algorithm}[H]
				\SetAlgoShortEnd
				\SetStartEndCondition{ (}{)}{)}
				\SetAlgoBlockMarkers{}{\}}%
				\SetKwFor{While}{while}{\ do\ }{}%
				#3
			\end{algorithm}
		\end{minipage}
	}}

\makeatletter \newcommand{\algorithmstyle}[1]{\renewcommand{\algocf@style}{#1}}
\makeatother \DontPrintSemicolon

\renewcommand{\arraystretch}{1.4}

\makeatletter \renewcommand*\env@matrix[1][\arraystretch]{%
	\edef\arraystretch{#1}%
	\hskip -\arraycolsep \let\@ifnextchar\new@ifnextchar \array{*\c@MaxMatrixCols c}}
\makeatother

\makeatletter \newcommand{\textlabel}[2]{%
	\protected@edef\@currentlabel{#1}
	\phantomsection
	#1\label{#2}
}
\makeatother

\newcounter{eq-term-closed-form}

\DeclareMathOperator{\depdeg}{depdeg}

\DeclareMathOperator{\coeffsop}{coefs}
\DeclareMathOperator{\signop}{sign}
\DeclareMathOperator{\unmarkop}{unmark}
\DeclareMathOperator{\baseop}{base}
\DeclareMathOperator{\lia}{red}
\DeclareMathOperator{\sic}{ic}

\DeclareMathOperator{\efoop}{Th_\exists}
\DeclareMathOperator{\qffoop}{Th_{\mathit{qf}}}

\DeclareMathOperator{\ld}{ld}

\DeclareMathOperator{\offset}{off}

\DeclareMathOperator{\idx}{idx}
\DeclareMathOperator{\spec}{spec}
\DeclareMathOperator{\diag}{diag}
\DeclareMathOperator{\coeff}{coeff}
\DeclareMathOperator{\act}{actVar}
\DeclareMathOperator{\blockz}{zero}

\DeclareMathOperator{\satis}{cndAssg}
\DeclareMathOperator{\positive}{1}
\DeclareMathOperator{\negative}{-1}
\DeclareMathOperator{\nonzero}{\star}
\DeclareMathOperator{\stir}{stir}

\newcommand{\iverson}[1]{\left[ {#1} \right]}

\newcommand{\ASSIGN}[2]{\ensuremath{#1 \assign #2}}

\DeclareFontFamily{U}{lasy}{}
\DeclareFontShape{U}{lasy}{m}{n}{
	<-5.5> lasy5 <5.5-6.5> lasy6 <6.5-7.5> lasy7 <7.5-8.5> lasy8 <8.5-9.5> lasy9 <9.5-> lasy10
}{}

\newcommand{\qiff}{\quad\text{iff}\quad}

\newcommand{\WHILEDO}[2]{
	\SetStartEndCondition{\ }{}{}
	\SetAlgoBlockMarkers{}{}%
	\SetKwFor{While}{while}{\ do\ }{}
	\text{\lWhile{$(\, {#1} \,)$}{$\, {#2} \,$}}
	\SetStartEndCondition{ (}{)}{)}
	\SetAlgoBlockMarkers{}{\}}%
	\SetKwFor{While}{while}{\ do\ }{}}

\crefname{definition}{Def.}{Def.}
\crefname{example}{Ex.}{Ex.}
\crefname{counterexample}{Counterex.}{Counterex.}
\crefname{appendix}{App.}{App.}
\crefname{ex}{Ex.}{Ex.}
\crefname{theorem}{Thm.}{Thm.}
\crefname{lemma}{Lemma}{Lemmas}
\crefname{remark}{Rem.}{Rem.}
\crefname{section}{Sect.}{Sect.}
\crefname{subsection}{Sect.}{Sect.}
\crefname{subsubsection}{Sect.}{Sect.}
\crefname{line}{Line}{Lines}
\crefname{corollary}{Cor.}{Cor.}
\crefname{figure}{Fig.}{Fig.}
\crefname{enumi}{}{}
\crefname{algorithm}{Alg.}{Alg.}

\newcommand{\eat}[1]{}

%% file: abstract.tex
We consider the problem of proving termination for triangular weakly non-linear loops (\twn-loops) over some ring $\ring$ like $\ZZ$, $\QQ$, or $\RR$.
The guard of such a loop is an arbitrary quantifier-free Boolean formula over (possibly non-linear) polynomial inequations, and the body is a single assignment of the form $
	\begin{sbmatrix}
		x_1\\
		\ldots \\
		x_d
	\end{sbmatrix}
	\assign
	\begin{sbmatrix}
		c_1 \cdot x_1 + p_1\\
		\ldots \\
		c_d \cdot x_d + p_d
	\end{sbmatrix}
$ where each $x_i$ is a variable, $c_i \in \ring$, and each $p_i$ is a (possibly non-linear) polynomial over $\ring$ and the variables $x_{i+1},\ldots,x_{d}$.

We show that the question of termination can be reduced to the existential fragment of the first-order theory of $\ring$.
For loops over $\RR$, our reduction implies decidability of termination.
For loops over $\ZZ$ and $\QQ$, it proves semi-decidability of non-termination.

Furthermore, we present a transformation to convert certain non-\twn-loops into \twn-form.
Then the original loop terminates iff the transformed loop terminates over a specific subset of $\RR$, which can also be checked via our reduction.
Moreover, we formalize a technique to \emph{linearize} (the updates of) \twn-loops in our setting and analyze its complexity.
Based on these results, we prove complexity bounds for the termination problem of \twn-loops as well as \emph{tight} bounds for two important classes of loops which can \emph{always} be transformed into \twn-loops.

Finally, we show that there is an important class of linear loops where our decision procedure results in an \emph{efficient} procedure for termination analysis, i.e., where the parameterized complexity of deciding termination is \emph{polynomial}.

%% file: introduction.tex
\section{Introduction}
\label{work_1:sec:introduction}
\emph{Termination} is one of the most important properties of a program.
In this work, we study \emph{complete} approaches for analyzing termination of certain classes of polynomial loops.
Compared to incomplete techniques, such approaches have the advantage that they always yield a definite result.
In particular, we investigate \emph{decidability} of termination for our classes of loops.

In the following, we give a short overview on the contributions of our article and highlight how it extends our earlier conference paper \cite{sas}.
There are already several decidability results for termination of linear loops \cite{dblp:conf/cav/tiwari04,li14,li-witnesses,dblp:journals/fac/xiayzz11,dblp:conf/cav/braverman06,dblp:conf/soda/ouakninepw15,bozga14,cav19,dblp:conf/icalp/hosseinio019}, but only few results on the decidability of termination for certain forms of non-linear loops \cite{li16,xiaz10,li17,dblp:conf/concur/neumanno020,dblp:conf/iccsa/wusbz10}.
Moreover, these previous works only deal with loops whose guards only contain conjunctions, besides \cite{dblp:conf/concur/neumanno020} which is restricted to guards defining compact sets.
In this work, we regard possibly non-linear loops with arbitrary guards, i.e., they may also contain disjunctions and define non-compact sets.
More precisely, we consider so-called \twn-loops, where the update is mildly restricted to be ``\underline{t}riangular'' and ``\underline{w}eakly \underline{n}on-linear'' (see \cref{sec:preliminaries} for a formal definition).
We study such loops over $\ZZ$, $\QQ$, $\RA$ (the real algebraic numbers), and $\RR$, whereas existing decidability results for non-linear loops are restricted to loops over the reals.

Most techniques for proving termination of loops rely on \emph{polynomial ranking functions}, see, e.g., \cite{dblp:conf/vmcai/podelskir04,dblp:conf/cav/bradleyms05,dblp:journals/jacm/ben-amramg14,dblp:conf/sas/ben-amramdg19,dblp:conf/cav/ben-amramg17,rank}.
However, such ranking functions are only \emph{sound} for proving termination, i.e.,
in general, they cannot refute termination.
In contrast to ranking functions, we use the computability of \emph{closed forms}
for the iterated update of the loop (\cref{sec:closed}).
In this way, we can reduce termination of a loop to (in)validity of a certain formula.
This reduction, which is a generalization of our earlier results for linear loops over $\ZZ$ with conjunctive guards \cite{cav19}, is sound and complete, i.e., validity of the resulting formula proves non-termination, whereas invalidity implies termination.
Moreover, our reduction is computable.
Analogously to our earlier work for linear loops \cite{cav19}, our decidability results on termination then follow from existing results on the decidability of certain theories.
In this way, we show that termination of \twn-loops is decidable over $\RA$ and $\RR$, and non-termination is semi-decidable over $\ZZ$ and $\QQ$ (\cref{sec:deciding}).

In \cref{sec:transf}
we use concepts from algebra to enlarge the classes of loops to which our reduction is applicable.
This is done by transforming (certain) non-\twn loops into \twn-form without affecting their termination behavior (\cref{subsec:transf}).
In \cref{subsec:automorphisms}, we discuss for which loops our transformation is applicable.
In this way, we generalize our results to a broader class of polynomial loops (\cref{subsec:adaptions}).

In contrast to our earlier conference paper \cite{sas}, in \cref{subsec:linearizing} we formalize the technique of \cite{oliveira16} to linearize (the updates of) \twn-loops in our setting.
Using this formalization, we develop novel results on the complexity of linearization.

Afterwards, based on our decision procedure for termination in \cref{sec:deciding}, on the transformation of \cref{sec:transf}, and on our complexity results for linearization from \cref{subsec:linearizing}, we study the complexity of deciding termination in \cref{sec:complexity}.

In \cref{subsec:linear-update} we show that deciding termination of linear loops with rational spectrum over $\ZZ$, $\QQ$, $\RA$, and $\RR$ is \cc{Co-NP}-complete.
Moreover, we show that deciding termination of linear-update loops (where the update is linear but the guard may be non-linear) with real spectrum over $\RA$ and $\RR$ is \cc{$\forall \RR$}-complete.
Here, a loop has \emph{rational} or \emph{real spectrum}, respectively, if its update matrix has rational or real eigenvalues only, and \cc{$\forall \RR$} is the complexity class of problems which can be reduced to validity of a universally quantified formula of polynomial inequations over the reals.
We also analyze the complexity of deciding termination for arbitrary \twn-loops (with possibly non-linear updates) in \cref{subsec:arbitrary-twn}.
In our conference paper \cite[Thm.\ 45]{sas}, we had only analyzed this case for a \emph{bounded} number of variables.
In contrast, we now extend our analysis to the general case where the number of variables is \emph{not} restricted (\cref{thm:three_exptime}).
To this end, we need our new results from \cref{subsec:linearizing} on the complexity of linearizing \twn-loops.

Finally, in contrast to \cite{sas}, we identify a class of linear loops (\emph{uniform loops}) where termination can be interpreted as a parameterized problem which is de\-cidable in polynomial time when fixing such a parameter (\cref{subsec:uniform}).
Based on the transformation of \cref{sec:transf}, we show that the closed forms arising from uniform loops have a special structure.
Therefore, here (in)validity of the formula\linebreak
from \cref{sec:deciding} which encodes termination can be checked in polynomial time.

Related work is discussed in \cref{part_1:sec:related_work} and all missing proofs can be found in \cref{app:proofs}.
So the current paper extends \cite{sas} by the following new material:
\begin{itemize}
	\item \cref{subsec:linearizing} on the linearization of \twn-loops and its complexity.
	\item \cref{thm:three_exptime} on the complexity of deciding termination for arbitrary \twn-loops where the number of variables is \emph{not} restricted.
	\item \cref{subsec:uniform} on \emph{uniform loops} where the parameterized complexity of deciding termination is \emph{polynomial}.
	\item Several additional explanations and remarks.
\end{itemize}

%% file: preliminaries.tex
\section{Preliminaries}
\label{sec:preliminaries}
A \emph{(polynomial) loop over a ring} $\ring$ has the form in \cref{fig:loop}, where $\ZZ \leq \ring \leq \RA$ and $\leq$ denotes the subring relation.
Here, $\vec{x}$ is a vector of $d \geq 1$ pairwise different variables that range over $\ring$ and $\update \in \left(\ring[\vec{x}]\right)^{d}$ where $\ring[\vec{x}]$ is the set of polynomials over $\vec{x}$ with coefficients in $\ring$.
To improve readability, we use row- and column-vectors interchangeably.
The guard $\cond$ is an arbitrary propositional (i.e., quantifier-free) formula over the atoms $\{p \triangleright 0 \mid p \in \ring[\vec{x}], {\triangleright}\in {\{\geq,} {>\}\}}$.
We denote the set of all such formulas by $\QFFO{\ring}$.
In our setting, negation is syntactic sugar as, e.g., $\neg(p > 0)$ is equivalent to $-p \geq 0$.
So w.l.o.g.\ the guard (or \emph{condition}) $\cond$ of a loop is built from atoms, $\land$, and $\lor$.

\begin{figure}[t]
	\hspace*{-.4cm}
	\begin{minipage}[t]{0.4\linewidth}
		\vspace*{.34cm}
		\hspace*{.2cm}
	        \minialg{\linewidth}{}{
			$\WHILEDO{\cond}{
					{\ASSIGN{\vec{x}}{\update}}}$}
		\vspace*{-.17cm}
		\captionof{figure}{Polynomial Loop}
		\label{fig:loop}
	\end{minipage}
	\hspace*{.3cm}
	\begin{minipage}[t]{0.6\linewidth}
		\vspace*{0pt}
		\minialg{\linewidth}{}{
		$\WHILEDO{{x_1}+ x_2^{2}>0}{
			{\ASSIGN{
						\begin{bmatrix}
							x_1 \\
							x_2 \\
							x_3
						\end{bmatrix}
					}{
						\begin{bmatrix}
							x_1+x_2^2 \cdot x_3 \\
							x_2 - 2 \cdot x_3^2 \\
							x_3
						\end{bmatrix}
					}}}$}
		\vspace*{-.3cm}
		\captionof{figure}{Loop $\LL_{\mathit{ex}}$ over $\ZZ$}
		\label{fig:loop:triangular}
	\end{minipage}
	\vspace*{-.3cm}
\end{figure}

We require $\ring \leq \RA$ instead of $\ring \leq \RR$, as it is unclear how to represent transcendental numbers on computers.
However, in \cref{sec:deciding} we will see that the loops considered in this work terminate over $\RR$ iff they terminate over $\RA$.
Thus, our results immediately carry over to loops where the variables range over $\RR$.
Hence, we sometimes also consider loops over $\ring = \RR$.
However, even then we restrict ourselves to loops where all constants in $\cond$ and $\update$ are algebraic.

We often represent a loop as in \cref{fig:loop} by the tuple $(\cond, \update)$ of the \emph{condition} $\cond$ and the \emph{update} $\update= (\updateElem_1,\ldots,\updateElem_d)$.
Unless stated otherwise, $(\cond, \update)$ is always a loop over $\ring$ using the variables $\vec{x} = (x_1,\ldots,x_d)$ where $\ZZ \leq \ring \leq \RA$.
We use the following notions for certain classes of loops:\ A \emph{linear-update loop} has the form $(\cond, A \cdot \vec{x} + \vec{b})$, and it has \emph{rational} or \emph{real spectrum}, respectively, if $A$ has rational or real eigenvalues only.
Then all eigenvalues of $A$ are real algebraic numbers, since $\update \in \left(\RA[\vec{x}]\right)^{d}$ implies that all constants in $\update$ (and thus in $A$) are algebraic.
A \emph{linear loop} is a linear-update loop where $\cond$ is linear (i.e., its atoms are only linear inequations).
Here, ``linear'' refers to ``linear polynomials'', i.e., of degree at most $1$, so it includes \emph{affine} functions.
A \emph{conjunctive loop} is a loop $(\cond,\update)$ where $\cond$ does not contain disjunctions.

For any entity $s$, $s[x / t]$ results from $s$ by replacing all free occurrences of $x$ by $t$.
Similarly, if $\vec{x} = (x_1,\ldots,x_d)$ and $\vec{t} = (t_1,\ldots, t_d)$, then $s[\vec{x} / \vec{t}]$ results from $s$ by replacing all free occurrences of $x_i$ by $t_i$, for each $1 \leq i \leq d$.

Any vector of polynomials $\update \in \left(\ring[\vec{x}]\right)^d$ can also be regarded as a function $\update:\left(\ring[\vec{x}]\right)^d \to \left(\ring[\vec{x}]\right)^d$ where for any $\vec{p} \in \left(\ring[\vec{x}]\right)^d$, $\update(\vec{p}) = \update[\vec{x}/\vec{p}]$ results from \emph{apply\-ing}
the polynomials $\update$ to the polynomials $\vec{p}$.
Similarly, we can also apply a formula to polynomials $\vec{p} \in \left(\ring[\vec{x}]\right)^d$.
To this end, we define $\psi(\vec{p}) = \psi[\vec{x}/\vec{p}]$ for first-order formulas $\psi$ with the free variables $\vec{x}$.
As usual, function application associates to the left, i.e., $\update (\vec{b}) (\vec{p})$ stands for $(\update(\vec{b})) (\vec{p})$.
However, obviously $(\update(\vec{b})) (\vec{p}) =
	\update (\vec{b} (\vec{p}))$ since applying polynomials only means that one instantiates variables.

\cref{def:term} formalizes the intuitive notion of termination for a loop $(\cond, \update)$ and the related notion of \emph{eventual} termination \cite{dblp:conf/cav/braverman06,dblp:conf/cav/tiwari04}.
Here, $\update^n$ denotes the $n$-fold application of $\update$, i.e., $\update^0(\vec{e}) = \vec{e}$ and $\update^{n+1}(\vec{e}) = \update(\update^n(\vec{e}))$.

\begin{definition}[Termination]
	\label{def:term}
	Let $(\cond,\update)$ be a loop over $\ring$ and $\witness \in \ring^{d}$.
	If $\forall n \in \NN.\ \cond(\update^n(\witness))$ holds, then $\witness$ is a \emph{witness for non-termination}.
	If $(\cond, \update)$ does not have any witnesses for non-termination, it \emph{terminates} (over $\ring$).

	If $\update^{n_0}(\witness)$ is a witness for non-termination for some $n_0 \in \NN$, then $\witness$ is called a \emph{witness for eventual non-termination}.

	$\mathit{(E)NT}_{(\cond, \update)}$ denotes the set of witnesses for (eventual) non-termination of $(\cond, \update)$ and we define $T_{(\cond,\update)} = \ring^d \setminus \mathit{NT}_{(\cond,\update)}$.
\end{definition}

For any entity $s$, $\VV(s)$ is the set of all its free variables.
Given an assignment $\vec{x} \assign \update$, the relation ${\succ_{\update}} \in \VV(\update) \times \VV(\update)$ is the transitive closure of $\{(x_i,x_j) \mid i,j \in \{1,\ldots,d\}, i \neq j, x_j \in \VV(\updateElem_i)\}$, i.e., $x_i \succ_{\update} x_j$ means that $x_i$ depends on $x_j$.
For example, if $\update = (x_1 + x_2^2, x_2 + 1)$ then we have ${\succ_{\update}} = \{ (x_1,x_2) \}$.

Now we can introduce the class of \twn-loops.
A loop $(\cond,\update)$ is \emph{triangular} if ${\succ_{\update}}$ is well founded.
It is \emph{weakly non-linear} if there is no $1 \leq i \leq d$ such that $x_i$ occurs in a non-linear monomial of $\updateElem_i$, i.e., $\updateElem_i = c_i \cdot x_i + p_i$ where $c_i \in \ring$ and $p_i \in \ring[\vec{x}]$ does \emph{not} contain $x_i$.
A \emph{twn-loop} is \underline{t}riangular and \underline{w}eakly \underline{n}on-linear.
We call a loop \emph{non-nega\-tive} if it is weakly non-linear and the coefficient $c_i$ of the mono\-mial $x_i$ in $\updateElem_i$ is non-negative for all $1 \leq i \leq d$.
A \emph{tnn-loop} is \underline{t}riangular and \underline{n}on-\underline{n}egative, i.e., a \tnn-loop is a special form of a \twn-loop.

The restriction to triangular loops prohibits ``cyclic dependencies'' of variables (e.g., where the new values of $x_1$ and $x_2$ both depend on the old values of $x_1$ and $x_2$).
For example, the loop whose body consists of the assignment $(x_1,x_2) \assign (x_1 + x_2^2,\, x_2+1)$ is triangular since ${\succ} = \{(x_1,x_2)\}$ is well founded, whereas a loop with the body $(x_1,x_2) \assign (x_1 + x_2^2,\,x_1+1)$ is not triangular.

By the restriction to \twn-loops we can compute a closed form for the $n$-fold application of the update $\update$ by handling one variable after the other.
A vector $\closed$ of $d$ arithmetic expressions over $\vec{x}$ and a distinguished variable $n$ is a \emph{closed form} for $\update \in (\ring[\vec{x}])^d$ if $\closed[\vec{x}/\witness,n/n'] = \update^{n'}(\witness)$ for all $\witness \in \ring^d$ and $n' \in \NN$, i.e., both vectors of expressions evaluate to the same element of $\ring^d$.
Thus, $\closed = \update^n$.

Triangular loops are very common in practice.
For example, in \cite{frohn20}, $1511$ polynomial loops were extracted from the \emph{Termination Problems Data Base (TPDB)} \cite{tpdb}, the benchmark collection used at the annual \emph{Termination and Complexity Competition} \cite{termcomp}, and only $26$ of them were non-triangular.

A loop with the body $(x_1,x_2) \assign (x_1 +x_2^2,\, x_2+1)$ is weakly non-linear, while\linebreak
a loop with $(x_1,x_2) \assign (x_1 \cdot x_2,\, x_2+1)$ is not.
In particular, weak non-linearity excludes assignments like $x_1 \assign x_1^2$ that need exponential space, as $x_1$ grows doubly exponentially.
By permuting variables, the update of every \twn-loop can be transformed to the following form where $c_i \in \ring$ and $p_i \in \ring[x_{i+1},\ldots,x_{d}]$:
\[
	\mat{
		x_1\\
		\ldots \\
		x_d
	}
	\assign \mat{
		c_1 \cdot x_1 + p_1\\
		\ldots \\
		c_d \cdot x_d + p_d
	}
\]

\begin{example}
	\label{ex:running_init}
	Consider the loop $\LL_{\mathit{ex}}$ over the ring $\ZZ$ in \cref{fig:loop:triangular}.
	This loop is triangular since ${\succ_{\update}} = \{(x_1,x_2),(x_1,x_3),(x_2,x_3)\}$ is well founded.
	Moreover, it is weakly non-linear.
	Since the coefficient of $x_i$ is $1$ in the update of $x_i$ for all $1 \leq i \leq 3$, this loop is also non-negative, i.e., $\LL_{\mathit{ex}}$ is \tnn.
\end{example}

Our \twn-loops are a special case of \emph{solvable loops}.
\begin{definition}[Solvable Loops \cite{solvable-maps}]
	\label{def:solvable}
	A loop $(\cond,\update)$ is \emph{solvable} if there is a partitioning $\mathcal{J} = \{J_1,\ldots,J_k\}$ of $\{1,\ldots,d\}$ such that for each $1 \leq i \leq k$ we have $\update_{J_i} = A_i \cdot \vec{x}_{J_i} + \vec{p}_i$, where $\update_{J_i}$ is the vector of all $\updateElem_j$ with $j \in J_i$ (and $\vec{x}_{J_i}$ is defined analogously), $d_i = \vert{}J_i\vert{}$, $A_i \in \ring^{d_i \times d_i}$, and $\vec{p}_i \in (\ring[\vec{x}_{J_{i+1}}, \ldots, \vec{x}_{J_k}])^{d_i}$.
	The eigenvalues of a solvable loop are the union of the eigenvalues of all $A_i$.
\end{definition}
So solvable loops allow for blocks of variables with linear dependencies, and \twn-loops correspond to the case that all blocks have size $1$.
While our approach could be generalized to solvable loops with real eigenvalues, \cref{thm:solvable} (\cref{sec:transf}) shows that this generalization does not increase its applicability.

For our decidability results in \cref{sec:deciding}, we reduce termination to the \emph{existential fragment} $\EFO{\ring}$ of the first-order theory of $\ring$ (see, e.g., \cite{renegar,existsr}).
$\EFO{\ring}$ consists of all formulas $\exists \vec{y} \in \ring^k. \, \psi$ where $k \in \NN$ and the propositional formula $\psi$ is built from $\land$ and $\lor$ over the atoms $\{p \triangleright 0 \mid p \in \QQ[\vec{y}, \vec{z}],\ {\triangleright} \in \{{\geq},
	{>}\}\}$.
Here, $\vec{y}$ and $\vec{z}$ are pairwise disjoint vectors of variables.
The \emph{free} variables $\vec{z}$ range over $\RA$ and they are needed in \cref{sec:transf} to characterize subsets of real (algebraic) numbers by formulas.

The \emph{existential fragment of the first-order theory of $\ring$ and $\RA$} is the set $\EFO{\ring,\RA}$ of all formulas $\exists \vec{y}\,' \in \RA^{k'}, \vec{y} \in \ring^k.\, \psi$, with a propositional formula $\psi$ over $\{p \triangleright 0 \mid p \in \QQ[\vec{y}\,', \vec{y}, \vec{z}],\ {\triangleright} \in \{{\geq}, {>}\}\}$ where $k', k \in \NN$ and the variables $\vec{y}\,'$, $\vec{y}$, and $\vec{z}$ are pairwise disjoint.
A formula without free variables is \emph{closed}.

In the following, we also consider formulas over inequations $p \triangleright 0$ where $p$'s coefficients are from $\RA$ to be elements of $\EFO{\RA}$ (resp.\ $\EFO{\ring,\RA}$).
The reason is that real algebraic numbers are $\EFO{\RA}$-definable.

Validity of closed formulas from $\EFO{\ring}$ or $\EFO{\ring,\RA}$ is decidable if $\ring \in \{\RA,\RR\}$ and semi-decidable if $\ring \in \{\ZZ,\QQ\}$ \cite{tarski_decidability,cohen69}.
By undecidability of Hilbert's Tenth Problem \cite{mr0258744}, it is undecidable for $\ring = \ZZ$.
While validity of full first-order formulas (i.e., also containing universal quantifiers) over $\ring = \QQ$ is undecidable \cite{robinson49}, it is still open whether validity of closed formulas from $\EFO{\QQ}$ or $\EFO{\QQ,\RA}$ is decidable.
However, validity of \emph{linear} closed formulas from $\EFO{\ring}$ or $\EFO{\ring,\RA}$ is decidable for all $\ring \in \{\ZZ,\QQ,\RA,\RR\}$ \cite{dantzig,kantorovich,branchbound,gomorymilp}.
Here, a formula is \emph{linear} if it only contains atoms $p \triangleright 0$ where $p$ is linear.

%% file: closed.tex
\section{Reducing Termination of \twn-Loops to Termination of \tnn-Loops}
\label{sec:closed}
For analyzing termination of \twn-loops, we can restrict ourselves to \tnn-loops, as
\twn-loops can be (automatically) transformed into \tnn-loops via \emph{chaining}.

\begin{definition}[Chaining]
	\label{def:chaining}
	\emph{Chaining} a loop $(\cond, \update)$ yields $(\cond \land \cond(\update), \update(\update))$.
\end{definition}

So for example, chaining the loop $(x_1 > 0, -x_1)$ (i.e., ``\textbf{while} $(x_1>0)$ \textbf{do}\linebreak
$x_1 \assign -x_1$'') yields $(x_1>0 \land -x_1 > 0, x_1)$.
Analogous to \cite{cav19} where chaining was used for triangular linear loops, we obtain the following theorem.

\begin{theorem}[Soundness of Chaining]
	\label{thm:ptill}
	Let $(\cond,\update)$ be a \twn-loop on $\ring^d$.
	\begin{enumerate}
		\item[(a)] $(\cond \land \cond(\update), \update(\update))$ is \tnn, i.e., it is triangular and the coefficient of each $x_i$ in $\updateElem_i(\update)$ is non-negative.
		\item[(b)] $(\cond,\update)$ terminates on $\witness \in \ring^d$ iff $(\cond \land \cond(\update), \update(\update))$ terminates on $\witness$.
	\end{enumerate}
\end{theorem}
\makeproof{thm:ptill}{
	\begin{enumerate}
		\item[(a)] By weak non-linearity, $\updateElem_i = c_i \cdot x_i + p_i$ with $x_i \notin \VV(p_i)$ for all $1 \leq i \leq d$.
			Then
			\[
				\updateElem_i(\update) = c_i \cdot (c_i \cdot x_i + p_i) + p_i(\update) = c_i^2 \cdot x_i + c_i \cdot p_i + p_i (\update).
			\]
			Assume $x_i \in \VV(p_i(\update))$.
			As $x_i \notin \VV(p_i)$ by weak non-linearity, there is an $x_j \in \VV(p_i)$ with\linebreak
			$x_j \neq x_i$ and $x_i \in \VV(\updateElem_j)$, which implies $x_j \succ_{\update} x_i$.
			But $x_j \in \VV(p_i)$ also implies $x_i \succ_{\update} x_j$, which violates well-foundedness of $\succ_{\update}$, i.e., it contradicts the triangularity of $(\cond, \update)$.
			Hence, $c_i^2$ is the coefficient of $x_i$ in $\updateElem_i(\update)$.
			Since $c_i^2 \geq 0$, $(\cond \land \cond(\update), \update(\update))$ is non-negative.

			Note that $x_i \succ_{\update(\update)} x_j$ implies $x_j \in \VV(p_i)$ (then we also have $x_i \succ_{\update} x_j$) or it implies that there is an $x_k \in \VV(p_i)$ with $x_j \in \VV(\updateElem_k)$ (then we have $x_i \succ_{\update} x_k$ and $x_k \succeq_{\update} x_j$).
			So in both cases, $x_i \succ_{\update(\update)} x_j$ implies $x_i \succ_{\update} x_j$.
			Thus, ${\succ_{\update(\update)}} \subseteq {\succ_{\update}}$.
			As ${\succ_{\update}}$ is well founded, this means that ${\succ_{\update(\update)}}$ is well founded, too.
			Hence, $(\cond \land \cond(\update), \update(\update))$ is triangular.

		\item[(b)] Now we prove that $(\cond, \update)$ does not terminate on $\witness \in \ring^d$ iff $(\cond \land \cond(\update), \update(\update))$ does not terminate on $\witness$.

			\vspace*{-.8cm}
			\[
				\begin{array}{rl}
					\hspace*{4cm} & (\cond, \update) \text{ does not terminate on } \witness                                                      \\[-.1cm]
					\iff          & \forall n \in \NN.\ \cond(\update^n(\witness)) \hspace*{3.2cm}
					\text{(by \cref{def:term})}                                                                                                   \\[-.1cm]
					\iff          & \forall n \in \NN.\ \cond(\update^{2 \cdot n}(\witness)) \land \cond(\update^{2 \cdot n + 1}(\witness))       \\[-.1cm]
					\iff          & \forall n \in \NN.\ \cond(\update^{2 \cdot n}(\witness)) \land \cond(\update) (\update^{2 \cdot n}(\witness)) \\[-.1cm]
					\iff          & \forall n \in \NN.\ (\cond\land \cond(\update)) \, (\update(\update))^n (\witness)                            \\[-.1cm]
					\iff          & (\cond \land \cond(\update), \update(\update)) \text{ does not terminate on } \witness
				\end{array}
			\]
	\end{enumerate}
	\vspace*{-.6cm}
}

As chaining is clearly computable, we get the following corollary.

\begin{corollary}
	\label{coro:ptill}
	Termination of \twn-loops is reducible to termination of \tnn-loops.
\end{corollary}

It is well known that closed forms for \tnn-loops are computable (see, e.g., \cite{kincaid19,xu13})
since their bodies correspond to \emph{C-finite recurrences}, which are known to be solvable \cite{tetrahedron}.
The resulting closed forms may contain polynomial arithmetic and exponentiation w.r.t.\ $n$ (as, e.g., $x_1 \assign 2 \cdot x_1$ has the closed form $x_1 \cdot 2^n$) as well as certain piecewise defined functions.
For example, the closed form $x_1^{(n)}$ of $x_1 \assign 1$ is $x_1^{(0)} = x_1$ and $x_1^{(n)} = 1$ for all $n \in \NN$ with $n > 0$.

We represent closed forms using poly-exponential expressions \cite{cav19}, where ins\-tead of handling piecewise defined functions via disjunctions (as in \cite{kincaid19}), we\linebreak
simulate them via Iverson brackets.
For a formula $\psi$ over $n$, its \emph{Iverson bracket} $\iverson{\psi}:\NN \to \{0,1\}$ evaluates to $1$ iff $\psi$ is satisfied (i.e., $\iverson{\psi}(e) = 1$ if $\psi[n/e]$ holds and $\iverson{\psi}(e) = 0$, otherwise).
Later, Iverson brackets can be replaced by the constants 0 or 1, as we only use them for formulas $\psi$ that are constantly false or true for large enough values of $n$, see \Cref{sec:deciding}.
\emph{Poly-exponential expressions} are sums of arithmetic terms over the variables $\vec{x}$ and the additional designated variable $n$, where it is always clear which addend determines the asymptotic growth of the expression when increasing $n$.
This is crucial for our reducibility proof in \cref{sec:deciding}.
\cref{def:poly-exp} slightly generalizes the poly-exponential expressions from \cite[Def.\ 9]{cav19} by allowing arbitrary polynomials over $\vec{x}$ (instead of just linear expressions) as coefficients.
In the following, for any set $X\subseteq \RR$, any $k \in X$, and ${\triangleright} \in \{ \geq, > \}$, let $X_{\triangleright k} = \setcomp{x \in X}{x \triangleright k}$.

\begin{definition}[Poly-Exponential Expressions]
	\label{def:poly-exp}
	Let $\mathcal{C}$ be the set of all finite conjunctions over $\{n = c, n \neq c \mid c \in \NN\}$ where $n$ is a designated variable and let $\Q_{\ring} = \setcomp{\tfrac{r}{s}}{r \in \ring, s \in \ring_{> 0}}$ be the quotient field of $\ring$.
	Then the set of all \emph{poly-ex\-po\-nen\-tial expressions} with the variables $\vec{x}$ over $\ring$ is
	\[
		\mbox{$\PE{\ring}[\vec{x}] = \{ \sum\nolimits_{j=1}^\ell \iverson{\psi_j} \cdot \alpha_j \cdot n^{a_j}
				\cdot b_j^n \mid \ell, a_j \in \NN, \, \psi_j \in \mathcal{C}, \, \alpha_j \in \Q_{\ring}[\vec{x}], \, b_j \in \ring_{> 0} \}$}.
	\]
\end{definition}
An example for a poly-exponential expression over $\ZZ$ (with $\Q_{\ZZ} = \QQ$) is
\[
	\iverson{n \neq 0 \land n \neq 1} \cdot \left(\tfrac{1}{2} \cdot {x_1^2} + \tfrac{3}{4} \cdot {x_2} -1 \right) \cdot n^3 \cdot 3^n + \iverson{n = 1} \cdot ({x_1}-{x_2}).
\]

The restriction to \tnn-loops ensures that for the closed form $\closed$ of the update we indeed have $\closed \in \left(\PE{\ring}[\vec{x}]\right)^d$.
For example, for arbitrary matrices $A \in \RA^{d\times d}$, the update $\vec{x} \assign A \cdot \vec{x}$ is known to admit a closed form as in \cref{def:poly-exp} with complex $b_j$'s, whereas real numbers suffice for triangular matrices.
Moreover, non-negativity is required to ensure $b_j > 0$ (e.g., a non-\tnn loop with the update $x_1 \assign -x_1$ has the closed form $x_1 \cdot (-1)^n$).

\begin{example}
	\label{ex:running_closed}
	For $\LL_{\mathit{ex}}$ in \cref{fig:loop:triangular},
	the closed form is $\closed \in \left(\PE{\ZZ}[\vec{x}]\right)^3$ with
	\[
		\closed=
		\begin{bmatrix}
			\tfrac{4}{3}\cdot x_3^{5}\cdot {n}^{3}+ \left(-2\cdot x_3^{5}-2\cdot x_2\cdot x_3^{3} \right)\cdot {n}^{2}+ \left( x_2^{2}\cdot x_3 +\tfrac{2}{3}\cdot x_3^{5}+2\cdot x_2\cdot x_3^{3} \right)\cdot n+x_1 \\
			-2\cdot x_3^{2}\cdot n+x_2                                                                                                                                                                                \\
			x_3
		\end{bmatrix}
		.
	\]
\end{example}

%% file: deciding.tex
\section{Reducing Termination of \tnn-Loops to $\EFO{\ring}$}
\label{sec:deciding}
It is known that the bodies of \tnn-loops can be linearized \cite{oliveira16}, i.e., one can reduce termination of a \tnn-loop $(\cond,\update)$ to termination of a linear-update \tnn-loop $(\cond',\update')$, where $\cond'$ may be \emph{non}-linear.
See \cref{subsec:linearizing} for a discussion of linearization and novel results on the linearization procedure.
Moreover, \cite{xiaz10,dblp:conf/iccsa/wusbz10} showed decidability of termination for certain classes of conjunctive linear-update loops over $\RR$, which cover conjunctive linear-update \tnn-loops.
So, by combining the results of \cite{oliveira16} and \cite{xiaz10,dblp:conf/iccsa/wusbz10}, one can conclude that termination of \emph{conjunctive} \tnn-loops over $\RR$ is decidable.

In contrast, we now present a reduction of termination of \tnn-loops to $\EFO{\ring}$ which applies to \tnn-loops over \emph{any} ring $\ZZ \leq \ring \leq \RR$ and which can also handle \emph{disjunctions} in the loop condition.
Moreover, our reduction yields tight complexity results on termination of linear loops over $\ZZ$, $\QQ$, $\RA$, and $\RR$, and on termination of linear-update loops over $\RA$ and $\RR$ (see \cref{subsec:linear-update}).

Similar to \cite{cav19}, our reduction exploits that for \tnn-loops $(\cond,\update)$ there is a closed form $\closed$ for $\update$ with $\closed \in\left(\PE{\ring}[\vec{x}]\right)^d$.
However, in \cite{cav19}
we only considered conjunctive linear loops over $\ZZ$.
In contrast, we now analyze loops over $\ring$ for any $\ZZ \leq \ring \leq \RA$ and allow \emph{non}-linearity and arbitrary propositional formulas in the condition.
Thus, the correctness proofs differ substantially from \cite{cav19}.

More precisely, we show that there is a function with the following specification that is computable in polynomial time:
\begin{equation}
	\label{eq:spec}
	\begin{array}{rl}
		\text{Input}:  & \quad \text{a \tnn-loop $(\cond,\update)$ over $\ring$ with closed form $\closed \in\left(\PE{\ring}[\vec{x}]\right)^d$ } \\[-.15cm]
		\text{Result}: & \quad \text{a closed formula $\chi \in \EFO{\ring}$ such that}                                                            \\[-.15cm]
		               & \quad \text{$\chi$ is valid iff $(\cond,\update)$ does not terminate on $\ring^d$}
	\end{array}
\end{equation}

We use the concept of \emph{eventual non-termination}, i.e., the condition of the loop may be violated finitely often, see \cref{def:term}.
Clearly, $(\cond,\update)$ is non-terminating iff it is eventually non-terminating \cite{dblp:conf/soda/ouakninepw15}.
The formula $\chi$ in \eqref{eq:spec} will encode the existence of a witness for eventual non-termination.
By the definition of $\closed$, eventual non-termination of $(\cond,\update)$ on $\ring^d$ is equivalent to validity of
\begin{equation}
	\label{eq:term-closed-form}
	\exists \vec{x} \in \ring^d, \, n_0 \in \NN.\ \forall n \in \NN_{>n_0}.\ \cond(\closed).
\end{equation}

\begin{example}
	\label{ex:running_ent}
	Continuing \cref{ex:running_closed}, $\LL_{\mathit{ex}}$ is eventually non-terminating over $\ZZ$ iff there is a corresponding witness $\witness \in \ZZ^3$, i.e., iff
	\begin{align}
		\label{ex:eventual_nonterm}
		                                                   & \exists x_1,x_2,x_3 \in \ZZ, \, n_0 \in \NN.\ \forall n \in \NN_{>n_0}.
		\; p>0                                                                                                                                                                                      \\
		\hspace*{-.3cm} \text{is valid where} \quad p = \; & (x_1 + x_2^{2} ) + ( x_2^{2}\cdot x_3 + \tfrac{2}{3}\cdot x_3^{5} +2\cdot x_2\cdot x_3^{3} -4\cdot x_2\cdot x_3^{2} )\cdot n \nonumber \\[-.1cm]
		                                                   & {} + ( -2\cdot x_3^{5}-2\cdot x_2\cdot x_3^{3}+4\cdot x_3^{4} )\cdot {n}^{2} + (\tfrac{4}{3}\cdot x_3^{5})\cdot {n}^{3}.\nonumber
	\end{align}
\end{example}

Let $\closednorm$ be like $\closed$, but each factor $\iverson{\psi}$ is replaced by $0$ if it contains an equation and by $1$, otherwise.
The reason is that for large enough $n$, equations in $\psi$ become $\FALSE$ and negated equations become $\TRUE$.
So \eqref{eq:term-closed-form} is equivalent to
\begin{equation}
	\label{eq:normalized}
	\exists \vec{x} \in \ring^d, \, n_0 \in \NN.\ \forall n \in \NN_{>n_0}.\ \cond(\closednorm).
\end{equation}
In this way, we obtain \emph{normalized} poly-exponential expressions.

\begin{definition}[Normalized $\PE{}$s]
	We call $p \in \PE{\ring}[\vec{x}]$ \emph{normalized} if it is in
	\[
		\mbox{$\PEN{\ring}[\vec{x}] = \{\sum\nolimits_{j=1}^\ell \alpha_j \cdot n^{a_j} \cdot b_j^n \mid \ell, a_j \in \NN, \; \alpha_j \in \Q_{\ring}[\vec{x}], \; b_j \in \ring_{> 0} \}$,}
	\]
	where w.l.o.g.\ $(b_i,a_i) \neq (b_j,a_j)$ if $i \neq j$.
	We define $\PEN{\ring} = \PEN{\ring}[\emptyset]$.
\end{definition}

As $\cond$ is a propositional formula over $\ring[\vec{x}]$-in\-equa\-tions, $\cond(\closednorm)$ is a propositional formula over $\PEN{\ring}[\vec{x}]$-inequations.
By \eqref{eq:normalized}, we need to check whether $\exists \vec{x} \in \ring^d.\ \cond(\closednorm)$ is valid for large enough $n$.
To this end, we will examine the dominant terms in the inequations of $\cond(\closednorm)$.

\begin{definition}[Asymptotic Domination \protect{\cite{landau}}]
	A function $g: \NN \to \RR$ \emph{dominates} a function $f: \NN \to \RR$ asymptotically ($f \in o(g)$) if for all $m>0$ there is an $n_0 \in \NN$ such that\footnote{Our definition is slightly more general than the original definition of \cite{landau}
		(which requires $\lim_{n \to \infty} \frac{f(n)}{g(n)} = 0$), but both definitions are equivalent if $g(n)$ is positive for large enough $n$.}
	$\vert{}f(n)\vert{} < m \cdot \vert{}g(n)\vert{}$ for all $n \in \NN_{>n_0}$.
\end{definition}
Now we can state the following lemma which generalizes \cite[Lemma 24]{cav19}.

\begin{lemma}
	\label{lem:domination}
	Let $b_1,b_2 \in \ring_{> 0}$ and $a_1, a_2 \in \NN$.
	If $(b_2, a_2) >_{lex} (b_1, a_1)$, then $n^{a_1} \cdot b_1^n \in o(n^{a_2} \cdot b_2^n)$, where $(b_2,a_2) >_{lex} (b_1,a_1)$ iff $b_2 > b_1$ or $b_2 = b_1 \land a_2 > a_1$.
\end{lemma}
\makeproof{lem:domination}{
	Recall that for $f,g: \NN \to \RR$, $f(n) \in o(g(n))$ means
	\[
		\forall m > 0.\ \exists n_0 \in \NN.\ \forall n \in \NN_{> n_0}.\ \vert{}f(n)\vert{} < m \cdot \vert{}g(n)\vert{}.
	\]
	First consider the case $b_2 > b_1$.
	We have $b_2^n = b_1^n \cdot (\tfrac{b_2}{b_1})^n$, where $\frac{b_2}{b_1} > 1$.
	As $n^{a_1} \in o((\tfrac{b_2}{b_1})^n)$, we obtain $n^{a_1} \cdot b_1^n \in o((\tfrac{b_2}{b_1})^n \cdot b_1^n) = o(b_2^n) \subseteq o(n^{a_2} \cdot b_2^n)$, i.e., $n^{a_1} \cdot b_1^n \in o(n^{a_2} \cdot b_2^n)$.

	Now consider the case $b_2 = b_1$ and $a_2 > a_1$.
	Then $n^{a_1} \cdot b_1^n \in o(n^{a_2} \cdot b_2^n)$ trivially holds.
}

In the following, let $p\geq 0$ or $p > 0$ occur in $\cond(\closednorm)$.
We can order the coefficients of $p$ according to the asymptotic growth of their addends w.r.t.\ $n$.
\begin{definition}[Ordering Coefficients]
	\label{def:marked}
	\emph{Marked coefficients} are of the form $\alpha^{(b,a)}$ where $\alpha \in \Q_{\ring}[\vec{x}], b \in \ring_{>0}$, and $a \in \NN$.
	We define $\unmarkop(\alpha^{(b,a)}) = \alpha$ and $\alpha_2^{(b_2,a_2)} \succc \alpha_1^{(b_1,a_1)}$ if $(b_2,a_2) >_{lex}
		(b_1,a_1)$.
	Let $p = \sum_{j=1}^\ell \alpha_j \cdot n^{a_j} \cdot b_j^n \in \PEN{\ring}[\vec{x}]$, where $\alpha_j \neq 0$ for all $1 \leq j \leq \ell$.
	Then the marked coefficients of $p$ are
	\[
		\coeffs{p} =
		\begin{cases}
			\{0^{(1,0)}\},                                    & \text{if } \ell = 0 \\
			\{\alpha_j^{(b_j,a_j)} \mid 1 \leq j \leq \ell\}, & \text{otherwise.}
		\end{cases}
	\]
\end{definition}

\begin{example}
	\label{ex:running_coeffs}
	Continuing \cref{ex:running_ent}, $\coeffs{p}$ is $\{ \alpha_1^{(1,0)}, \alpha_2^{(1,1)}, \alpha_3^{(1,2)}, \alpha_4^{(1,3)} \}$ where
	\[\resizebox{\textwidth}{!}{$
		\begin{array}{l@{\,}c@{\,}l@{\quad}l@{\,}c@{\,}l}
			\alpha_1 & = & x_1 + x_2^{2}                                          & \alpha_2 & = & x_2^{2}\cdot x_3 + \tfrac{2}{3}\cdot x_3^{5}+ 2\cdot x_2\cdot x_3^{3} -4\cdot x_2\cdot x_3^{2} \\[-.1cm]
			\alpha_3 & = & -2\cdot x_3^{5}-2\cdot x_2\cdot x_3^{3}+4\cdot x_3^{4} & \alpha_4 & = & \tfrac{4}{3}\cdot x_3^{5}
		\end{array}
	$}\]
\end{example}

Note that $p(\witness) \in \PEN{\ring}$ for any $\witness \in \ring^d$, i.e., the only variable in $p(\witness)$ is $n$.
Now the $\succc$-maximal addend determines the asymptotic growth of $p(\witness)$:
\begin{equation}
	\label{cor:theta}
	o(p(\witness)) = o(k \cdot n^a \cdot b^n) \quad \text{where } k^{(b,a)} = \maxsucc{\coeffs{p(\witness)}}.
\end{equation}
\makeproof{cor:theta}{
If $p(\witness) = 0$, then $k = 0$ by \cref{def:marked} and hence $o(p(\witness)) = o(k \cdot n^a \cdot b^n) = o(0)$.
Otherwise, $p(\witness)$ has the form
\[
	\mbox{$k \cdot n^a \cdot b^n + \sum_{i=1}^m k_i \cdot n^{a_i} \cdot b_i^n$}
\]
for $k \neq 0$ and $m \geq 0$.
We have $k_i^{(b_i,a_i)} \in \coeffs{p(\witness)}$ and hence $(b,a) >_{lex} (b_i,a_i)$ for all $1 \leq i \leq m$.
Thus, \cref{lem:domination} implies $n^{a_i} \cdot b_i^n \in o(n^a \cdot b^n)$ and we get
\[
	\begin{array}{l}
		o(p(\witness)) = o\left(k \cdot n^a \cdot b^n + \sum_{i=1}^m k_i \cdot n^{a_i} \cdot b_i^n\right) = o(n^a \cdot b^n)= o(k \cdot n^a \cdot b^n).
	\end{array}
\]

\vspace*{-.6cm}

}

\vspace*{-.3cm}

Note that \eqref{cor:theta} would be incorrect for the case $k = 0$ if we replaced $o(p(\witness)) = o(k \cdot n^a \cdot b^n)$ with $o(p(\witness)) = o(n^a \cdot b^n)$ as $o(0) = \emptyset \neq o(1)$.
Obviously, \eqref{cor:theta} implies
\begin{equation}
	\label{eq:asym}
	\exists n_0 \in \NN.\ \forall n \in \NN_{> n_0}.\ \sign{p(\witness)} = \sign{k},
\end{equation}
where $\sign{0} = 0$, $\sign{k} = 1$ if $k > 0$, and $\sign{k} = -1$ if $k < 0$.
This allows us to reduce eventual non-termination to $\EFO{\ring}$ if $\cond$ is an atom.
In the following, let $\coeffs{p} = \{\alpha_1^{(b_1,a_1)}, \ldots, \alpha^{(b_{\ell},a_{\ell})}_{\ell}\}$, where $\alpha^{(b_i,a_i)}_i \preccc \alpha^{(b_{j},a_{j})}_{j}$ for all $1 \leq i < j \leq \ell$. Then we define
\begin{equation}
	\label{lia-def}
	\begin{array}{l@{\qquad}r@{\;\;}c@{\;\;}l}
		           & \lia(p > 0)    & = & \bigvee_{i=1}^\ell (\alpha_i > 0 \land \bigwedge_{j=i+1}^{\ell} \alpha_j = 0) \\
		\text{and} & \lia(p \geq 0) & = & \lia(p>0) \vee \bigwedge_{i=1}^{\ell} \alpha_i = 0.
	\end{array}
\end{equation}
\makeproof{eq:asym}{
	If $k = 0$, the claim is trivial, so assume $k \neq 0$, i.e., $p(\witness) = k \cdot b^n \cdot n^a + p'$ for some $p' \in \PEN{\ring}$.
	By \cref{lem:domination}
	we have
	\[
		\begin{array}{ll}
			         & p' \in o(k \cdot b^n \cdot n^a)                                                                                                  \\[-.1cm]
			\iff     & \forall m > 0.\ \exists n_0 \in \NN.\ \forall n \in \NN_{> n_0}.\ \vert{}p'\vert{} < m \cdot \vert{}k \cdot b^n \cdot n^a\vert{} \\[-.1cm]
			\implies & \exists n_0 \in \NN.\ \forall n \in \NN_{> n_0}.\ \vert{}p'\vert{} < \vert{}k \cdot b^n \cdot n^a\vert{}.
		\end{array}
	\]
	Assume $k > 0$.
	Then

	\vspace*{-.3cm}

	\[
		\begin{array}{ll}
			         & \exists n_0 \in \NN.\ \forall n \in \NN_{> n_0}.\ \vert{}p'\vert{} < \vert{}k \cdot b^n \cdot n^a\vert{} \\[-.1cm]
			\implies & \exists n_0 \in \NN.\ \forall n \in \NN_{> n_0}.\ -p' < \vert{}k \cdot b^n \cdot n^a\vert{}              \\[-.1cm]
			\iff     & \exists n_0 \in \NN.\ \forall n \in \NN_{> n_0}.\ -p' < k \cdot b^n \cdot n^a                            \\[-.1cm]
			\iff     & \exists n_0 \in \NN.\ \forall n \in \NN_{> n_0}.\ 0 < k \cdot b^n \cdot n^a + p'                         \\[-.1cm]
			\iff     & \exists n_0 \in \NN.\ \forall n \in \NN_{> n_0}.\ 0 < p(\witness)                                        \\[-.1cm]
			\iff     & \exists n_0 \in \NN.\ \forall n \in \NN_{> n_0}.\ \sign{p(\witness)} = \sign{k}.
		\end{array}
	\]
	If $k < 0$, then

	\vspace*{-.3cm}

	\[
		\begin{array}{ll}
			         & \exists n_0 \in \NN.\ \forall n \in \NN_{> n_0}.\ \vert{}p'\vert{} < \vert{}k \cdot b^n \cdot n^a\vert{} \\[-.1cm]
			\implies & \exists n_0 \in \NN.\ \forall n \in \NN_{> n_0}.\ p' < \vert{}k \cdot b^n \cdot n^a\vert{}               \\[-.1cm]
			\iff     & \exists n_0 \in \NN.\ \forall n \in \NN_{> n_0}.\ p' < -k \cdot b^n \cdot n^a                            \\[-.1cm]
			\iff     & \exists n_0 \in \NN.\ \forall n \in \NN_{> n_0}.\ k \cdot b^n \cdot n^a + p' < 0                         \\[-.1cm]
			\iff     & \exists n_0 \in \NN.\ \forall n \in \NN_{> n_0}.\ p(\witness) < 0                                        \\[-.1cm]
			\iff     & \exists n_0 \in \NN.\ \forall n \in \NN_{> n_0}.\ \sign{p(\witness)} = \sign{k}.
		\end{array}
	\]

	\vspace*{-.5cm}
}

\begin{lemma}
	\label{lem:eventual_positiveness}
	Given $p \in \PEN{\ring}[\vec{x}]$ and ${\triangleright} \in \{{\geq}, {>}\}$, validity of
	\begin{equation}
		\label{eq:decidable}
		\begin{array}{l}
			\exists \vec{x} \in \ring^d, \, n_0 \in \NN.\ \forall n \in \NN_{>n_0}.\ p \triangleright 0
		\end{array}
	\end{equation}
	can be reduced	to validity of the closed formula $\exists \vec{x} \in \ring^d.\ \lia(p \triangleright 0)$ from $\EFO{\ring}$.
	This reduction takes polynomially many steps in the size of $p$.
\end{lemma}
\makeproof{lem:eventual_positiveness}{We have $p \in \PEN{\ring}[\vec{x}]$, so $p(\witness) \in \PEN{\ring}$ for any $\witness \in \ring^d$.
Hence,
\[
	\exists n_0\!\in\!\NN. \forall n\!\in\!\NN_{>n_0}. p(\witness) \, \triangleright \, 0 \qiff \unmarkop(\maxsucc{\coeffs{p(\witness)}}) \, \triangleright \, 0 \ \text{(by \eqref{eq:asym}).}
\]

Let $\coeffs{p} = \{\alpha_1^{(b_1,a_1)} ,\ldots,\alpha^{(b_{\ell},a_{\ell})}_{\ell}\}$, where $\alpha^{(b_i,a_i)}_i \preccc \alpha^{(b_{j},a_{j})}_{j}$ for all $1 \leq i < j \leq \ell$.
If $p(\witness) = 0$, then $\alpha_1(\witness) = \ldots = \alpha_{\ell}(\witness) = 0$ and thus $\coeffs{p(\witness)} = \{ 0^{(1,0)} \}$\linebreak
and $\unmarkop(\maxsucc{\coeffs{p(\witness)}}) = 0$.
Otherwise, there is an $1 \leq i \leq \ell$ with
\[
	\unmarkop(\maxsucc{\coeffs{p(\witness)}}) = \alpha_i(\witness) \neq 0 \text{ and } \alpha_j(\witness) = 0 \text{ for all } i+1 \leq j \leq \ell.
\]
So when defining $\lia(p > 0)$ and $\lia(p \geq 0)$ as in \eqref{lia-def},
we obviously have
\[
	\unmarkop(\maxsucc{\coeffs{p(\witness)}}) \triangleright 0 \qiff (\lia(p \triangleright 0)) \, (\witness) \text{ holds}.
\]
Hence, \eqref{eq:decidable} is equivalent to
\begin{equation}
	\label{eq:atoms}
	\exists \vec{x} \in \ring^d.\ \lia(p \triangleright 0).
\end{equation}

The time needed to compute and sort $\coeffs{p}$ is polynomial.
Furthermore, $\lia(p\triangleright 0)$ is a disjunction of at most $\ell + 1$ subformulas, where each subformula is a conjunction of at most $\ell$ (in-)equations over $\coeffs{p}$.
Thus, the time needed to compute $\lia(p\triangleright 0)$ resp.\ \eqref{eq:atoms} is polynomial in the size of $p$.}
To gain intuition for the formula $\lia(p \triangleright 0)$, note that by \eqref{eq:asym}, we have $p(\witness) > 0$ for large enough values of $n$ iff the coefficient of the asymptotically fastest growing addend $\alpha(\witness) \cdot n^a \cdot b^n$ that does not vanish (i.e., where $\alpha(\witness) \neq 0$) is \emph{positive}.
Similarly, we have $p(\witness) < 0$ for large enough $n$ iff $\alpha(\witness) < 0$.
If \emph{all} addends of $p$ vanish when instantiating $\vec{x}$ with $\witness$, then $p(\witness) = 0$.
In other words, \eqref{eq:decidable} holds iff there is an $\witness \in \ring^d$ such that $\unmark{\maxsucc{\coeffs{p(\witness)}}} \, \triangleright \, 0$.
The formula $\lia(p\triangleright 0)$ expresses the latter in $\EFO{\ring}$.

\begin{example}
	\label{ex:running_ent_finish}
	We continue \cref{ex:running_coeffs}.
	By the construction in \cref{lem:eventual_positiveness}, \eqref{ex:eventual_nonterm} is valid iff $\exists x_1,x_2,x_3 \in \ZZ.\ \lia(p > 0)$ is valid, where $\lia(p > 0)$ is
	\[
		\begin{array}{r@{\;\;}l@{\;\;}c@{\;\;}l}
			     & \left(\alpha_1 > 0 \land \alpha_2 = 0 \land \alpha_3 = 0 \land \alpha_4 = 0\right) & \lor & \left( \alpha_2 > 0 \land \alpha_3 = 0 \land \alpha_4 = 0\right) \\[-.1cm]
			\lor & \left(\alpha_3 > 0 \land \alpha_4 = 0 \right)                                      & \lor & \alpha_4 > 0.
		\end{array}
	\]
	For example, $[x_1/-4, x_2/2,x_3/1]$ satisfies $\alpha_4 >0$ as $\left(\frac{4}{3}
		\cdot x_3^5\right)[x_1/-4, x_2/2,\linebreak
			x_3/1] > 0$.
	Thus, $(-4,2,1)$ witnesses eventual non-termination of $\LL_{\mathit{ex}}$ over $\ZZ$.
\end{example}

Now we lift our reduction to propositional formulas.
To handle disjunctions, the proof of \cref{thm:eventual_positiveness} exploits the crucial insight that a \tnn-loop $(\cond \lor \cond', \update)$ terminates iff $(\cond, \update)$ and $(\cond', \update)$ terminate, which is not true in general (as, e.g., witnessed by the loop $(x_1>0 \lor -x_1>0, -x_1)$).
In the following, the formula $\lia(\xi)$ results from $\xi$ by replacing each atom $p \triangleright 0$ in $\xi$ by $\lia(p \triangleright 0)$.

\begin{theorem}[Reducing Eventual Non-Termination]
	\label{thm:eventual_positiveness}
	For a propositio\-nal formula $\xi$ over the atoms $\{p \triangleright 0 \mid p \in \PEN{\ring}[\vec{x}], \triangleright \in \{\geq, >\}\}$, validity of
	\begin{equation}
		\label{eq:decidable_full}
		\exists \vec{x} \in \ring^d, n_0 \in \NN.\ \forall n \in \NN_{>n_0}.\ \xi
	\end{equation}
	can be reduced to validity of the closed formula $\exists \vec{x} \in \ring^d.\lia(\xi) \in \EFO{\ring}$.
	This reduction takes polynomially many steps in the size of $\xi$.
\end{theorem}
\makeproof{thm:eventual_positiveness}{
	We have to prove
	\begin{equation}
		\label{eq:main_equivalence_fundamental}
		\eqref{eq:decidable_full} \iff \exists \vec{x} \in \ring^d.\ \lia(\xi)
	\end{equation}
	where $\lia(\xi)$ results from replacing each atom $p \triangleright 0$ in $\xi$ by $\lia(p \triangleright 0)$.
	Since each $\lia(p \triangleright 0)$ can be computed in polynomial time due to \cref{lem:eventual_positiveness}, the computation of the formula ``$\exists \vec{x} \in \ring^d.\ \lia(\xi)$'' clearly works in polynomial time, too.

	To prove \eqref{eq:main_equivalence_fundamental}, we introduce the notion of a \emph{fundamental set}.
	Let $p_1 \triangleright_1 0, \ldots, p_{k}\triangleright_{k} 0$ denote the atoms in $\xi$.
	We call a subset $I \subseteq\{1,\ldots, k\}$ \emph{fundamental} if $\bigwedge_{i \in I}
		p_i \triangleright_i 0\implies \xi$. Recall that w.l.o.g., we can assume that $\xi$ does not contain any connectives except $\land$ and $\lor$. Thus, whenever $\xi \neq \FALSE$, the formula $\xi$ must have fundamental sets. Clearly, we have
	\[
		\mbox{$\exists \vec{x} \in \ring^d.\ \lia(\xi) \iff \exists \text{ fundamental set } I.\; \exists \vec{x} \in \ring^d.\ \bigwedge\nolimits_{i \in I} \lia(p_i\triangleright_i 0)$.}
	\]
	Thus, to prove \eqref{eq:main_equivalence_fundamental}, it suffices to show the following:
	\begin{equation}
		\label{eq:equivalence_fundamental}
		\mbox{$\eqref{eq:decidable_full} \iff \exists \text{ fundamental set } I.\; \exists \vec{x} \in \ring^d.\ \bigwedge\nolimits_{i \in I} \lia(p_i \triangleright_i 0)$.}
	\end{equation}

	For the ``${\impliedby}$''-direction, assume there is such a fundamental set and $\vec{e} \in \ring^d$, i.e.,
	\[
		\mbox{$\bigwedge\nolimits_{i \in I} \lia(p_i\triangleright_i 0)(\vec{e})$}
	\]
	is valid.
	Then as in the proof of \cref{lem:eventual_positiveness}, for each $i \in I$, there is an $n_i \in \NN$ such that
	\[
		\mbox{$\exists \vec{x} \in \ring^d.\ \forall n \in \NN_{>n_i}.\ p_i \triangleright_i 0$.}
	\]
	As $I$ is finite, $n_{\max} = \max\setcomp{n_i}{i \in I}$ exists.
	Hence, we get
	\[
		\mbox{$\exists \vec{x} \in \ring^d.\ \forall n \in \NN_{>n_{max}}.\ \bigwedge\nolimits_{i \in I} p_i \triangleright_i 0$.}
	\]
	Since $I$ is fundamental, this implies \eqref{eq:decidable_full}.

	For the ``${\implies}$''-direction, assume \eqref{eq:decidable_full}.
	Then there is an $\witness \in \ring^d$ and an $n_0 \in \NN$ such that for each $n \in \NN_{>n_0}$, there is a fundamental set $I_n$ such that $\bigwedge_{i \in I_n} p_i(\witness) \triangleright_i 0$ holds.
	As there are only finitely many fundamental sets, there is some fundamental set $I$ that occurs infinitely often in $(I_n)_{n \in \NN_{>n_0}}$.
	Hence we get
	\begin{equation}
		\label{eq:infty}
		\mbox{$\exists n_0 \in \NN.\ \exists^\infty n \in \NN_{>n_0}.\ \bigwedge\nolimits_{i \in I} p_i(\witness) \triangleright_i 0$.}
	\end{equation}
	By definition of poly-exponential expressions, each $p_i(\witness)$ is weakly monotonic in $n$ for large enough $n$.
	Thus, \eqref{eq:infty} implies\footnote{This corresponds to the observation that if the loop condition is a disjunction (and hence also $\xi$ is a disjunction of the form $\xi_1 \vee \xi_2$), then non-termination of the original loop implies non-termination of one of the loops where instead of the disjunction one only takes one of the disjuncts as the loop guard.
		The reason is that for every fundamental set $I$ and every $\vec{e} \in \ring^d$, $\bigwedge_{i \in I} p_i(\witness) \triangleright_i 0 \implies (\xi_1(\witness) \vee \xi_2(\witness))$ implies $\bigwedge_{i \in I} p_i(\witness) \triangleright_i 0 \implies \xi_1(\witness)$ or $\bigwedge_{i \in I} p_i(\witness) \triangleright_i 0 \implies \xi_2(\witness)$.
		The restriction to \tnn-loops implies that the closed forms are poly-exponential expressions and hence, that the $p_i(\witness)$ are weakly monotonic in $n$ for large enough $n$.
		Therefore, the above argumentation in the proof shows that there is a fundamental set $I$ such that $\bigwedge_{i \in I} p_i(\witness) \triangleright_i 0$ holds for all large enough $n$ and thus, for some $j \in \{1,2\}$, $\xi_j(\witness)$ holds for all large enough $n$ as well.
		Hence, the loop would also be non-terminating if one only takes the corresponding disjunct as the loop guard.}
	\[
		\mbox{$\exists n_0 \in \NN.\ \forall n \in \NN_{>n_0}.\ \bigwedge\nolimits_{i \in I} p_i(\witness) \triangleright_i 0$.}
	\]
	As $\witness \in \ring^d$, there is a fundamental set $I$ such that $\exists \vec{x} \in \ring^d.\ \bigwedge_{i \in I} \lia(p_i\triangleright_i 0)$ holds.
}
The time needed to compute the formula \eqref{eq:decidable_full} is polynomial in the sum of the sizes of all poly-exponential expressions in $\xi$.
So the function \eqref{eq:spec} is computable in polynomial time w.r.t.\ the size of its input: $\closednorm$ can clearly be computed in polynomial time from $\closed$ and we can then apply \cref{thm:eventual_positiveness} to $\cond(\closednorm)$.
Combining \cref{coro:ptill}, \eqref{eq:normalized}, and \cref{thm:eventual_positiveness} leads to the main result of this section.
\begin{theorem}[Reducing Termination]
	\label{thm:decidable}
	Termination of \tnn-loops (resp.\ \twn-loops) on $\ring^d$ is reducible to $\EFO{\ring}$.
\end{theorem}
\makeproof{thm:decidable}{
	By \cref{coro:ptill}, termination of \twn-loops is reducible to termination of \tnn-loops.
	Given a \tnn-loop $(\cond,\update)$, we obtain $\closednorm \in \left(\PEN{\ring}[\vec{x}]\right)^{d}$ such that $(\cond,\update)$ is (eventually) non-terminating iff \eqref{eq:normalized} holds, where $\cond$ is a propositional formula over the atoms $\{\alpha \geq 0, \alpha > 0 \mid \alpha \in \ring[\vec{x}]\}$.
	Hence, $\cond(\closednorm)$ is a propositional formula over the atoms $\{p \triangleright 0 \mid p \in \PEN{\ring}[\vec{x}], {\triangleright} \in \{ \geq, > \}\}$.
	Thus, by \cref{thm:eventual_positiveness}, validity of \eqref{eq:normalized}
	resp.\ \eqref{eq:decidable_full}
	is reducible to $\EFO{\ring}$.
}

However, if the update contains \emph{non-linear} terms, then its closed form and hence this reduction are not always computable in polynomial space (and thus, also not in polynomial time).
Consider the following \tnn-loop $\LL_{non-pspace}$:
\begin{equation}
	\label{fig:loop:non-poly-ex}
	\hspace*{-2cm}	\minialg{0.9\linewidth}{}{
		$\WHILEDO{\TRUE}{
				{\ASSIGN{\left(x_1^{\phantom{d}},\, x_2^{\phantom{d}}, \ldots, \,x_{d-1}^{\phantom{d}},\, x_d^{\phantom{d}}\right)}{\left(x_2^d,\, x_3^d, \ldots,\, x_d^d, \, d \cdot x_d\right)}}}$}
\end{equation}
The closed form for $x_i$ (i.e., the value of $x_i$ after $n$ loop iterations) is $q_i = d^{(d^{d-i}\cdot (n-d+i))}\cdot x_d^{(d^{d-i})}$ for all $n \geq d$.
Thus, the closed form $q_1$ for $x_1$ contains constants like $d^{(d^{d-1})}$ whose logarithm grows faster than any polynomial in $d$.
Hence, $q_1$ cannot be computed in polynomial space.

Instead of computing closed forms directly, one could first linearize the loop (see \cite{oliveira16}
and \cref{subsec:linearizing}) and then compute the closed form for the resulting linear-update loop.
However, this approach cannot be computed in polynomial space either, because the linearization of $\LL_{non-pspace}$ contains the constant $d^{(d^{d-1})}$ as well (see \cref{non-pspace linearization}
in \cref{subsec:linearizing}).
We refer to \cref{sec:complexity} for an analysis of the complexity of deciding termination for \twn-loops.

Our reduction also works if $\ring = \RR$, i.e., termination over $\RR$ is reducible to $\EFO{\RR}$, since $\RR$ and $\RA$ are \emph{elementary equivalent} (i.e., a first-order formula is valid over $\RR$ iff it is valid over $\RA$, see, e.g., \cite{complexityexistentialreal}). Thus, we get the following corollary by using that validity of closed formulas from $\EFO{\ring}$ is decidable for $\ring \in \{\RA,\RR\}$ and semi-decidable for $\ring \in \{\ZZ,\QQ\}$ \cite{tarski_decidability,cohen69}.
\begin{corollary}[(Semi-)Deciding (Non-)Termination]
	\label{coro:decidability_reals}
	Let $(\cond,\update)$ be a \twn-loop.
	\begin{enumerate}
		\item[(a)] The loop $(\cond,\update)$ terminates over $\RA$ iff it terminates over $\RR$.
		\item[(b)] Termination of $(\cond,\update)$ on $\ring^d$ is decidable if $\ring = \RA$ or $\ring = \RR$.
		\item[(c)] Non-termination of $(\cond,\update)$ on $\ring^d$ is semi-decidable if $\ring = \ZZ$ or $\ring = \QQ$.
	\end{enumerate}
\end{corollary}
\makeproof{coro:decidability_reals}{
	Again, by \cref{coro:ptill}, termination of \twn-loops is reducible to termination of \tnn-loops.
	By \cref{thm:decidable}, termination of \tnn-loops is reducible to invalidity of a closed formula $\chi \in \EFO{\ring}$.
	If $\ring = \RA$, then validity of $\chi$ is decidable, and if $\ring = \ZZ$ or $\ring = \QQ$, then validity of $\chi$ is semi-decidable \cite{tarski_decidability,cohen69}.
	But $\chi$ is valid iff the loop is non-terminating.
	Hence, non-termination is decidable for $\ring = \RA$ and semi-decidable if $\ring = \ZZ$ or $\ring = \QQ$.
	The claim (b) for $\ring = \RA$ follows since \emph{deciding} non-termination is equivalent to deciding termination.
	Finally, (a) and the claim (b) for $\ring = \RR$ follow due to elementary equivalence of $\RA$ and $\RR$.
}

Our technique does not yield witnesses for non-termination, but the formula constructed by \cref{thm:eventual_positiveness} describes the set of \emph{all} witnesses for \emph{eventual} non-termination.
So this set can be characterized by a formula from $\EFO{\ring}$ (i.e., it is \emph{$\EFO{\ring}$-definable}, see \cref{subsec:transf}), while in general the set of witnesses for non-termination cannot be characterized in this way (see \cite{dblp:conf/ictac/daix12}).
\begin{lemma}
	\label{lem:witnesses}
	Let $\xi = \cond(\closednorm)$.
	Then $\witness \in \ring^d$ witnesses eventual non-termination of $(\cond,\update)$ on $\ring^d$ iff $\lia(\xi) (\witness)$ holds.
\end{lemma}
\makeproof{lem:witnesses}{
	We have:

	\vspace*{-1cm}

	\[
		\begin{array}{@{\hspace*{2cm}}rl}
			     & \witness \text{ witnesses eventual non-termination of $(\cond,\update)$}                                                                     \\[-.1cm]
			\iff & \exists n_0 \in \NN.\ \forall n \in \NN_{>n_0}.\ \left(\cond(\closednorm) \right) (\witness) \hspace*{2cm} \text{(by \eqref{eq:normalized})} \\[-.1cm]
			\iff & \exists n_0 \in \NN.\ \forall n \in \NN_{>n_0}.\ \xi(\witness)                                                                               \\[-.1cm]
			\iff & \lia(\xi)(\witness) \hspace*{\fill} \text{(as in the proof of \cref{thm:eventual_positiveness})}
		\end{array}
	\]

	\vspace*{-.6cm}

}
In \cite{lpar20}, we show how to compute witnesses for non-termination from witnesses for eventual non-termination of \twn-loops.
Thus, combining \cref{lem:witnesses} with \cite{lpar20} shows that $\mathit{NT}_{(\cond, \update)}$ is recursively enumerable for \twn-loops over $\ZZ \leq\linebreak
	\ring \leq \RA$. \cref{alg:deciding} summarizes our technique to check termination of \twn-loops.

\algorithmstyle{boxruled}
\begin{algorithm}
	\caption{Checking Termination}
	\label{alg:deciding}
	\KwIn{a \twn-loop $(\cond,\update)$}
	\KwResult{{$\top$ resp.\ $\bot$ if (non-)termination of $(\cond,\update)$ on $\ring^d$ is
				proven, $?$ otherwise}}
	\lIf{$(\cond, \update)$ is not \tnn}{$(\cond, \update) \assign (\cond \land \cond(\update),\update(\update))$}
	$\closed \assign \text{closed form for }\update$\;
	\lIf{(in)validity of { $\exists \vec{x} \in \ring^d.\ \lia(\cond(\closednorm))$} cannot be
		proven}{\Return $?$}
	\leIf{$\exists \vec{x} \in \ring^d.\ \lia(\cond(\closednorm))$ {is valid}}{\Return $\bot$}{\Return $\top$}
\end{algorithm}

%% file: transformations.tex
\section{Transformation to Triangular Weakly Non-Linear Form}
\label{sec:transf}
In this section, we show how to handle loops that are not yet \twn.
To this end, we introduce a transformation of loops via \emph{polynomial automorphisms} in \cref{subsec:transf}
and show that our transformation preserves (non-)termination (\cref{thm:termination_conjugation}).
In \cref{subsec:automorphisms}, we use results from algebraic geometry to show that the question whether a loop can be transformed into \twn-form is reducible to validity of $\EFO{\RA}$-formulas (\cref{thm:triangularizability_with_fixed_degree}).
Moreover, we show that it is decidable whether a \emph{linear} automorphism can transform a loop into a special case of the \twn-form (\cref{thm:strongly_nilpotent_jacobian}).
Finally, based on the transformation of \cref{subsec:transf,subsec:automorphisms}
we generalize our results from \cref{sec:deciding} to certain non-\twn loops in \cref{subsec:adaptions}.

\subsection{Transforming Loops}
\label{subsec:transf}
Clearly, the \emph{polynomials} $x_1,\ldots,x_d$ are \emph{generators} of the $\ring$-algebra $\ring[\vec{x}]$, i.e., every polynomial from $\ring[\vec{x}]$ can be obtained from $x_1,\ldots,x_d$ and the operations of the algebra (i.e., addition and multiplication).
So far, we have implicitly chosen a special ``representation'' of the loop based on the generators $x_1,\ldots,x_d$.

We now change this representation, i.e., we use $d$ different polynomials which are also generators of $\ring[\vec{x}]$.
Then the loop has to be modified accordingly to adapt it to this new representation.
This modification does not affect the loop's termination behavior, but it may transform a non-\twn-loop into \twn-form.
This change of representation is described by \emph{$\ring$-automorphisms} of $\ring[\vec{x}]$.

\begin{definition}[$\ring$-Endomorphisms]
	A mapping $\eta:\ring[\vec{x}] \to \ring[\vec{x}]$ is an \emph{$\ring$-endomorphism} of $\ring[\vec{x}]$ if it is $\ring$-linear and multiplicative, i.e., for all $c,c' \in \ring$ and all $p,p' \in \ring[\vec{x}]$ we have $\eta(c \cdot p + c' \cdot p') = c \cdot \eta (p) + c' \cdot \eta(p')$, $\eta(1)=1$, and $\eta(p \cdot p') = \eta(p)\cdot \eta(p')$.
	We denote the set of all $\ring$-endomorphisms of $\ring[\vec{x}]$ by 	$\End{\ring}$.
	The set $\Aut{\ring}$ of \emph{$\ring$-automorphisms} of $\ring[\vec{x}]$ consists of those $\eta \in \End{\ring}$ which are \emph{invertible}, i.e., there exists an $\eta^{-1} \in \End{\ring}$ with $\eta \circ \eta^{-1}
		= \eta^{-1} \circ \eta = id_{\ring[\vec{x}]}$, where $id_{\ring[\vec{x}]}$ is the identity function on $\ring[\vec{x}]$. $\Aut{\ring}$ is a group under function composition $\circ$ with identity $id_{\ring[\vec{x}]}$.
\end{definition}

\begin{example}
	\label{ex:automorphism}
	Let $\eta \in \mathrm{End}_{\ring}\left(\ring[x_1,x_2]\right)$ with $\eta(x_1) = x_2$ and $\eta(x_2) = x_1 - x_2^2$.
	Then $\eta \in \mathrm{Aut}_{\ring}\left(\ring[x_1,x_2]\right)$, where $\eta^{-1}(x_1) = x_1^2 + x_2$ and $\eta^{-1}(x_2) = x_1$.
\end{example}

As $\ring[\vec{x}]$ is free on the generators $\vec{x}$, an endomorphism $\eta \in \End{\ring}$ is uniquely determined by the images of the variables, i.e., by $\eta(x_1), \ldots, \eta(x_d)$.
Hence, we have a one-to-one correspondence between elements of $\left(\ring[\vec{x}]\right)^d$ and $\End{\ring}$.
In par\-ticu\-lar, every tuple $\update = (\updateElem_1,\ldots,\updateElem_d) \in \left(\ring[\vec{x}]\right)^d$ corresponds to the unique endomorphism $\widetilde{\updateElem} \in \End{\ring}$ which is defined as follows:
\[
	\widetilde{\updateElem}(x_i) = \updateElem_i \qquad \text{for all } 1 \leq i \leq d
\]
So for any $p \in \ring[\vec{x}]$ we have $\widetilde{\updateElem}(p) = p(\update)$.
Thus, the update of a loop induces an endomorphism which operates on polynomials.

\begin{example}
	\label{ex:update}
	Consider the loop $\LL_{\mathit{aut}} = (\cond, \update)$ where $\update = (\updateElem_1, \updateElem_2)$:
	\begin{equation}
		\label{fig:loop:aut}
		\hspace*{-5.5cm} \minialg{0.6\linewidth}{}{
		$\WHILEDO{x_2^3 + x_1-x_2^2 > 0}{
			{\ASSIGN{\mat{x_1\\
				x_2}}{\mat{((-x_2^{2}+x_1)^{2}+x_2)^{2}-2\cdot x_2^{2}+2\cdot x_1\\
			(-x_2^{2}+x_1 )^{2}+x_2}}}}$}
	\end{equation}
	Then $\update$ induces the endomorphism $\widetilde{\updateElem}$ with $\widetilde{\updateElem}(x_1) = \updateElem_1$ and $\widetilde{\updateElem}(x_2) = \updateElem_2$.
	So we have $\widetilde{\updateElem}(2 \cdot x_1 + x_2^3) = (2 \cdot x_1 + x_2^3)(\update) = 2 \cdot \updateElem_1 + \updateElem_2^3$.

	Therefore, for a tuple of numbers like $\witness = (5,2)$, the induced endomorphism $\widetilde{e}$ is $\widetilde{e}(x_1) = 5$ and $\widetilde{e}(x_2) = 2$.
	Thus, we have $\widetilde{e}(x_2^3 + x_1 - x_2^2) = (x_2^3 + x_1 - x_2^2) (5,2) = 2^3 + 5 - 2^2 = 9$.
\end{example}

We extend the application of endomorphisms $\eta: \ring[\vec{x}] \to \ring[\vec{x}]$ to vectors of polynomials $\update = (\updateElem_1,\ldots,\updateElem_d)$ by defining $\eta(\update) = (\eta(\updateElem_1),\ldots, \eta(\updateElem_d))$ and to formulas $\cond \in \EFO{\ring,\RA}$ by defining $\eta(\cond) = \cond(\eta(\vec{x}))$, i.e., $\eta(\cond)$ results from $\cond$ by applying $\eta$ to all polynomials that occur in $\cond$.
This allows us to transform $(\cond, \update)$ into a new loop $\Tr{\eta}{\cond}{\update}$ using any automorphism $\eta \in \Aut{\ring}$.

\begin{definition}[$\TrOp$]
	\label{def:action_loops}
	For $\eta \in \Aut{\ring}$,
	we define $\Tr{\eta}{\cond}{\update} = (\cond', \update\,')$ where
	\[
		\cond' = \eta^{-1}(\cond) \qquad \text{and} \qquad \update\,' = (\eta^{-1}\circ\widetilde{\updateElem}\circ \eta)(\vec{x}).
	\]
\end{definition}
\noindent
In other words, we have $\update\,' = (\eta(\vec{x})) \, (\update) \, (\eta^{-1}(\vec{x}))$ since $(\eta^{-1}\circ\widetilde{\updateElem}\circ \eta)(\vec{x}) = \eta^{-1}(\eta(\vec{x})[\vec{x}/\update]) = \eta(\vec{x})[\vec{x}/\update][\vec{x} / \eta^{-1}(\vec{x})] = (\eta(\vec{x}))(\update)(\eta^{-1}(\vec{x}))$.
\begin{example}
	\label{ex:transformation}
	We transform the loop $\LL_{\mathit{aut}}$ in \eqref{fig:loop:aut} with the automorphism $\eta$ from \cref{ex:automorphism}.
	We obtain $\Tr{\eta}{\cond}{\update} = (\cond', \update\,')$ where
	\[
		\begin{array}{rcccc}
			\cond'     & = & \multicolumn{3}{l}{\eta^{-1}(\cond) = ((\eta^{-1}(x_2))^3 + \eta^{-1}(x_1) - (\eta^{-1}(x_2))^2 > 0 )}                                                                                                           \\[-.1cm]
			           & = & \multicolumn{3}{l}{(x_1^3 + x_1^2 + x_2 - x_1^2 > 0) = (x_1^3 + x_2 > 0) \hfill \text{and} \hfill}
			\\
			\update\,' & = & ((\eta^{-1}\circ\widetilde{\updateElem}\circ \eta)(x_1), (\eta^{-1}\circ\widetilde{\updateElem}\circ \eta)(x_2)) & = & (\eta^{-1}(\widetilde{\updateElem}(x_2)), \eta^{-1}(\widetilde{\updateElem}(x_1- x_2^2))) \\[-.1cm]
			           & = & (\eta^{-1}(\updateElem_2), \eta^{-1}(\updateElem_1 - \updateElem_2^2))                                           & = & (x_1 + x_2^2, 2 \cdot x_2).
		\end{array}
	\]
	So the resulting transformed loop is $(x_1^3 +x_2 > 0,\ (x_1 + x_2^2, 2 \cdot x_2))$.
	Note that while the original loop $(\cond, \update)$ is neither triangular nor weakly non-linear, the resulting transformed loop is \twn.
	Also note that we used a \emph{non-linear} automorphism with $\eta(x_2) = x_1 - x_2^2$ for the transformation.
\end{example}

While the above example shows that our transformation can indeed transform non-\twn-loops into \twn-loops, it remains to prove that this transformation preserves (non-)termination.
Then we can use our techniques for termination analysis of \twn-loops for \emph{\twn-transformable}-loops as well, i.e., for all loops $(\cond, \update)$ where $\Tr{\eta}{\cond}{\update}$ is \twn for some automorphism $\eta$.
(The question how to find such automorphisms will be addressed in \cref{subsec:automorphisms}.)

As a first step,
by \cref{lem:action}, our transformation is ``compatible'' with the operation $\circ$ of the group $\Aut{\ring}$, i.e., it is an \emph{action}.

\begin{lemma}
	\label{lem:action}
	$\Aut{\ring}$ \emph{acts} via $\TrOp$ on loops, i.e., for $\eta_1, \eta_2 \in \Aut{\ring}$, we have $\Tr{id_{\ring[\vec{x}]}}{\cond}{\update} = (\cond, \update)$ and	$\Tr{\eta_1 \circ \eta_2}{\cond}{\update} = \mathit{Tr}_{\eta_2}(\Tr{\eta_1}{\cond}{\update})$.
\end{lemma}
\makeproof{lem:action}{
	Let $(\cond, \update)$ be a loop.
	Since $id_{\ring[\vec{x}]}^{-1} = id_{\ring[\vec{x}]}$, we obtain $\Tr{id_{\ring[\vec{x}]}}{\cond}{\update} = (\cond', \update\,')$ with
	\[
		\mbox{$
				\begin{array}{rcl}
					\cond'     & = & id_{\ring[\vec{x}]}^{-1} \left(\cond\right) = \cond                                                                                  \\[-.1cm]
					\update\,' & = & (id_{\ring[\vec{x}]}^{-1}\circ\widetilde{\updateElem}\circ id_{\ring[\vec{x}]})(\vec{x}) =\widetilde{\updateElem}(\vec{x}) = \update
				\end{array}
			$}
	\]

	Now we take $\eta_1, \eta_2 \in \Aut{\ring}$.
	Note that $(\eta_1 \circ \eta_2)^{-1}
		= \eta_2^{-1} \, \circ \, \eta_1^{-1}$. Let $\Tr{\eta_1 \circ \eta_2}{\cond}{\update} = (\cond',\update\,')$, $\Tr{\eta_1}{\cond}{\update} = (\cond'',\update\,'')$, and $\mathit{Tr}_{\eta_2}(\cond'',\update\,'') = (\cond''',\update\,''')$. We have
	\[
		\mbox{$
				\begin{array}{rcl}
					\cond'   & = & (\eta_2^{-1} \circ \eta_1^{-1}) (\cond) \\[-.1cm]
					\cond''  & = & \eta_1^{-1} (\cond)                     \\[-.1cm]
					\cond''' & = & \eta_2^{-1}(\cond'')                    \\[-.1cm]
					         & = & \eta_2^{-1}(\eta_1^{-1} (\cond))        \\[-.1cm]
					         & = & (\eta_2^{-1} \circ \eta_1^{-1}) (\cond) \\[-.1cm]
					         & = & \cond'
				\end{array}
			$}
	\]

	Moreover, we have

	\vspace*{-.7cm}

	\[
		\begin{array}{rcl}
			\update\,'   & = & (\eta_2^{-1} \circ \eta_1^{-1} \circ \widetilde{\updateElem} \circ \eta_1 \circ \eta_2 )(\vec{x})       \\[-.1cm]
			             & = & (\eta_2(\vec{x})) \; (\eta_1(\vec{x})) \; (\update) \; (\eta_1^{-1}(\vec{x}))\; (\eta_2^{-1}(\vec{x}))  \\[-.1cm]
			\update\,''  & = & (\eta_1^{-1} \circ \widetilde{\updateElem} \circ \eta_1)(\vec{x})                                       \\[-.1cm]
			             & = & (\eta_1(\vec{x})) \; (\update) \; (\eta_1^{-1}(\vec{x}))                                                \\[-.1cm]
			\update\,''' & = & (\eta_2^{-1} \circ \widetilde{\update\,''} \circ \eta_2)(\vec{x})                                       \\[-.1cm]
			             & = & (\eta_2(\vec{x})) \; (\eta_1(\vec{x})) \; (\update) \; (\eta_1^{-1}(\vec{x})) \; (\eta_2^{-1}(\vec{x})) \\[-.1cm]
			             & = & \update\,'
		\end{array}
	\]
	\vspace*{-1cm}
}
The following lemma will enable us to generalize our results on witnesses for (eventual) non-termination to loops which can be transformed into \twn-form.
\begin{lemma}
	\label{lem:update_transf}
	Let $\Tr{\eta}{\cond}{\update}\!=\!(\cond', \update\,')$ and let $\widehat{\eta}\!:\!\ring^d\!\to\!\ring^d$ map $\witness$ to $\widehat{\eta}(\witness) = \widetilde{e} (\eta(\vec{x})) = (\eta(\vec{x})) (\witness)$.
	Then 	$\cond(\update^n(\witness)) = \cond'((\update\,')^n(\widehat{\eta}(\witness)))$ for all $\witness \in \ring^d$ and $n \in \NN$.
\end{lemma}
\makeproof{lem:update_transf}{
	Let $\witness \in \ring^d$ and $n \in \NN$.
	Then \vspace*{-.65cm}
	\[
		\begin{array}{rcl}
			\hspace*{4.4cm} &      & \cond'((\update\,')^n((\eta(\vec{x})) \, (\witness)))                                                                                                                                                 \\[-.1cm]
			                & = \; & \eta^{-1}(\cond) \; ((\update\,')^n((\eta(\vec{x})) \, (\witness)))                                                                                                                                   \\[-.1cm]
			                & =\;  & \cond \, [\vec{x}/\eta^{-1}(\vec{x})] \, \underbrace{[\vec{x}/\update\,']}_{n \text{ times}}
			[\vec{x}/\eta(\vec{x})] \, [\vec{x}/\witness]                                                                                                                                                                                  \\[-.1cm]
			                & =\;  & \cond \, [\vec{x}/\eta^{-1}(\vec{x})] \, [\vec{x}/\eta(\vec{x})] \, \underbrace{[\vec{x}/\update]}_{n \text{ times}} \, [\vec{x}/\eta^{-1}(\vec{x})] \, [\vec{x}/\eta(\vec{x})] \, [\vec{x}/\witness] \\[-.1cm]
			                & =\;  & \cond \, \underbrace{[\vec{x}/\update]}_{n \text{ times}} \, [\vec{x}/\witness]                                                                                                                       \\[-.1cm]
			                & =\;  & \cond(\update^n(\witness))
		\end{array}
	\]

	\vspace*{-.6cm}
}

\cref{lem:update_transf} yields the following corollary which shows that $\eta(\vec{x})$ transforms witnesses for (eventual) non-termination of $(\cond,\update)$ into witnesses for $\Tr{\eta}{\cond}{\update}$.

\begin{corollary}
	\label{coro:witnesses_nontermination}
	If $\witness$ witnesses (eventual) non-termination of $(\cond, \update)$, then $\widehat{\eta}(\witness)$ witnesses (eventual) non-termination of $\Tr{\eta}{\cond}{\update}$.
\end{corollary}

\begin{example}
	\label{ex:transforming_witnesses}
	For the tuple $\witness = (5,2)$ from \cref{ex:update} and the automorphism $\eta$ from \cref{ex:automorphism} with $\eta(x_1) = x_2$ and $\eta(x_2) = x_1 - x_2^2$, we obtain
	\[
		\widehat{\eta}(\witness) =(\eta(x_1), \eta(x_2)) \, (\witness) = (2, 5 - 2^2) = (2,1).
	\]
	As $\witness = (5,2)$ witnesses non-termination of the loop $\LL_{\mathit{aut}} = (\cond,\update)$ in \eqref{fig:loop:aut}, $\widehat{\eta}(\witness) =(2,1)$ witnesses non-termination of $\Tr{\eta}{\cond}{\update}$ due to \cref{coro:witnesses_nontermination}.
\end{example}

Finally, \cref{thm:termination_conjugation} states that transforming loops preserves \mbox{(non-)}ter\-mi\-na\-tion.

\begin{theorem}[$\TrOp$ Preserves Termination]
	\label{thm:transf_preserves_termination}
	\label{thm:termination_conjugation}
	If $\eta \in \Aut{\ring}$, then $\widehat{\eta}$ is a bijection between the respective sets of witnesses for (eventual) non-termination, i.e., for $\Tr{\eta}{\cond}{\update}=(\cond',\update\,')$ we have
	$\witness \in \mathit{(E)NT}_{(\cond, \update)}$ iff $\widehat{\eta}(\witness) \in \mathit{(E)NT}_{(\cond', \update\,')}$.
	Therefore, $(\cond, \update)$ terminates iff $\Tr{\eta}{\cond}{\update}=(\cond',\update\,')$ terminates.
\end{theorem}

\makeproof{thm:transf_preserves_termination}{
In \cref{coro:witnesses_nontermination}, we have seen that if $\witness$ is a witness for (eventual) non-termination of $(\cond, \update)$, then $\widehat{\eta}(\witness)$ witnesses (eventual) non-termination of $\Tr{\eta}{\cond}{\update}$.
Now let $\witness\,'$ be a witness for (eventual) non-termination of $\Tr{\eta}{\cond}{\update}$.
Then by \cref{coro:witnesses_nontermination}, $\widehat{\eta^{-1}}(\witness\,')$ witnesses (eventual) non-termination of $\mathit{Tr}_{\eta^{-1}}(\Tr{\eta}{\cond}{\update}) \overset{\textnormal{\cref{lem:action}}}{=}
	\Tr{\eta \circ \eta^{-1}}{\cond}{\update} = (\cond,\update)$.
Hence, $\widehat{\eta}$ maps witnesses for (eventual) non-termination of $(\cond, \update)$ to witnesses for (eventual) non-termination of $\Tr{\eta}{\cond}{\update}$ and $\widehat{\eta^{-1}}$ maps witnesses for (eventual) non-termination of $\Tr{\eta}{\cond}{\update}$ to witnesses for $(\cond, \update)$.
These two mappings are inverse to each other:
For $\witness\,' \in \ring^d$ we have
\[
	\begin{array}{cl}
		      & \widehat{\eta}(\widehat{\eta^{-1}}(\witness\,'))                                                                \\[-.1cm]
		{}={} & \widehat{\eta}((\eta^{-1}(\vec{x}))(\witness\,')) \hspace*{2cm}
		\text{(by definition of $\widehat{\eta^{-1}}$)}                                                                         \\[-.1cm]

		{}={} & (\eta(\vec{x}))((\eta^{-1}(\vec{x}))(\witness\,')) \hspace*{\fill} \text{(by definition of $\widehat{\eta}$)}   \\[-.1cm]
		{}={} & \eta(\vec{x})[\vec{x}/\eta^{-1}(\vec{x})][\vec{x}/\witness\,']                                                  \\[-.1cm]
		{}={} & \witness\,'                                                                                                     \\[-.1cm]
		\\[-.1cm]
		      & \widehat{\eta^{-1}}(\widehat{\eta}(\witness))                                                                   \\[-.1cm]
		{}={} & \widehat{\eta^{-1}}((\eta(\vec{x}))(\witness)) \hspace*{\fill} \text{(by definition of $\widehat{\eta}$)}       \\[-.1cm]
		{}={} & (\eta^{-1}(\vec{x}))((\eta(\vec{x}))(\witness)) \hspace*{\fill} \text{(by definition of $\widehat{\eta^{-1}}$)} \\[-.1cm]
		{}={} & \eta^{-1}(\vec{x})[\vec{x}/\eta(\vec{x})][\vec{x}/\witness]                                                     \\[-.1cm]
		{}={} & \witness.
	\end{array}
\]
Hence, $\widehat{\eta}$ is indeed a bijection with inverse mapping $\widehat{\eta^{-1}}$.
}
Up to now, we only transformed a loop $(\cond, \update)$ on $\ring^d$ using elements of $\Aut{\ring}$.
To see that this is a limitation, consider a linear-update loop where $\update = A \cdot \vec{x}$ and $A$ only has real eigenvalues.
In \Cref{subsec:linear-update} we will show that these loops can always be transformed into \twn-form and a suitable automorphism $\eta$ can be obtained by computing the Jordan normal form of $A$.
This automorphism $\eta$ is only an element of $\Aut{\ring}$ if the eigenvalues of $A$ are from $\ring$.
So if $\ring = \ZZ$, then this transformation is only applicable if all eigenvalues of $A$ are integers.

However, we can also transform $(\cond, \update)$ into the loop $\Tr{\eta}{\cond}{\update}$ on $\RA^d$ using an automorphism $\eta \in \Aut{\RA}$.
Nevertheless, our goal remains to prove termination on $\ring^d$ instead of $\RA^d$, which is not equivalent in general.
Thus, in \cref{subsec:adaptions} we will show how to analyze termination of loops on certain subsets $F$ of $\RA^d$.
This allows us to analyze termination of $(\cond, \update)$ on $\ring^d$ by checking termination of $\Tr{\eta}{\cond}{\update}$ on the subset $\widehat{\eta}(\ring^d) \subseteq \RA^d$ instead.

By our definition of loops over a ring $\ring$, we have $\update(\witness) \in \ring^d$ for all $\witness \in \ring^d$, i.e., $\ring^d$ is \emph{$\update$-invariant}.
This property is preserved by our transformation.

\begin{definition}[$\update$-Invariance]
	\label{def:update_invariant}
	Let $(\cond,\update)$ be a loop on $\ring^d$ and let $F \subseteq \ring^d$.
	We call $F$ \emph{$\update$-invariant} or \emph{update-invariant} if for all $\witness \in F$ we have $\update(\witness) \in F$.
\end{definition}

\begin{lemma}
	\label{lem:action_preserves_invariance}
	Let $(\cond,\update)$ be a loop on $\ring^d$, $F\subseteq \ring^d$ be $\update$-invariant, $\eta \in \Aut{\RA}$, and
	let $\Tr{\eta}{\cond}{\update} = (\cond', \update\,')$.
	Then $\widehat{\eta}(F)$ is $\update\,'$-invariant.
\end{lemma}
\makeproof{lem:action_preserves_invariance}{
	Let $\witness\,' \in \widehat{\eta}(F)$.
	Then $\witness\,' = \widehat{\eta}(\witness)$ for some $\witness\in F$.
	As $F$ is $\update$-invariant, we have $\update(\witness) \in F$.
	We obtain

	\vspace*{-.6cm}

	\[
		\begin{array}{rcl}
			\update\,'(\witness\,') & = & (\eta(\vec{x})) \, (\update) \, (\eta^{-1}(x)) \, (\witness\,')                 \\[-.1cm]
			                        & = & (\eta(\vec{x})) \, (\update) \, (\eta^{-1}(x)) \, (\eta(\vec{x})) \, (\witness) \\[-.1cm]
			                        & = & (\eta(\vec{x})) \, (\update) \, (\witness)                                      \\[-.1cm]
			                        & = & \widehat{\eta}(\update(\witness)) \in \widehat{\eta}(F).
		\end{array}
	\]

	\vspace*{-.3cm}

}

Our goal is to reduce termination to a $\EFO{\ring, \RA}$-formula.
Clearly, \emph{termination on $F$} cannot be encoded with such a formula if $F$ cannot be defined via $\EFO{\ring, \RA}$.
Thus, we require that $F$ is \emph{$\EFO{\ring,\RA}$-definable}, i.e., that there is a $\psi \in \EFO{\ring,\RA}$ with free variables $\vec{x}$ such that we have $\witness \in F$ iff $\psi(\witness)$ is valid.
Then we also say that $\psi$ \emph{defines} $F$.
An example for a $\EFO{\ZZ,\RA}$-definable set is $\left\{ (a, 0, a) \mid a \in \ZZ\right\}$, which is cha\-rac\-te\-rized by the formula $\exists a \in \ZZ.\ x_1 = a \land x_2 = 0 \land x_3 = a$.

To analyze termination of $(\cond,\update)$ on $\ring^d$, we can analyze termination of $\Tr{\eta}{\cond}{\update}$ on $\widehat{\eta} (\ring^d)\subseteq \RA^d$ instead.
The reason is that $\witness \in \ring^d$ is a witness for (eventual) non-termination of $(\cond,\update)$ iff $\widehat{\eta}(\witness)$ is a witness for $\Tr{\eta}{\cond}{\update}$ due to \cref{coro:witnesses_nontermination}, i.e., $\ring^d$ contains a witness for (eventual) non-termination of $(\cond,\update)$ iff $\widehat{\eta}(\ring^d)$ contains a witness for
$\Tr{\eta}{\cond}{\update}$.
While $\ring^d$ is clearly $\EFO{\ring,\RA}$-definable, \cref{lem:action_preserves_definability} shows that $\widehat{\eta} (\ring^d)$ is $\EFO{\ring,\RA}$-definable, too.
More precisely, $\EFO{\ring,\RA}$-definability is preserved by polynomial endomorphisms.

\begin{lemma}
	\label{lem:action_preserves_definability}
	Let $\ZZ \leq \ring \leq \RA$ and let $\eta \in \End{\RA}$.
	If $F\subseteq \RA^d$ is $\EFO{\ring,\RA}$-definable then so is $\widehat{\eta}(F)$.
\end{lemma}
\makeproof{lem:action_preserves_definability}{
	Let $F$ be defined by $\psi_F \in \EFO{\ring,\RA}$.
	Consider the following formula $\psi \in \EFO{\ring,\RA}$:
	\[
		\exists \vec{y} \in \RA^d.\ \psi_F(\vec{y}) \land \vec{x} = \left(\eta(\vec{x})\right)(\vec{y}).
	\]
	Then $\psi(\witness)$ holds for $\witness \in \RA^d$ iff $\witness = \widehat{\eta}(\vec{u})$ for some $\vec{u} \in \RA^d$ with $\psi_F(\vec{u})$, i.e., with $\vec{u} \in F$.
}
\begin{example}
	The set $\ZZ^2$ is $\EFO{\ZZ,\RA}$-definable as we have $(x_1,x_2)\in \ZZ^2$ iff
	\[
		\exists a,b \in \ZZ. \; x_1=a \land x_2= b.
	\]
	Let $\eta \in \End{\RA}$ with $\eta(x_1) = \frac{1}{2}\cdot x_1^2 + x_2^2$ and $\eta(x_2) = x_2^2$.
	Then $\widehat{\eta}(\ZZ^2)$ is also $\EFO{\ZZ,\RA}$-definable because for $x_1, x_2 \in \RA$ we have $(x_1,x_2) \in \eta(\ZZ^2)$ iff
	\[
		\exists y_1,y_2 \in \RA, \; a,b \in \ZZ. \;\; y_1=a \land y_2 = b \; \land \; x_1 = \tfrac{1}{2}\cdot y_1^2 + y_2^2 \land x_2 = y_2^2.
	\]
\end{example}

\cref{thm:solvable} shows that instead of regarding \emph{solvable loops}
\cite{solvable-maps}, w.l.o.g.\ we can restrict ourselves to \twn-loops.
The reason is that every solvable loop with real eigenvalues can be transformed into a \twn-loop by a \emph{linear} automorphism $\eta$, i.e., the degree $\deg(\eta)$ of $\eta$ is 1, where $\deg(\eta)=\max \{ \deg(\eta(x_i)) \mid 1 \leq i \leq d \}$.

\begin{theorem}[\twn-Transformability of Solvable Loops]
	\label{thm:solvable}
	Let $(\cond,\update)$ be a solvable loop with real eigenvalues.
	Then one can compute a linear automorphism $\eta \in \Aut{\RA}$ such that $\Tr{\eta}{\cond}{\update}$ is \twn.
\end{theorem}
\makeproof{thm:solvable}{
As $(\cond,\update)$ is solvable, there is a partitioning $\mathcal{J} = \{J_1 , \ldots, J_k\}$ as in \cref{def:solvable}, i.e., $\{1,\ldots, d\} = \biguplus_{i = 1}^k J_i$ and $\update_{J_i} = A_i \cdot \vec{x}_{J_i} + \vec{p}_i$ for all $1 \leq i \leq k$, where $\vec{p}_i \in (\ring[\vec{x}_{J_{i+1}}, \ldots, \vec{x}_{J_k}])^{d_i}$.
W.l.o.g., $\vec{x}$ is ordered according to $\mathcal{J}$, i.e., if $x_{i_1} \in J_{j_1}$ and $x_{i_2} \in J_{j_2}$ for $j_1 < j_2$, then $i_1 < i_2$.

For each $A_i$, let $Q_i = T_i \cdot A_i \cdot T_i^{-1}$ be its Jordan normal form, where $T_i$ is the corresponding transformation matrix.
Since $A_i$ has only real eigenvectors, this means that the entries of $Q_i$, $T_i$, and $T_i^{-1}$ are real algebraic numbers.
Let $\eta$ be the endomorphism defined by $\eta(\vec{x}_{J_i}) = T_i \cdot \vec{x}_{J_i}$.
This means that $\eta$ is induced by the block diagonal matrix $\mathit{Diag}(T_1,T_2,\ldots,T_k)$.
Then $\eta$ is an automorphism and its inverse satisfies $\eta^{-1}(\vec{x}_{J_i}) = T^{-1}_i \cdot \vec{x}_{J_i}$.
Furthermore, the degree of $\eta$ is obviously $1$.
Moreover, $\eta$ and $\eta^{-1}$ are compatible with the partition, i.e., the images of the variables in $\vec{x}_{J_i}$ under $\eta$ and $\eta^{-1}$ are polynomials only using the variables $\vec{x}_{J_i}$.
For each $1 \leq i \leq k$ we have:

\vspace*{-.8cm}

\[
	\begin{array}{rcl}
		\hspace*{4cm} &      & (\eta^{-1} \circ \widetilde{\updateElem} \circ \eta)(\vec{x}_{J_i})                                                                           \\[-.1cm]
		              & = {} & \eta(\vec{x_{J_i}}) [\vec{x}/\update][\vec{x} / \eta^{-1}(\vec{x})]                                                                           \\[-.1cm]
		              & = {} & \left(T_i \cdot \vec{x}_{J_i}\right)[\vec{x}/\update][\vec{x} / \eta^{-1}(\vec{x})]                                                           \\[-.1cm]
		              & = {} & (T_i \cdot \update_{J_i})[\vec{x} / \eta^{-1}(\vec{x})]                                                                                       \\[-.1cm]
		              & = {} & \left(T_i \cdot \left(A_i \cdot \vec{x}_{J_i} + \vec{p}_i\right)\right)[\vec{x} / \eta^{-1}(\vec{x})]                                         \\[-.1cm]
		              & = {} & \left(T_i \cdot A_i \cdot \vec{x}_{J_i} + T_i \cdot \vec{p}_i \right)[\vec{x} / \eta^{-1}(\vec{x})]                                           \\[-.1cm]
		              & = {} & \left(T_i \cdot A_i \cdot \vec{x}_{J_i}\right)[\vec{x} / \eta^{-1}(\vec{x})] + \left(T_i \cdot \vec{p}_i\right)[\vec{x} / \eta^{-1}(\vec{x})] \\[-.1cm]
		              & = {} & \left(T_i \cdot A_i \cdot T_i^{-1} \cdot \vec{x}_{J_i}\right) + \left(T_i \cdot \vec{p}_i\right)[\vec{x} / \eta^{-1}(\vec{x})]                \\[-.1cm]
		              & = {} & \left(Q_i \cdot \vec{x}_{J_i}\right) + \left(T_i \cdot \vec{p}_i\right)[\vec{x} / \eta^{-1}(\vec{x})]
	\end{array}
\]
We have $T_i \cdot \vec{p}_i \in \ring[\vec{x}_{J_{i+1}}, \ldots, \vec{x}_{J_k}]^{d_i}$.
Therefore, $\left(T_i \cdot \vec{p}_i\right)[\vec{x} / \eta^{-1}(\vec{x})] \in \ring[\vec{x}_{J_{i+1}}, \ldots, \vec{x}_{J_k}]^{d_i}$ as well, since $\eta^{-1}$ is compatible with the partitioning.
This implies that $\Tr{\eta}{\cond}{\update}$ is weakly non-linear.
As we assumed that $\vec{x}$ is ordered w.r.t.\ the partitioning and each $Q_i$ is triangular, $\Tr{\eta}{\cond}{\update}$ is triangular, too.
Thus, $\Tr{\eta}{\cond}{\update}$ is in \twn-form.
}
We recapitulate our most important results on $\TrOp$ in the following corollary.
Here, we generalize the result of \cref{thm:transf_preserves_termination} to the setting where we consider termination on some update-invariant and $\EFO{\ring,\RA}$-definable set.

\begin{corollary}[Properties of $\TrOp$]
	\label{coro:conclusion_transformations}
	Let $(\cond, \update)$ be a loop, $\eta \in \Aut{\RA}$, $\Tr{\eta}{\cond}{\update} = (\cond',\update\,')$, and $F\subseteq \ring^d$ be $\update$-invariant and $\EFO{\ring,\RA}$-definable.
	\begin{enumerate}
		\item[(a)] $\widehat{\eta} (F)\subseteq \RA^d$ is $\update\,'$-invariant and $\EFO{\ring,\RA}$-definable.
		\item[(b)] $(\cond, \update)$ terminates on $F$ iff $(\cond',\update\,')$ terminates on $\widehat{\eta} (F)$.
		\item[(c)] $\witness \in F$ witnesses (eventual) non-termination of $(\cond, \update)$ iff\\
			$\widehat{\eta}(\witness) \in \widehat{\eta}(F)$ witnesses (eventual) non-termination of $(\cond',\update\,')$.
	\end{enumerate}
\end{corollary}

%% file: automorphisms.tex
\subsection{Finding Automorphisms to Transform Loops into \twn-Form}
\label{subsec:automorphisms}

The goal of $\TrOp_\eta$ from \cref{subsec:transf} is to transform $(\cond,\update)$ into \twn-form such that our techniques from \cref{sec:deciding} can be used to decide termination.
So the two remaining challenges are to find a suitable automorphism $\eta \in \Aut{\RA}$ such that $\Tr{\eta}{\cond}{\update}$ is \twn, and to adapt our techniques from \cref{sec:deciding} such that they can be applied to \twn-loops where one only wants to show termination on an update-invariant $\EFO{\ring,\RA}$-definable subset.
We discuss the latter challenge in \cref{subsec:adaptions}.
In this section, we present two techniques to check the existence of automorphisms for the transformation into \twn-form \emph{constructively}, i.e., these techniques can also be used to compute such automorphisms.

The search for suitable automorphisms is closely related to the question if a polynomial automorphism can be conjugated into a ``de Jonquiéres''-automorphism, a difficult question from algebraic geometry \cite{polyautomorphisms}.
So future advances in this field may help to improve the results of the current section.

The first technique (\cref{thm:triangularizability_with_fixed_degree}) reduces the search for a suitable automorphism \emph{of bounded degree} to $\EFO{\RA}$.
For any automorphism, the degree of its inverse is bounded in terms of the length $d$ of $\vec{x}$.

\begin{theorem}[Degree of Inverse  \protect{\cite[Cor.\ 2.3.4]{polyautomorphisms}}]
	\label{thm:deg_inverse}
	Let $\eta \in \Aut{\RA}$.
	Then we have $\deg(\eta^{-1}) \leq (\deg(\eta))^{d-1}$.
\end{theorem}

By \cref{thm:deg_inverse}, checking if an endomorphism is an automorphism can be reduced to $\EFO{\RA}$.
To do so, one encodes the existence of coefficients for the polynomials $\eta^{-1}(x_1), \ldots, \eta^{-1}(x_d)$, which all have at most degree $\left(\deg(\eta)\right)^{d-1}$.

\begin{lemma}
	\label{lem:deciding_automorphism}
	Let $\eta \in \End{\RA}$.
	Then the question if $\eta\in \Aut{\RA}$ holds is reducible to $\EFO{\RA}$.
\end{lemma}

\makeproof{lem:deciding_automorphism}{
	Let $\delta = \deg(\eta)$.
	For any $k \in \NN$, there is only a finite number of monomials over $\vec{x}$ of degree $k$.
	(The number of monomials of exactly degree $k$ is $\binom{d+k-1}{k}$, see the proof of \cref{thm:alg_depdeg} (b).) Hence, for any $1 \leq i \leq d$ we can construct the following term that stands for $\eta^{-1}(x_i)$:
	\[
		\mbox{$\sum\nolimits_{m \text{ is a monomial of (at most) degree $\delta^{d-1}$}} \; a_{i,m} \cdot m$}
	\]
	Here, the monomials $m$ contain the variables $\vec{x}$ and the $a_{i,m}$ are variables that stand for the unknown coefficients of the polynomial $\eta^{-1}(x_i)$.

	Hence, for any $1 \leq i \leq d$ we now build a formula $\rho_{r,i}$ which stands for the requirement ``$\left(\eta \circ \eta^{-1}\right)(x_i) = \left(\eta^{-1}(x_i)\right)(\eta(\vec{x})) = x_i$'' (i.e., that $\eta^{-1}$ is a right inverse of $\eta$):
	\[
		\mbox{$\rho_{r,i}: \qquad \sum\nolimits_{m \text{ is a monomial of (at most) degree $\delta^{d-1}$}} \; a_{i,m} \cdot \eta(m) = x_i$}
	\]

	Similarly, for any $1 \leq i \leq d$ we construct a formula $\rho_{l,i}$ which stands for the requirement ``$\left(\eta^{-1} \circ \eta\right)(x_i) = (\eta(x_i))\left(\eta^{-1}(\vec{x})\right) = x_i$'' (i.e., that $\eta^{-1}$ is a left inverse of $\eta$):
	\[
		\mbox{$\rho_{l,i}: \qquad \eta(x_i) \, \left( \sum\nolimits_{m \text{ is a monomial of (at most) degree $\delta^{d-1}$}} \; a_{i,m} \cdot m \right) = x_i$}
	\]

	Thus, the formula
	\begin{equation}
		\mbox{$\forall \vec{x} \in \RA^d.\quad \bigwedge\nolimits_{i=1}^d \rho_{r,i} \wedge \bigwedge\nolimits_{i=1}^d \rho_{l,i}$} \label{eqn:inverse}
	\end{equation}
	is valid iff $\eta$ has an inverse of degree at most $\delta^{d-1}$.
	By \cref{thm:deg_inverse}, this is equivalent to the question whether $\eta$ has an inverse, i.e., whether $\eta$ is an automorphism.
	Unfortunately, $\eqref{eqn:inverse} \not\in \EFO{\RA}$.
	However, $\bigwedge_{i=1}^d \rho_{r,i} \wedge \bigwedge_{i=1}^d \rho_{l,i}$ has to hold for all $\vec{x} \in \RA^d$.
	So, we can reduce this formula to a system of equations: one simply has to check whether there is an instantiation of the unknown coefficients $a_{i,m}$ such that all monomials in $\rho_{r,i}$ and $\rho_{l,i}$ except $x_i$ get the coefficient $0$ and the monomial $x_i$ gets the coefficient 1.
	When building the conjunction of these equations and existentially quantifying the unknown coefficients $a_{i,m}$, one indeed obtains a formula from $\EFO{\RA}$.
}

Based on \cref{lem:deciding_automorphism}, we now present our first technique to find an automorphism $\eta$ that transforms a loop into \twn-form.

\begin{theorem}[$\TrOp$ With Automorphisms of Bounded Degree]
	\label{thm:triangularizability_with_fixed_degree}
	For any $\delta \geq 0$, the question whether there exists an $\eta \in \Aut{\RA}$ with $\deg(\eta) \leq \delta$ such that $\Tr{\eta}{\cond}{\update}$ is \twn is reducible to $\EFO{\RA}$.
\end{theorem}
\makeproof{thm:triangularizability_with_fixed_degree}{
	For every $1 \leq i \leq d$, let
	\[
		\mbox{$\eta(x_i)= \sum\nolimits_{m \text{ is a monomial of (at most) degree $\delta$}} \; b_{i,m} \cdot m,$}
	\]
	where the $b_{i,m}$ are variables that stand for unknown coefficients.
	By \cref{lem:deciding_automorphism} there is a $\EFO{\RA}$-formula that contains both $b_{i,m}$ and the variables $a_{i,m}$ (for the coefficients of $\eta^{-1}$) which expresses that $\eta$ is an automorphism.

	Furthermore, using these coefficients we can construct a formula from $\EFO{\RA}$ which expresses that the update $\update' = (\updateElem_1', \ldots, \updateElem_d') = (\eta^{-1} \circ \widetilde{u}\circ \eta) (\vec{x})$ is \twn:
	We have $\deg(\update') = \deg((\eta^{-1} \circ \widetilde{u}\circ \eta) (\vec{x})) \leq \deg(\eta^{-1}) \cdot \deg(\widetilde{u}) \cdot \deg(\eta) \leq \delta^{d-1}\cdot \deg(\widetilde{u})\cdot \delta$.
	So there is a bound on the degree of the polynomials in the transformed loop $\Tr{\eta}{\cond}{\update}$.
	For every $1 \leq i \leq d$, let
	\[
		\mbox{$\updateElem'_i = \sum\nolimits_{m \text{ is a monomial of (at most) degree $\delta^{d-1}\cdot \deg(\widetilde{u})\cdot \delta$}} \; c_{i,m}
				\cdot m,$}
	\]
	where the variables $c_{i,m}$ stand again for unknown coefficients.
	Now we can build a $\EFO{\RA}$-formula which is valid iff $\update'$ is in \twn-form by requiring that certain coefficients $c_{i,m}$ are zero.
	Moreover, we can construct a $\EFO{\RA}$-formula which is valid iff $\update' = (\eta^{-1} \circ \widetilde{u}\circ \eta) (\vec{x})$.
}
\noindent
So if the degree of $\eta$ is bounded a priori, it is decidable whether there exists an $\eta\in \Aut{\RA}$ such that $\Tr{\eta}{\cond}{\update}$ is \twn since $\EFO{\RA}$ is decidable.

We call a loop \emph{\twn-transformable} if there is an $\eta\in \Aut{\RA}$ such that $\Tr{\eta}{\cond}{\update}$ is \twn.
By \cref{thm:triangularizability_with_fixed_degree}, \twn-transformability is \emph{semi-decidable} since one can increment $\delta$ until a suitable automorphism is found.
So in other words, any loop which is transformable to a \twn-loop can be transformed via \cref{thm:triangularizability_with_fixed_degree}.

We call our transformation $\TrOp$ \emph{complete} for a class of loops if \emph{every} loop from this class is \twn-transformable.
For such classes, a suitable automorphism is \emph{computable} by \cref{thm:triangularizability_with_fixed_degree}.
Together with \cref{thm:solvable}, we get \cref{cor:solvable}.
\begin{corollary}
	\label{cor:solvable}
	$\TrOp$ is complete for solvable loops with real eigenvalues.
\end{corollary}

Note that for solvable loops $(\cond,\update)$, instead of computing $\eta$ using \cref{thm:triangularizability_with_fixed_degree}, the proof of \cref{thm:solvable} yields a more efficient way to compute a linear automorphism $\eta$ such that $\Tr{\eta}{\cond}{\update}$ is \twn.
For this, one computes the Jordan normal form of each $A_i$ (see \cref{def:solvable}), which is possible in polynomial time (see, e.g., \cite{jordan,dblp:journals/siamcomp/giesbrecht95}).

Our second technique to find suitable automorphisms for our transformation is restricted to \emph{linear} automorphisms.
In this case, it is decidable whether a loop can be transformed into a \twn-loop $(\cond',\update')$ where the monomial for $x_i$ has the coefficient 1 in each $\updateElem'_i$.
The decision procedure checks if a certain Jacobian matrix is \emph{strongly nilpotent}, i.e., it is not based on a reduction to $\EFO{\RA}$.

\begin{definition}[Strong Nilpotence]
	Let $J \in \left(\RA[\vec{x}]\right)^{d \times d}$ be a matrix of polynomials.
	For all $1 \leq i \leq d$, let $\vec{y}^{(i)}$ be a vector of fresh variables.
	$J$ is \emph{strongly nilpotent} if $\prod\nolimits_{i=1}^d \; J[\vec{x}/\vec{y}^{(i)}]=0^{d\times d}$, where $0^{d\times d}$ is the zero matrix.
\end{definition}

Our second technique is formulated in the following theorem which follows from an existing result in linear algebra \cite[Thm.\ 1.6.]{vandenessen1996121}.

\begin{theorem}[$\TrOp$ With Linear Automorphisms \cite{vandenessen1996121}]
	\label{thm:strongly_nilpotent_jacobian}
	Let $(\cond,\update)$ be a loop.
	The Jacobian matrix $\left(\frac{\partial (\updateElem_i - x_i)}{\partial x_j}\right)_{1 \leq i,j \leq d} \in \left(\RA[\vec{x}]\right)^{d \times d}$ is strongly nilpotent iff there exists a linear automorphism $\eta \in \Aut{\RA}$ with
	\begin{equation}
		\label{eq:trans-jacobi}
		\Tr{\eta}{\cond}{\update} = (\cond',(x_1 + p_1,\ldots,x_d + p_d))
	\end{equation}
	and $p_i \in \RA[x_{i+1},\ldots,x_{d}]$ for all $1 \leq i \leq d$.
	Thus, $\Tr{\eta}{\cond}{\update}$ is \twn.
\end{theorem}

As strong nilpotence of the Jacobian matrix is clearly decidable, \cref{thm:strongly_nilpotent_jacobian} gives rise to a decision procedure for the existence of a linear automorphism that transforms $(\cond, \update)$ to the form \eqref{eq:trans-jacobi}.

\begin{figure}
	\centering
	\minialg{0.9\linewidth}{}{
		$\WHILEDO{4\cdot{x_2}^{2}+x_1+x_2+x_3 > 0}{
				{\ASSIGN{(x_1,x_2,x_3)}{(\updateElem_1,\updateElem_2,\updateElem_3)}}}$

		\vspace*{-.4cm}

		\[
			\begin{array}{rcl}
				\text{with}\ \updateElem_1 & = & x_1+8\cdot x_1\cdot{x_2}^{2}+16\cdot {x_2}^{3}+16\cdot {x_2}^{2} \cdot x_3                                  \\[-.1cm]
				\updateElem_2              & = & x_2-{x_1}^{2}-4\cdot x_1\cdot x_2-4\cdot x_1\cdot x_3-4\cdot {x_2}^{2}-8\cdot x_2\cdot x_3-4\cdot {x_3}^{2} \\[-.1cm]
				\updateElem_3              & = & x_3 -4 \cdot x_1\cdot {x_2}^{2}-8\cdot {x_2}^{3}-8\cdot {x_2}^{2}\cdot x_3+{x_1}^{2}+4\cdot x_1\cdot x_2\,+ \\[-.1cm]
				                           &   & 4\cdot x_1\cdot x_3 + 4\cdot {x_2}^{2}+8\cdot x_2\cdot x_3+4\cdot
				{x_3}^{2}
			\end{array}
		\]
	}
	\vspace*{-.6cm}
	\caption{Loop $\LL_{\mathit{non-twn}}$}
	\label{fig:loop:horrible}
	\vspace*{-.3cm}
\end{figure}

\begin{example}
	\label{ex:linear_transf}
	The loop $\LL_{\mathit{non-twn}}$ on $\ring^3$ in \cref{fig:loop:horrible} shows how our results enlarge the class of loops where termination is reducible to $\EFO{\ring, \RA}$.
	This loop is clearly \emph{not} in \twn-form.
	To transform it, we use \cref{thm:strongly_nilpotent_jacobian}.
	The Jacobian matrix $J$ of $(\updateElem_1 - x_1, \updateElem_2 - x_2, \updateElem_3 - x_3)$ is:
        \[\resizebox{\textwidth}{!}{$
		\left[
			\begin{smallmatrix}
				8\cdot x_2^2 & 16\cdot x_1 \cdot x_2 +48\cdot x_2^2 + 32 \cdot x_2 \cdot x_3 &16\cdot x_2^2 \\
				-2 \cdot x_1 - 4 \cdot x_2 - 4 \cdot x_3 & -4 \cdot x_1 - 8 \cdot x_2 - 8 \cdot x_3 & -4 \cdot x_1 - 8 \cdot x_2 - 8 \cdot x_3 \\
				-4 \cdot x_2^2+2 \cdot x_1 + 4 \cdot x_2 +4 \cdot x_3 \, & \, -8 \cdot x_1 \cdot x_2 -24 \cdot x_2^2-16 \cdot x_2 \cdot x_3 +4 \cdot x_1 +8 \cdot x_2 +8 \cdot x_3 \, &\, -8 \cdot x_2^2+4 \cdot x_1 +8 \cdot x_2 +8 \cdot x_3
			\end{smallmatrix}
			\right]
	$}\]
	One easily checks that $J$ is strongly nilpotent.
	Thus, by \cref{thm:strongly_nilpotent_jacobian} the loop can be transformed into \twn-form by a linear automorphism.
	Indeed, consider the linear automorphism $\eta\!\in\!\Aut{\RA}$ induced by the matrix $M\!=\!\left[
			\begin{smallmatrix}
				1 & 1 & 1 \\
				0 & 2 & 0 \\
				1 & 2 & 2
			\end{smallmatrix}
			\right]$, i.e.,
	\[
		\begin{array}{l@{\;\;}l@{\quad}l@{\quad}l@{\;\;}l}
			\eta:      & x_1 \mapsto x_1 + x_2 + x_3,   & x_2 \mapsto 2 \cdot x_2,           & x_3 \mapsto x_1 + 2 \cdot x_2 + 2 \cdot x_3   & \text{and} \\[-.1cm]
			\eta^{-1}: & x_1 \mapsto 2 \cdot x_1 - x_3, & x_2 \mapsto \tfrac{1}{2}\cdot x_2, & x_3 \mapsto -x_1-\tfrac{1}{2}\cdot x_2 + x_3. &
		\end{array}
	\]
	If we transform $\LL_{\mathit{non-twn}}$ with $\eta$, we obtain the \twn-loop $\LL_{\mathit{ex}}$ in \cref{fig:loop:triangular}.
	If $\ring = \RA$, then $\LL_{\mathit{ex}}$ terminates on $\RA^3$ iff $\LL_{\mathit{non-twn}}$ terminates on $\RA^3$ by \cref{thm:termination_conjugation}.
	Thus, as seen in \cref{ex:running_ent_finish}, $\LL_{\mathit{non-twn}}$ does not terminate on $\RA^3$.
	Now assume $\ring = \ZZ$, i.e., we analyze termination of $\LL_{\mathit{non-twn}}$ on $\ZZ^3$ instead of $\RA^3$.
	Note that $\widehat{\eta}$ maps $\ZZ^3$ to the set of all $\ZZ$-linear combinations of columns of $M$, i.e.,
	\[
		\mbox{$\widehat{\eta}(\ZZ^3)=\left\{a \cdot (1,0,1) + b \cdot (1,2,2) + c \cdot (1,0,2) \relmiddle{\vert{}} a,b,c \in \ZZ \right\}$.}
	\]
	By \cref{coro:conclusion_transformations}, $\LL_{\mathit{ex}}$ terminates on $\widehat{\eta}(\ZZ^3)$ iff $\LL_{\mathit{non-twn}}$ terminates on $\ZZ^3$.
	Moreover, $\widehat{\eta}(\ZZ^3)$ is $\EFO{\ZZ,\RA}$-definable: We have $(x_1,x_2,x_3) \in \widehat{\eta}(\ZZ^3)$ iff
	\begin{equation}
		\label{etaZexp}
		\exists a,b,c\in \ZZ.~ x_1= a \cdot 1 + b \cdot 1 + c \cdot 1 \land x_2 = b \cdot 2 \land x_3 = a\cdot 1 + b \cdot 2 + c \cdot 2.
	\end{equation}

	We will discuss how to analyze termination of \tnn-loops like $\LL_{\mathit{ex}}$ on update-invariant and $\EFO{\ZZ,\RA}$-definable sets like $\widehat{\eta}(\ZZ^3)$ in \cref{subsec:adaptions}.
\end{example}

To summarize, if a loop is \twn-transformable, then we can \emph{always} find a suitable automorphism via \cref{thm:triangularizability_with_fixed_degree}.
So whenever \cref{thm:strongly_nilpotent_jacobian} is applicable, a suitable linear automorphism can also be obtained by \cref{thm:triangularizability_with_fixed_degree}.
Hence, our first technique from \cref{thm:triangularizability_with_fixed_degree}
subsumes our second one from \cref{thm:strongly_nilpotent_jacobian}.
However, while \cref{thm:triangularizability_with_fixed_degree} is \emph{always} applicable, \cref{thm:strongly_nilpotent_jacobian} is \emph{easier} to apply.
The reason is that for \cref{thm:triangularizability_with_fixed_degree} one has to check
validity of a possibly \emph{non-linear} formula over the reals, where the degree of the
occurring polynomials depends on $\delta$ and the update $\update$ of the loop, and the
number of variables can be exponential in $d$, see \cref{thm:deg_inverse,lem:deciding_automorphism}.
So even when searching for a linear automorphism, one may obtain a non-linear formula if the loop is non-linear.
In contrast, \cref{thm:strongly_nilpotent_jacobian} only requires linear algebra.
So it is preferable to first check whether the loop can be transformed into a \twn-loop $(\cond',(x_1 + p_1, \ldots, x_d + p_d))$ with $x_i \notin \VV(p_i)$ via a linear automorphism.
This check is \emph{decidable} by \cref{thm:strongly_nilpotent_jacobian}.

Note that the proof of \cref{thm:triangularizability_with_fixed_degree} is constructive.
Moreover, the proof of \cite[Thm.\ 1.6.]{vandenessen1996121} which implies \cref{thm:strongly_nilpotent_jacobian}
is also constructive:\ the idea is to use basic results from linear algebra to compute an invertible matrix $T \in \RA^{d \times d}$ such that $T\cdot J \cdot T^{-1}$ is triangular where $J$ is the Jacobian matrix $\left(\frac{\partial (\updateElem_i - x_i)}{\partial x_j}\right)_{1 \leq i,j \leq d}$.
Then $\eta$ with $\eta(\vec{x}) = T \cdot \vec{x}$ transforms the loop into the form \eqref{eq:trans-jacobi}.
Hence, \cref{thm:strongly_nilpotent_jacobian} is also constructive.
Thus, we can not only check the existence of a suitable automorphism, but we can also compute it whenever it exists.

%% file: adaptions.tex
\subsection{Analyzing \twn-Transformable Loops}
\label{subsec:adaptions}
In this section, we generalize our results from \cref{sec:deciding} to \twn-transformable loops.
Our transformation from \cref{subsec:transf,subsec:automorphisms}
transforms \twn-transformable loops over update-invariant and $\EFO{\ring,\RA}$-definable sets into \twn-loops over update-invariant and $\EFO{\ring,\RA}$-definable sets.
Thus, in this section we fix a \twn-loop $(\cond,\update)$ and such a set $F \subseteq \RA^d$.
Let $\psi_F \in \EFO{\ring,\RA}$ define $F$.

While in \cref{sec:deciding} we were concerned with the termination of loops on a set $\ring^d$ for a ring $\ring$, we now show that termination of $(\cond,\update)$ \emph{on $F$}
can also be reduced to an existential formula (from $\EFO{\ring,\RA}$).
Here, we indeed rely on the update-invariance of $F$ as otherwise eventual non-termination and non-termination of $(\cond,\update)$ on $F$ would not be equivalent.
In \cref{sec:deciding}, this equivalence is crucial since we reduce non-termination via eventual non-termination to $\EFO{\ring,\RA}$.

Again, let $\closednorm$ be the normalized closed form of $\update$.
Similar to \eqref{eq:normalized}, $(\cond,\update)$ is eventually non-terminating on $F$ iff
\begin{equation}
	\label{eq:normalized-F}
	\exists \vec{x} \in F, \, n_0 \in \NN.\ \forall n \in \NN_{>n_0}.\ \cond(\closednorm).
\end{equation}
In \cref{thm:eventual_positiveness}, we have seen that given a propositional formula $\xi$ over the atoms $\{p \triangleright 0\mid p \in \PEN{\ring}[\vec{x}], \triangleright \in \{\geq, >\}\}$, one can reduce validity of $\exists \vec{x} \in \ring^d, n_0 \in \NN.\ \forall n \in \NN_{>n_0}.\ \xi$ to validity of $\exists \vec{x} \in \ring^d.\lia(\xi) \in \EFO{\ring}$ and the resulting formula can be computed in polynomial time from $\xi$.
Thus, by using the formula $\exists \vec{x} \in \RA^d.\ \psi_F \land \lia(\xi) \in \EFO{\ring,\RA}$ instead, we obtain \cref{coro:eventual_positiveness-F}.

\begin{corollary}[Reducing Eventual Non-Termination on a Set]
	\label{coro:eventual_positiveness-F}
	For a propositio\-nal formula $\xi$ over $\{p \triangleright 0\mid p \in \PEN{\RA}[\vec{x}], \triangleright \in \{\geq, >\}\}$, validity of
	\[
		\exists \vec{x} \in F, n_0 \in \NN.\ \forall n \in \NN_{>n_0}.\ \xi
	\]
	can be reduced to validity of a closed formula in $\EFO{\ring, \RA}$ in polynomial time.
\end{corollary}

\noindent
Combining \eqref{eq:normalized-F} and \cref{coro:eventual_positiveness-F} yields the following refined version of \cref{thm:decidable}.

\vspace*{.01cm}

\begin{corollary}[Reducing Termination on Sets]
	\label{coro:decidable}
	Termination of \tnn- or \twn-loops, respectively, on update-invariant $\EFO{\ring, \RA}$-definable sets is reducible to $\EFO{\ring, \RA}$.
\end{corollary}

\begin{example}
	\label{ex:linear_transf_deciding}
	Reconsider \cref{ex:linear_transf}, where we have seen that $\LL_{\mathit{non-twn}}$ (see \cref{fig:loop:horrible}) terminates on $\ZZ^3$ iff $\LL_{\mathit{ex}}$ (see \cref{fig:loop:triangular}) terminates on the update-invariant and $\EFO{\ZZ,\RR}$-definable set $\widehat{\eta}(\ZZ^3) = F$ defined by the formula \eqref{etaZexp}.
	In \cref{ex:running_ent_finish}, we showed that $(-4,2,1)$ witnesses eventual non-termination of $\LL_{\mathit{ex}}$.
	As $\widehat{\eta}(-9,\linebreak 1,4) = (-9+1+4, 1 \cdot 2, -9+1\cdot 2 + 4 \cdot 2) = (-4,2,1)$, we have $(-4,2,1) \in F$.
	Furthermore, $(-9,1,4)$ witnesses eventual non-termination of $\LL_{\mathit{non-twn}}$ on $\ZZ^3$ by \cref{coro:conclusion_transformations} (c).
	Hence, $\LL_{\mathit{non-twn}}$ does \emph{not} terminate on $\ZZ^3$.
\end{example}
In addition, we get the following refined version of \cref{coro:decidability_reals}.

\begin{corollary}[(Semi-)Deciding (Non-)Termination on a Set]
	\label{coro:decidability_reals-F}
	Let $(\cond,\update)$ be a \twn-loop and let $F \subseteq \RA^d$ be update-invariant and $\EFO{\ring,\RA}$-definable.
	\begin{enumerate}
		\item[(a)] The loop $(\cond,\update)$ terminates over $\RA$ iff it terminates over $\RR$.
		\item[(b)] Termination of $(\cond,\update)$ on $F$ is decidable if $\ring = \RA$ or $\ring = \RR$.
		\item[(c)] Non-termination of $(\cond,\update)$ on $F$ is semi-decidable if $\ring = \ZZ$ or $\ring = \QQ$.
	\end{enumerate}
\end{corollary}
Of course, \cref{lem:witnesses} also holds in this setting.

\begin{corollary}
	\label{coro:witnesses-F}
	Let $\xi = \cond(\closednorm)$.
	Then $\witness \in \RA^d$ witnesses eventual non-termination of $(\cond,\update)$ on $F$ iff $\psi_F(\witness) \land \lia(\xi) (\witness)$ holds.
\end{corollary}
Finally, \cref{alg:deciding-F} generalizes \cref{alg:deciding} to \twn-transformable loops.

\vspace*{-.3cm}

\algorithmstyle{boxruled}
\begin{algorithm}
	\caption{Checking Termination on Sets}
	\label{alg:deciding-F}
	\KwIn{a \twn-transformable-loop $(\cond,\update)$ and $\psi_F \in \EFO{\ring, \RA}$}
	\KwResult{$\top$ resp.\ $\bot$ if (non-)termination of $(\cond,\update)$ on $F$ is
		proven, $?$ otherwise}
	$(\cond,\update) \assign \Tr{\eta}{\cond}{\update}$, $\psi_F \assign \psi_{\widehat{\eta}(F)}$, such
	that $(\cond,\update)$ becomes \twn\;
	\lIf{$(\cond, \update)$ is not \tnn}{$(\cond, \update) \assign (\cond \land \cond(\update), \update(\update))$}
	$\closed \assign \text{closed form for }\update$\;
	\lIf{(in)validity of $\exists \vec{x} \in \RA^d.\ \psi_F\land\lia(\cond(\closednorm))$ cannot be
		proven}{\Return $?$}
	\leIf{$\exists \vec{x} \in \RA^d.\ \psi_F \land \lia(\cond(\closednorm))$ is valid}{\Return $\bot$}{\Return $\top$}
\end{algorithm}

\vspace*{-.5cm}

%% file: linearizing.tex
\section{Linearization of \tnn-Loops}
\label{subsec:linearizing}

In \cite{oliveira16}, a technique was proposed to linearize polynomial loops.
As mentioned, by combining the linearization with existing decidability results \cite{xiaz10,dblp:conf/iccsa/wusbz10}, one can conclude decidability of termination for \emph{conjunctive} \twn-loops over $\RR$ (whereas \cref{coro:decidability_reals} (b) extends this result also to non-conjunctive loops).

In this section, we adapt the linearization technique of \cite{oliveira16} to our setting and formalize it.
This allows us to obtain novel results on the complexity of li\-nearization which we use to analyze the complexity of deciding termination for\linebreak
arbitrary \twn-loops in \cref{subsec:arbitrary-twn}.
We start with the definition of \emph{linearization}.

\begin{definition}[Linearization]
	\label{def:linearization}
	Let $\update \in (\ring[\vec{x}])^d$ and $\vec{y}$ be a vector of $d'$ fresh variables with $d' \geq d$.
	Let $\update\,' \in (\ring[\vec{y}])^{d'}$ be \emph{linear} and $\vec{w}
		\in (\ring[\vec{x}])^{d'}$ with $w_i = x_i$ for all $1 \leq i \leq d$. Then $\update\,'$ is a \emph{linearization of $\update$ via $\vec{w}$} if $\vec{w}(\update(\witness)) = \update\,'(\vec{w}(\witness))$ holds for all $\witness \in \ring^d$, where $\update\,'(\vec{w}(\witness))$ stands for $\update\,'[\vec{y}/\vec{w}(\witness)]$.
	Instead of $y_i$ we often write $y_{w_i}$ for all $1 \leq i \leq d'$.
\end{definition}
So $y_1,\ldots,y_d$ (i.e., $y_{w_1}, \ldots, y_{w_d}$) correspond to the variables $x_1, \ldots, x_d$, whereas $y_{d+1},\ldots,y_{d'}$ are used to mimic the non-linear part of $\update$ in a linear way in $\update\,'$.
This non-linear behavior is captured by the polynomials $w_{d+1},\ldots,w_{d'}$.

\begin{example}
	\label{ex:linearization}
	Let $\update=(x_2^2,x_3^2,x_3) \in (\ZZ[x_1,x_2,x_3])^3$.
	Then $\update\,' = (y_{x_2^2}, y_{x_3^2}, y_{x_3},\linebreak
		y_{x_3^4}, y_{x_3^2}, y_{x_3^4})$ over the variables $(y_{x_1},y_{x_2},y_{x_3},y_{x_2^2}, y_{x_3^2}, y_{x_3^4})$ is a linearization of $\update$ via $\vec{w} = (x_1, x_2, x_3, x_2^2,x_3^2,x_3^4)$, since for all $\witness = (e_1,e_2,e_3) \in \ZZ^3$ we have:
	\[
		\vec{w}(\update(\witness)) = (e_2^2,e_3^2,e_3,e_3^4,e_3^2,e_3^4) = \update\,'(e_1,e_2,e_3,e_2^2,e_3^2,e_3^4) = \update\,'(\vec{w}(\witness)).
	\]
	Here, the non-linear part of $\update$ is mimicked by the variables $y_{x_2^2}$, $y_{x_3^2}$, and $y_{x_3^4}$.
\end{example}

The linearization of \cref{def:linearization} also works when applying the update repeatedly.

\begin{corollary}[Iterated Update of Linearization]
	\label{coro:lin_closed_forms}
	Let $\update \in (\ring[\vec{x}])^d$ and $\update\,' \in (\ring[\vec{y}])^{d'}$ be its linearization via $\vec{w} \in (\ring[\vec{x}])^{d'}$.
	Then for all $\witness \in \ring^d$ and all $n \in \NN$ we have $\vec{w}(\update^n(\witness)) = (\update\,')^n(\vec{w}(\witness))$.
\end{corollary}
\makeproof{coro:lin_closed_forms}{
	The proof is by induction on $n$.
	The induction base $n=0$ is trivial.
	In the induction step $n > 0$ we obtain
	\begin{align*}
		     & \vec{w}(\update^{n+1}(\witness))                                                                                 \\[-.1cm]
		= \; & \vec{w}(\update^{n}(\update(\witness)))                                                                          \\[-.1cm]
		= \; & (\update\,')^n(\vec{w}(\update(\witness))) \tag{by the induction hypothesis for $\update(\witness) \in \ring^d$} \\[-.1cm]
		= \; & (\update\,')^n( \update\,'(\vec{w}(\witness))) \tag{by \cref{def:linearization}}                                 \\[-.1cm]
		= \; & (\update\,')^{n+1}(\vec{w}(\witness))
	\end{align*}

	\vspace*{-.5cm}
}

We now define the linearization of a loop to be a linearization of its update where the loop guard is extended to ensure that the fresh variables $y_{w_{d+1}}, \ldots, y_{w_{d'}}$ indeed correspond to $w_{d+1}, \ldots, w_{d'}$.

\begin{definition}[Linearization of a Loop]
	\label{def:linearization_loop}
	Let $(\cond,\update)$ be a loop on $\ring^d$ using the variables $\vec{x}$.
	A loop $(\cond',\update\,')$ on $\ring^{d'}$ using the variables $\vec{y}$ is a \emph{linearization of $(\cond,\update)$ via $\vec{w}\in (\ring[\vec{x}])^{d'}$}
	if both
	\begin{enumerate}
		\item[(a)] $\update\,'$ is a linearization of $\update$ via $\vec{w}$
		\item[(b)] $\cond' = \cond[x_1/y_{x_1}, \ldots, x_d/y_{x_d}] \land \bigwedge_{i = d+1}^{d'} \left(y_{w_i} - w_i[x_1/y_{x_1}, \ldots, x_d/y_{x_d}] = 0\right)$.\label{lin:cond}
	\end{enumerate}
\end{definition}

\begin{example}
	\label{ex:linearization_loop}
	Consider the loop $(\cond,\update)$ on $\ZZ^3$ where $\cond$ is $x_2 > x_3$ and $\update=(x_2^2,x_3^2,x_3)$.
	Then the linearization of $(\cond,\update)$ via $\vec{w}$ is $(\cond',\update\,')$ where $\update\,'$ is as in \cref{ex:linearization} and $\cond'$ is
        \begin{samepage}
	  \begin{align*}
            & \cond[x_1/y_{x_1}, x_2/y_{x_2}, x_3/y_{x_3}] & \land \; y_{x_2^2} - y_{x_2}^2 = 0 \land y_{x_3^2} - y_{x_3}^2 = 0 \land y_{x_3^4} - y_{x_3}^4 = 0 \\
            {} = {} & \hspace*{1cm} y_{x_2} > y_{x_3}              & \land \; y_{x_2^2} - y_{x_2}^2 = 0 \land y_{x_3^2} - y_{x_3}^2 = 0 \land y_{x_3^4} - y_{x_3}^4 = 0
          \end{align*}
        \end{samepage}
	To illustrate the correspondence between $(\cond,\update)$ and $(\cond',\update\,')$, consider the initial value $\vec{e} = (1,3,2)$. Here, the original loop yields the trace $(\vec{e}, \, \vec{u}(\vec{e}), \, \vec{u}^2(\vec{e}),\linebreak \ldots)
		= ((1,3,2), \, (9,4,2), \, (16,4,2), \,\ldots)$. The linearized loop
                operates over the variables $(y_{x_1},y_{x_2},y_{x_3},y_{x_2^2},
                y_{x_3^2}, y_{x_3^4})$. Thus, the first three variables correspond to
                $x_1, x_2, x_3$ and the latter ones correspond to $x_2^2, x_3^2,
                x_3^4$. So the corresponding initial value is $\vec{e}\,' =
                (1,3,2,9,4,16)$ and the resulting trace is $(\vec{e}\,', \,
                \vec{u}\,'(\vec{e}\,'),\linebreak (\vec{u}\,')^2(\vec{e}), \,
		\ldots) = ((1,3,2,9,4,16), \; (9,4,2,16,4,16), \; (16,4,2,16,4,16), \;\ldots)$.
\end{example}

\cref{lemma:linearizing_equivalent_behavior} shows that linearization preserves the termination behavior.

\begin{lemma}
	\label{lemma:linearizing_equivalent_behavior}
	Let $(\cond',\update\,')$ on $\ring^{d'}$ be a linearization of $(\cond,\update)$ on $\ring^d$ via $\vec{w}$.
	\begin{enumerate}
		\item[(a)] $(\cond',\update\,')$ terminates on $\witness\,' \in \ring^{d'}$ if there is no $\witness \in \ring^d$ such that $\witness\,'= \vec{w}(\witness)$.
		\item[(b)] The loop $(\cond,\update)$ terminates on $\witness \in \ring^d$ iff $(\cond',\update\,')$ terminates on $\vec{w}(\witness)$.
	\end{enumerate}
\end{lemma}
\makeproof{lemma:linearizing_equivalent_behavior}{
	\begin{enumerate}
		\item[(a)] If there is no $\witness \in \ring^d$ such that $\witness\,'= \vec{w}(\witness)$, then $\cond'[\vec{y}/\witness\,']$ is $\FALSE$ by \cref{def:linearization_loop}.
		\item[(b)] For any $n \in \NN$, we have
			\[
				\begin{array}{rll}
					                                                                                                  & \cond'[\vec{y}/(\update\,')^n(\vec{w}(\witness))]                                                                                        \\[-.1cm]
					\iff                                                                                              & \cond'[\vec{y}/\vec{w}(\update^n(\witness))]                                                  & \text{(by \cref{coro:lin_closed_forms})} \\[-.1cm]
					\iff                                                                                              & \cond[\vec{x}/\update^n(\witness)] \land \bigwedge\nolimits_{i = d+1}^{d'}
					\left(y_{w_i}[\vec{y}/\vec{w}(\update^n(\witness))] - w_i[\vec{x}/\update^n(\witness)] = 0\right) & \text{(by \cref{def:linearization_loop})}                                                                                                \\[-.1cm]
					\iff                                                                                              & \cond[\vec{x}/\update^n(\witness)] \land \bigwedge\nolimits_{i = d+1}^{d'}
					\left(w_i(\update^n(\witness)) - w_i(\update^n(\witness)) = 0\right)                                                                                                                                                                         \\[-.1cm]
					\iff                                                                                              & \cond[\vec{x}/\update^n(\witness)] \land \bigwedge\nolimits_{i = d+1}^{d'} \left(0 = 0\right)                                            \\[-.1cm]
					\iff                                                                                              & \cond[\vec{x}/\update^n(\witness)].
				\end{array}
			\]
			Hence, $(\cond,\update)$ terminates on $\witness \in \ring^d$ iff $(\cond',\update\,')$ terminates on $\vec{w}(\witness)$.
	\end{enumerate}

	\vspace*{-.3cm}

}

While \cref{lemma:linearizing_equivalent_behavior} proves the soundness of linearization, we now show how to
find $\update\,'$ and $\vec{w}$ automatically, where it suffices to only use monomials (instead of arbitrary polynomials) in $\vec{w}$.
A \emph{monomial} over $\vec{x}$ has the form $x_1^{z_1} \cdot \ldots \cdot x_d^{z_d}$ with $z_i \in \NN$ for all $1 \leq i \leq d$.
Let $\vec{x}^{\vec{z}}$ with $\vec{z} = (z_1,\ldots,z_d)$ abbreviate $x_1^{z_1} \cdot \ldots \cdot x_d^{z_d}$.

The original update $\update$ consists of polynomials $u_i$ to update the variable $x_i$, for all $1 \leq i \leq d$.
The linearized update $\update\,'$ consists of polynomials $u_{m}'$ to update the variables $y_m$ for all monomials $m$ in $\vec{w}$.
Here, for any monomial $m = x_1^{z_1} \cdot \ldots \cdot x_d^{z_d}$, the polynomial $u'_m$ results from
\emph{replacing} each monomial $p$ in $u_1^{z_1} \cdot \ldots \cdot
u_d^{z_d}$ (i.e., the monomial $p$ in each addend
 $c \cdot p$ of the polynomial $u_1^{z_1} \cdot \ldots \cdot
u_d^{z_d}$) by the variable $y_{p}$. More precisely, if $u_1^{z_1} \cdot \ldots \cdot
u_d^{z_d}$ has the form $c_1 \cdot p_1 +\ldots+ c_k \cdot p_k$ for monomials
$p_1,\ldots,p_k$ and numbers $c_1,\ldots,c_k \in \ring$, then $u_m' = c_1 \cdot y_{p_1}
+\ldots+ c_k \cdot y_{p_k}$.

\cref{alg:linearization} summarizes the linearization procedure.
The vector $\vec{v}$ always contains those monomials $p$ for which we still have to define the linearized update $u'_p$.
So initially, $\vec{v}$ consists of the original variables, i.e., $\vec{v} = \vec{x}$.
Whenever a new variable $y_p$ is introduced in the linearized update, $p$ is inserted into $\vec{v}$ at the end.

\begin{algorithm}
	\caption{Linearizing \tnn-Loops}
	\label{alg:linearization}
	\SetAlgoLined
	\LinesNumbered
	\KwIn{\tnn-loop $(\cond,\update)$ using the variables $\vec{x}$}
	\KwOut{\parbox[t]{7cm}{linear-update \tnn-loop $(\cond',\update')$
			and $\vec{w}$\\such that $(\cond',\update')$ is a linearization of
			$(\cond,\update)$ via $\vec{w}$}
	}
	$\vec{v} \assign (x_1, \ldots, x_d)$ and
	$\vec{w} \assign (x_1, \ldots, x_d)$\label{line:V}\;
	\While{$\vec{v} \neq (\,)$}{
	remove the first monomial
	$m = x_1^{z_1} \cdot \ldots \cdot x_d^{z_d}$
	from $\vec{v}$ and insert $m$ at the end of $\vec{w}$\;
	$u'_{m} \assign u_1^{z_1} \cdot
		\ldots \cdot u_d^{z_d}$, where the monomials $p$ in non-constant addends of $u_1^{z_1} \cdot
		\ldots \cdot u_d^{z_d}$ are replaced by $y_p$ and we insert $p$ at the end of
	$\vec{v}$ if $p$ is not yet contained in $\vec{w}$\label{line:update}
	\;
	}
	\Return{$\vec{w}$ and $( \cond[x_1/y_{x_1}, \ldots, x_d/y_{x_d}] \land
		\qquad \bigwedge \limits_{\mathclap{\substack{\text{$m = \vec{x}^{\vec{z}}$ is contained in $\vec{w}$}\\
		\text{where $m \notin \{x_1,\ldots,x_d\}$}}}} \qquad
		(y_m - \prod_{i = 1}^d y_{x_i}^{z_i} = 0),\quad \update\,')$
	}\;
\end{algorithm}
\begin{example}
	\label{ex:alg:linearization}
	We apply \cref{alg:linearization} to linearize the loop $(\cond,\update)$ from \cref{ex:linearization_loop} where $\cond$ is $x_2 > x_3$ and $\update = (x_2^2, x_3^2, x_3)$.
	In the beginning, we have $\vec{v}=(x_1, x_2, x_3)$.
	We start with $x_1$ and remove it from $\vec{v}$.
	In $u_1 = x_2^2$ we have to replace the monomial $x_2^2$ by the fresh variable $y_{x_2^2}$ when constructing $u'_{x_1}$.
	Hence, $u'_{x_1} = y_{x_2^2}$ and we obtain $\vec{v} = (x_2, x_3, x_2^2)$.

	Next, we consider $x_2$, where $u_2 =x_3^2$.
	Thus, we obtain $u'_{x_2} = y_{x_3^2}$ and $\vec{v} = (x_3, x_2^2, x_3^2)$.
	Then we take $x_3$, where $u_3 = x_3$.
	Hence, $u'_{x_3} = y_{x_3}$, but we do not insert $x_3$ into $\vec{v}$ again, since we just computed $u'_{x_3}$.
	So we have $\vec{v} = (x_2^2, x_3^2)$.

	We now handle $x_2^2$.
	For the linearized update, we take $u_2^2 = x_3^4$ but replace the monomial $x_3^4$ by a fresh variable $y_{x_3^4}$.
	Hence, $u'_{x_2^2} = y_{x_3^4}$ and $\vec{v} = ( x_3^2, x_3^4)$.

	Next, we take the monomial $x_3^2$ and in $u_3^2 = x_3^2$ we have to replace the monomial by $y_{x_3^2}$.
	This leads to the linearized update $u'_{x_3^2} = y_{x_3^2}$, but we do not insert $x_3^2$ into $\vec{v}$ again, since we just computed $u'_{x_3^2}$.
	Hence, we have $\vec{v} = ( x_3^4)$.

	Finally, we treat $x_3^4$ and in $u_3^4 = x_3^4$ we replace the monomial by $y_{x_3^4}$, i.e., $u'_{x_3^4} = y_{x_3^4}$.
	Now $\vec{v}=(\,)$. Hence, \cref{alg:linearization} terminates and returns the loop with the guard $\cond'$ from \cref{ex:linearization_loop} and the (linear) update $(y_{x_2^2}, y_{x_3^2}, y_{x_3}, y_{x_3^4}, y_{x_3^2}, y_{x_3^4})$ over the variables $(y_{x_1}, y_{x_2}, y_{x_3}, y_{x_2^2}, y_{x_3^2}, y_{x_3^4})$, as in \cref{ex:linearization}.
\end{example}

Now we infer an upper bound on \cref{alg:linearization}'s complexity.
To this end, we will show that the degrees of the monomials in $\vec{w}$ which are used for the linearization can be bounded by the maximal \emph{dependency degree} of the loop's update $\update$.
For $1 \leq i \leq d$, the dependency degree $\depdeg_{\update}(x_i)$ is the degree of $u_i$, but this degree is expressed in terms of those variables that are minimal w.r.t.\ $\succ_{\update}$.
Recall that $u_i$ has the form $c_i \cdot x_i + p_i$ where $p_i$ is a polynomial which only contains variables that are smaller than $x_i$ w.r.t.\ $\succ_{\update}$.
W.l.o.g.\ we may assume that $x_i \succ_{\update} x_j$ implies $i > j$ for all $1 \leq i,j \leq d$.
Then for every monomial $x_{i+1}^{z_{i+1}} \cdot \ldots \cdot x_d^{z_d}$ in $p_i$, the corresponding dependency degree is $z_{i+1} \cdot \depdeg_{\update}(x_{i+1}) + \ldots + z_{d} \cdot \depdeg_{\update}(x_{d})$.
The dependency degree of $p_i$ is the maximal dependency degree of its monomials.

\begin{definition}[Dependency Degree]
	\label{def:depdeg}
	Let $(\cond,\update)$ be a \twn-loop with $u_i = c_i \cdot x_i + p_i$ for all $1 \leq i \leq d$, where $p_i$ only contains variables that are smaller than $x_i$ w.r.t.\ $\succ_{\update}$.
	We define the \emph{dependency degree} w.r.t.\ $\update$ as follows:
	\begin{itemize}
		\item $\depdeg_{\update}(x_i) = \max \{ 1, \depdeg_{\update}(p_i) \}$ for all $1 \leq i \leq d$.
		\item $\depdeg_{\update}(p) = \max \{ \depdeg_{\update}(m) \mid m \text{ is a monomial in } p\}$ for every non-zero $p \in \ring[\vec{x}]$ and $\depdeg_{\update}(0) = -\infty$.
		\item $\depdeg_{\update}(x_{1}^{z_{1}} \cdot \ldots \cdot x_d^{z_d}) = \sum\nolimits_{i=1}^d z_i \cdot \depdeg_{\update}(x_i)$ for all $z_1,\ldots,z_d \in \NN$.
	\end{itemize}
\end{definition}

Since $\succ_{\update}$ is well founded by the triangularity of the loop, $\depdeg_{\update}$ is well defined: for the variables $x_i$ which are minimal w.r.t.\ $\succ_{\update}$, $p_i$ is a constant and thus, $\depdeg_{\update}(x_i) = 1$.
For other variables $x_i$ with $p_i \neq 0$, we can compute $\depdeg_{\update}(p_i)$ because $\depdeg_{\update}(x_j)$ is already known for all variables $x_j$ occurring in $p_i$.
\cref{depdegLemma}
states three easy observations on $\depdeg$.
Here, $\deg$ denotes the degree of monomials or polynomials, i.e., $\deg(x_1^{z_1} \cdot \ldots \cdot x_d^{z_d}) = z_1 + \ldots + z_d$.

\begin{lemma}
	\label{depdegLemma}
	Let $(\cond,\update)$ be a \twn-loop.
	\begin{itemize}
		\item[(a)] For every monomial $m$ over $\vec{x}$, we have $\deg(m) \leq \depdeg_{\update}(m)$.
		\item[(b)] If $\mathit{deg}$ is the maximum of 1 and the highest degree of any polynomial in $\update$, then for any $1 \leq i \leq d$ we have $\depdeg_{\update}(x_i) \leq \mathit{deg}^{d-i}$.
		\item[(c)] For $\mathit{mdepdeg} = \max \{ \depdeg_{\update}(x_i) \mid 1 \leq i \leq d \}$, we have $\mathit{mdepdeg} \leq \mathit{deg}^{d-1}$.
	\end{itemize}
\end{lemma}
\makeproof{depdegLemma}{The claim (a) is obvious.
	The claim (b) is proved by induction on $i$.
	In the induction base, let $i = d$.
	Since $x_d$ is minimal w.r.t.\ $\succ_{\update}$, we have $\depdeg_{\update}(x_d) = 1 \leq \mathit{deg}^0 = \mathit{deg}^{d-d}$.

	In the induction step $i<d$, the claim is obviously true if $p_i = 0$.
	Otherwise, we obtain:
	\[
		\begin{array}{rl}
			     & \depdeg_{\update}(x_i)                                                                                                                                     \\[-.1cm]
			=    & \max \{ 1, \depdeg_{\update}(p_i) \}                                                                                                                       \\[-.1cm]
			=    & \max (\{1\} \cup \{ \depdeg_{\update}(m) \mid m \text{ is a monomial in } p_i\})                                                                           \\[-.1cm]
			=    & \max (\{1\} \cup \{ \sum\nolimits_{j=i+1}^d z_j \cdot \depdeg_{\update}(x_j) \mid x_{i+1}^{z_{i+1}} \cdot \ldots \cdot x_d^{z_d} \text{ occurs in } p_i\}) \\[-.1cm]
			\leq & \max (\{1\} \cup \{ \sum\nolimits_{j=i+1}^d z_j \cdot \mathit{deg}^{d-j}
			\mid x_{i+1}^{z_{i+1}} \cdot \ldots \cdot x_d^{z_d} \text{ occurs in }
			p_i\}) \hspace*{.2cm} \text{(by induction hypothesis)}                                                                                                            \\[-.1cm]
			\leq & \max (\{1\} \cup \{ \mathit{deg}^{d-i-1} \cdot \sum\nolimits_{j=i+1}^d z_j \mid x_{i+1}^{z_{i+1}} \cdot \ldots \cdot x_d^{z_d} \text{ occurs in } p_i\})   \\[-.1cm]
			\leq & \max (\{1\} \cup \{ \mathit{deg}^{d-i-1} \cdot \mathit{deg} \}) \hspace*{\fill} \text{(as $\sum\nolimits_{j=i+1}^d z_j \leq \mathit{deg}$)}                \\[-.1cm]
			=    & \mathit{deg}^{d-i}
		\end{array}
	\]
	The claim in (c) immediately follows from (b).}

\begin{example}
	\label{ex:running_depdeg}
	Again, we consider a loop as in \cref{ex:alg:linearization} with update $\update = (x_2^2, x_3^2, x_3)$.
	Then, $\depdeg_{\update}(x_3) = \max \{1,\depdeg_{\update}(0) \} = \max \{1,-\infty \} = 1$, $\depdeg_{\update}(x_2) = \depdeg_{\update}(x_3^2) = 2 \cdot \depdeg_{\update}(x_3) = 2\cdot 1 = 2$, and $\depdeg_{\update}(x_1) = \depdeg_{\update}(x_2^2) = 2 \cdot \depdeg_{\update}(x_2) = 2\cdot 2 = 4$.
	So intuitively, the update of $x_1$ has degree $4$ in terms of the $\succ_{\update}$-minimal variable $x_3$ since the update of $x_2$ is quadratic in $x_3$ and $x_2^2$ then has degree $4$ w.r.t.\ $x_3$.
	Here, $\mathit{mdepdeg} = 4$ and for the maximal degree $\mathit{deg}=2$ occurring in the update, we indeed have $\mathit{mdepdeg} = \mathit{deg}^{d-1} = 2^{3-1}$.
	So the bound on $\mathit{mdepdeg}$ in \cref{depdegLemma} (c) is tight.

	As another example, consider the update $\update = (3 \cdot x_1 + 5 \cdot x_2^4 \cdot x_3^6 + 7 \cdot x_3^8,\linebreak
		x_3^2, 9)$.
	Now we have $\depdeg_{\update}(x_3) = \max \{1, \depdeg_{\update}(9)\} = \max \{1, 0\} = 1$, $\depdeg_{\update}(x_2) =\depdeg_{\update}(x_3^2) = 2 \cdot \depdeg_{\update}(x_3) = 2 \cdot 1 = 2$, and $\depdeg_{\update}(x_1) = \depdeg_{\update}(5 \cdot x_2^4 \cdot x_3^6 + 7 \cdot x_3^8) = \max \{
		\depdeg_{\update}(x_2^4 \cdot x_3^6), \depdeg_{\update}(x_3^8) \} = \max \{
		4 \cdot \depdeg_{\update}(x_2) + 6 \cdot \depdeg_{\update}(x_3), 8 \cdot \depdeg_{\update}(x_3) \}
		= \max \{
		4 \cdot 2 + 6 \cdot 1, 8 \cdot 1 \} = 14$.
\end{example}

Now we prove that \cref{alg:linearization} only constructs updates $u'_m$ for monomials $m$ with $\depdeg(m) \leq \mathit{mdepdeg}$.
Hence, this also proves termination of the algorithm since there are only finitely many such monomials, and it allows us to give a bound on the number of iterations of the algorithm's \textbf{while}-loop.

\begin{theorem}[Dependency Degree Suffices for Linearization]
	\label{thm:alg_depdeg}
	\mbox{}
	\begin{itemize}
		\item[(a)] \cref{alg:linearization} only computes $u'_m$ for monomials $m$ with $\depdeg(m) \leq \mathit{mdepdeg}$.
		\item[(b)] The
			{\normalfont{\textbf{while}}}-loop of \cref{alg:linearization} is executed at most $\binom{d + \mathit{mdepdeg}}{\mathit{mdepdeg}} - 1$ times.
		\item[(c)] \cref{alg:linearization}
			terminates.
	\end{itemize}
\end{theorem}
\makeproof{thm:alg_depdeg}{Let $z_1,\ldots,z_d \in \NN$.
We first show that for all monomials $m$ in $u_1^{z_1} \cdot \ldots \cdot u_d^{z_d}$ we have
\begin{equation}
	\label{claimMonomialDegree}
	\mbox{$\depdeg(m) \leq \depdeg(x_1^{z_1} \cdot \ldots \cdot x_d^{z_d})$.}
\end{equation}
To prove \eqref{claimMonomialDegree}, note that $m$ must have the form $m_{1,1}\cdot \ldots \cdot m_{1,z_1} \cdot \ldots \cdot m_{d,1}\cdot \ldots \cdot m_{d,z_d}$ where the monomial $m_{i,j}$ occurs in $u_i$ for all $1 \leq i \leq d$ and $1 \leq j \leq z_i$.
Therefore, we have $\depdeg(m_{i,j}) \leq \depdeg(x_i)$.
This is clear for $m_{i,j} = x_i$ and for $m_{i,j} \neq x_i$ it follows directly from the definition of the dependency degree in \cref{def:depdeg}.
Hence, we can now prove \eqref{claimMonomialDegree}:
\[
	\begin{array}{rcl}
		\depdeg(m) & =    & \depdeg(m_{1,1})+ \ldots + \depdeg(m_{d,z_d})           \\[-.1cm]
		           & \leq & z_1 \cdot \depdeg(x_1)+ \ldots + z_d \cdot \depdeg(x_d) \\[-.1cm]
		           & =    & \depdeg(x_1^{z_1} \cdot \ldots \cdot x_d^{z_d}).
	\end{array}
\]

Now assume that (a) were not true.
Then consider the first execution of \cref{line:update}
where we compute an update $u'_m$ for a monomial $m$ with $\depdeg_{\update}(m) > \mathit{mdepdeg}$.
The monomial $m$ resulted from taking
a monomial $x_1^{z_1}\cdot \ldots \cdot x_d^{z_d}$ from $\vec{v}$ and constructing $u_1^{z_1} \cdot \ldots \cdot u_d^{z_d}$, where $m$ occurs in $u_1^{z_1}\cdot \ldots \cdot u_d^{z_d}$.
Since $m$ is the first monomial whose dependency degree is greater than $\mathit{mdepdeg}$, we have $\depdeg_{\update}(x_1^{z_1}\cdot \ldots \cdot x_d^{z_d}) \leq \mathit{mdepdeg}$.
But by \eqref{claimMonomialDegree}, this implies $\depdeg_{\update}(m) \leq \depdeg_{\update}(x_1^{z_1} \cdot \ldots \cdot x_d^{z_d}) \leq \mathit{mdepdeg}$, which contradicts our assumption.

For (b), since we do not build any update $u'_m$ for the constant monomial $m$, (a) implies that the number of non-constant monomials $m$ over the variables $\vec{x}$ with $\depdeg(m) \leq \mathit{mdepdeg}$ is an upper bound on the number of executions of the \textbf{while}-loop.
As we have $\deg(m) \leq \depdeg(m)$ for any monomial by \cref{depdegLemma} (a), this number is bounded by the number of monomials over the variables $\vec{x}$ with a degree between 1 and $\mathit{mdepdeg}$.

The number of monomials over $d$ variables with the exact degree $\mathit{mdepdeg}$ is $\binom{d + \mathit{mdepdeg} -1}{\mathit{mdepdeg}}$ (this is the number of so-called \emph{weak compositions} of $\mathit{mdepdeg}$ into $d$ parts) and the number of monomials over $d$ variables with a degree between 1 and $\mathit{mdepdeg}$ is $\binom{d}{1} + \binom{d+1}{2} + \ldots + \binom{d + \mathit{mdepdeg}
		-1}{\mathit{mdepdeg}} = \binom{d + \mathit{mdepdeg}}{\mathit{mdepdeg}} -1$.

\noindent
Termination of \cref{alg:linearization} follows from (b) since the \textbf{while}-loop is only executed finitely often.}

The following theorem summarizes the main properties of \cref{alg:linearization}.

\begin{theorem}[Soundness of \cref{alg:linearization}]
	\label{thm:twn_linearizable}
	\mbox{}
	\begin{itemize}
		\item[(a)] For any \tnn-loop $(\cond,\update)$, \cref{alg:linearization} computes a linearization $(\cond',\update')$ via $\vec{w} = (m_1,\ldots,m_{d'})$, where $\deg(m_i) \leq \depdeg(m_i) \leq \mathit{mdepdeg}$ for all $1 \leq i \leq d'$.
		\item[(b)] The loop $(\cond,\update)$ terminates iff $(\cond',\update')$ does.
		\item[(c)] The loop $(\cond',\update')$ is a linear-update \tnn-loop.
	\end{itemize}
\end{theorem}
\makeproof{thm:twn_linearizable}{The statement in (a) is obvious from \cref{alg:linearization} and \cref{thm:alg_depdeg}, the claim in (b) follows from \cref{lemma:linearizing_equivalent_behavior}, and the linearity of the update in (c) is again obvious from \cref{alg:linearization}.

It remains to show that $(\cond',\update')$ is triangular and non-negative (i.e., it is weakly non-linear and the coefficient of the monomial $y_m$ in $u_m'$ is always non-negative).

For triangularity, we again assume that $x_i \succ_{\update} x_j$ implies $i > j$ for all $1 \leq i,j \leq d$.
Then we show that $y_{m'} \succ_{\update'} y_{m}$ implies $m' \succ m$, where $\succ$ is the lexicographic ordering on monomials.
Thus, if $m' =\vec{x}^{\vec{z}\,'}$ and $m = \vec{x}^{\vec{z}}$, then $m' \succ m$ holds iff $\vec{z}\,' >_{lex} \vec{z}$.
Since $\succ$ is well founded, this implies the well-foundedness of $\succ_{\update'}$, i.e., the loop $(\cond',\update')$ is triangular.

So let $m' \neq m$ and $y_m$ occur in $u'_{m'}$.
If $m' = x_1^{z_1}\cdot \ldots \cdot x_d^{z_d}$, then this means that $m$ occurs in $u_1^{z_1}\cdot \ldots \cdot u_d^{z_d}$.
Thus, $m = m_{1,1}\cdot \ldots \cdot m_{1,z_1} \cdot \ldots \cdot m_{d,1}\cdot \ldots \cdot m_{d,z_d}$ where $m_{i,j}$ occurs in $u_i$ for all $1 \leq i \leq d$ and $1 \leq j \leq z_i$.
Hence, we have $m_{i,j} = x_i$ or $m_{i,j}$ only contains variables $x_j$ with $x_i \succ_{\update}
	x_j$. Thus, $x_i \succeq m_{i,j}$, where $\succeq$ is the reflexive closure of $\succ$. Hence, this implies $m' = x_1^{z_1}\cdot \ldots \cdot x_d^{z_d} \succeq m_{1,1}\cdot \ldots \cdot m_{1,z_1} \cdot \ldots \cdot m_{d,1}\cdot \ldots \cdot m_{d,z_d} = m$. Since $m' \neq m$, we have $m' \succ m$. As $\succ_{\update'}$ is the transitive closure of $\{(y_{m'},y_{m}) \mid m \text{ occurs in }
	u'_{m'} \}$, the claim follows.

For non-negativity, note that $u'_{x_1^{z_1}\cdot \ldots \cdot x_d^{z_d}}$ results from $u_1^{z_1}\cdot \ldots \cdot u_d^{z_d}$, where $u_i = c_i \cdot x_i + p_i$ for all $1 \leq i \leq d$, and where $c_i \geq 0$ since $(\cond,\update)$ is a \tnn-loop.
Hence, $y_{x_1^{z_1}\cdot \ldots \cdot x_d^{z_d}}$ only occurs in $u'_{x_1^{z_1}\cdot \ldots \cdot x_d^{z_d}}$ in the addend $c_1 \cdot \ldots \cdot c_d \cdot y_{x_1^{z_1}\cdot \ldots \cdot x_d^{z_d}}$.
Since this is a linear monomial and since $c_1 \cdot \ldots \cdot c_d \geq 0$, this implies non-negativity of $(\cond',\update')$.}

As mentioned, the technique in this section is based on the linearization method of \cite{oliveira16}, where instead of \tnn-loops as in \cref{thm:twn_linearizable}, \cite{oliveira16} works in the setting of solvable loops (\cref{def:solvable}).
But \cite{oliveira16}
has no notion like the dependency degree of \cref{def:depdeg}.
Instead they only consider the degree of the polynomials in the update $\vec{u}$.
However, \cref{ex:alg:linearization} shows that the polynomials in $\vec{w}$ that are used for the linearization may have a higher degree than the ones in $\vec{u}$.
Here, the polynomials in $\vec{u} = (x_2^2, x_3^2, x_3)$ only have degree 2.
However, $x_1$ is (eventually) updated to $x_3^4$.
Thus, to linearize this loop, polynomials up to degree 2 do \emph{not}
suffice, but $\vec{w}$ must contain a polynomial of degree 4 like $x_3^4$.

As we showed in \cref{thm:alg_depdeg} (a), the dependency degree (and hence, also the degree) of the polynomials in $\vec{w}$ is bounded by $\mathit{mdepdeg} = \max \{ \depdeg_{\update}(x_i) \mid 1 \leq i \leq d \}$.
Indeed, in \cref{ex:alg:linearization} we have $\depdeg_{\update}(x_1) = 4$.
Hence, our new concept of the dependency degree was needed for the upper bound on the number of iterations of the linearization algorithm in \cref{thm:alg_depdeg} (b).
Based on this, we can now infer the asymptotic complexity of \cref{alg:linearization}.
As mentioned, we will need this in \cref{subsec:arbitrary-twn} to analyze the complexity of deciding termination of \twn-loops.

By \cref{thm:alg_depdeg} (b), the \textbf{while}-loop of \cref{alg:linearization} is executed at most $\binom{d + \mathit{mdepdeg}}{\mathit{mdepdeg}} - 1$ times.
Since $\binom{n}{k} \in \OO(n^k)$ for any natural numbers $n\geq k$, we have
\[
	\mbox{$\binom{d + \mathit{mdepdeg}}{\mathit{mdepdeg}} = \binom{d + \mathit{mdepdeg}}{d} \in \OO((d + \mathit{mdepdeg})^{\mathit{mdepdeg}}) \cap \OO((d + \mathit{mdepdeg})^{d})$.}
\]
By \cref{depdegLemma} (c), we have $\mathit{mdepdeg} \leq \mathit{deg}^{d-1}$ where $\mathit{deg}$ is the maximum of 1 and the highest degree of any polynomial in the update $\update$.
Hence,
\[
	\mbox{$\binom{d + \mathit{mdepdeg}}{\mathit{mdepdeg}} \in \OO((d + \mathit{deg}^{d-1})^d) \cap \OO((d + \mathit{deg}^{d-1})^{\mathit{deg}^{d-1}}) \subseteq \OO((d + \mathit{deg}^{d-1})^d)$.}
\]
For the expression $(d + \mathit{deg}^{d-1})^d$ we have (see also \cref{eq:expspace_proof}):
\begin{equation}
	(d + \mathit{deg}^{d-1})^d \leq 2 \cdot 2^{d + \ld(\mathit{deg})\cdot (d-1)\cdot d} \label{eq:expspace}
\end{equation}
Here, as usual, $\ld$ denotes the logarithm to the base $2$.
\makeproof{eq:expspace}{
	We have
	\[
		\begin{array}{rl}
			          & (d+ \mathit{deg}^{d-1})^d                                                                 \\[-.1cm]
			{}={}     & \sum_{i=0}^d \binom{d}{i} \cdot \left(\mathit{deg}^{d-1}\right)^i \cdot d^{d-i}           \\[-.1cm]
			{}\leq{}  & \sum_{i=0}^d \binom{d}{i} \cdot \max\left\{\mathit{deg}^{d-1}, d\right\}^d                \\[-.1cm]
			{}={}     & \sum_{i=0}^d \binom{d}{i} \cdot \max\left\{\left(\mathit{deg}^{d-1}\right)^d, d^d\right\} \\[-.1cm]
			{}\leq {} & \sum_{i=0}^d \binom{d}{i} \cdot \left(\left(\mathit{deg}^{d-1}\right)^d + d^d\right)      \\[-.1cm]
			{}={}     & 2^d \cdot \left(\left(\mathit{deg}^{d-1}\right)^d + d^d\right)                            \\[-.1cm]
			{}={}     & 2^{d + \ld(\mathit{deg})\cdot (d-1)\cdot d} + 2^{d + \ld(d)\cdot d}                       \\[-.1cm]
			{}\leq{}  & 2 \cdot 2^{d + \ld(\mathit{deg})\cdot (d-1)\cdot d}
		\end{array}
	\]

	\vspace*{-.5cm}

}

Thus, $(d+\mathit{deg}^{d-1})^d$ is at most \emph{exponential} in $d$, i.e., the number of iterations of \cref{alg:linearization} is at most exponential in $d$.
In each such iteration, one has to com\-pute a new polynomial $u_1^{z_1}
	\cdot \ldots \cdot u_d^{z_d}$. By \cref{thm:alg_depdeg} (a), this polynomial only
        contains monomials $m$ with $\depdeg(m) \leq \mathit{mdepdeg}$ and there are
        $\binom{d + \mathit{mdepdeg}}{\mathit{mdepdeg}} \in \OO((d +
        \mathit{deg}^{d-1})^d)$ many such monomials (see \cref{thm:alg_depdeg} (b)). To
        compute their coefficients, one has to multiply up to $z_1 + \ldots + z_d$
        factors, where $z_1 + \ldots + z_d \leq \mathit{mdepdeg} \leq
        \mathit{deg}^{d-1}$. This corresponds to a nested multiplication of two factors,
        where the result of one multiplication step is the input to the next
        multiplication, and the depth of the nesting is exponential in $d$. So the results
        and the factors of the multiplications grow at most doubly exponentially in
        $d$. Therefore, this proves \cref{lem:linearizing-complexity} (a), i.e., the
        runtime of \cref{alg:linearization} is at most double exponential.

However, if the number of variables $d$ is bounded by a constant $D$, then the number of iterations of \cref{alg:linearization}
and the number of monomials in the linearized updates is bounded by $\binom{d + \mathit{mdepdeg}}{\mathit{mdepdeg}} \in \OO((\mathit{deg}^D + D)^D)$, which is polynomial in $\mathit{deg}$.
For their coefficients, one has to multiply up to $\mathit{mdepdeg} \leq
\mathit{deg}^{D-1}$ (i.e., polynomially) many factors, i.e., this corresponds to a nested
multiplication where the depth of the nesting is polynomial in $\mathit{deg}$. So the
results and the factors of the multiplications grow at most exponentially in
$\mathit{deg}$. Therefore, then linearization can be computed in exponential time.
This proves \cref{lem:linearizing-complexity} (b).
\begin{lemma}
	\label{lem:linearizing-complexity}
	Let $D\in\NN$ be fixed.
	The linearization of a \tnn-loop
	\begin{enumerate}
	\item[(a)] can be computed in double exponential time.\label{it:general_linearization}
	\item[(b)] can be computed in exponential time
          if the number of variables $d$ is at most $D$.\label{it:bounded_vars_linearization}
	\end{enumerate}
\end{lemma}

\begin{example}
	\label{non-pspace linearization}
	While \cref{lem:linearizing-complexity} only gives upper bounds on the complexity of linearization, the loop $\LL_{\mathit{non-pspace}}$ from \eqref{fig:loop:non-poly-ex} can be used to infer lower bounds.
	Here, the linearized loop operates on the variables
	\[
		y_{x_1},\ldots,y_{x_d}, \;\; y_{x_2^d},\ldots,y_{x_d^d}, \;\; y_{x_3^{(d^2)}},\ldots,y_{x_d^{(d^2)}}, \;\; \ldots \;\; y_{x_{d-1}^{(d^{d-2})}}, y_{x_d^{(d^{d-2})}}, \;\; y_{x_{d}^{(d^{d-1})}}
	\]
	and the corresponding linearized update $\update'$ instantiates
	\begin{itemize}
		\item $y_{x_1}$ by $y_{x_2^d}$,
		\item $y_{x_2^{(d^i)}}$ with $y_{x_3^{(d^{(i+1)})}}$ for all $0 \leq i \leq 1$,
		\item \ldots
		\item $y_{x_{d-1}^{(d^i)}}$ with $y_{x_d^{(d^{(i+1)})}}$ for all $0 \leq i \leq d-2$, and
		\item $y_{x_d^{(d^i)}}$ with $d^{(d^i)} \cdot y_{x_d^{(d^i)}}$ for all $0 \leq i \leq d-1$.
	\end{itemize}
	So in particular, the update contains the constant $d^{(d^{d-1})}$ which shows
        that this linearization requires exponential space
        if $d$ is not bounded.
\end{example}

%% file: complexity.tex
\section{Complexity of Deciding Termination}
\label{sec:complexity}

In this section, we study the complexity of deciding termination for different classes of loops by using our results from \cref{sec:deciding,sec:transf,subsec:linearizing}.
We first regard \emph{linear-update} loops in \cref{subsec:linear-update}, where the update is of the form $\vec{x} \leftarrow A \cdot \vec{x} + \vec{b}$ with $A \in\linebreak
	\ring^{d \times d}$ and $\vec{b} \in \ring^d$.
The reason for this restriction is that such loops can always be transformed into \twn-form by our transformation $\TrOp$ from \cref{sec:transf}.
More precisely, we show that termination of linear loops with rational spectrum is \cc{Co-NP}-complete if $\ring \in \{ \ZZ, \QQ, \RA \}$ and that termination of linear-update loops with real spectrum is $\forall\RR$-complete if $\ring = \RA$.
Since our proof is based on a reduction to $\EFO{\ring,\RA}$, and $\RA$ and $\RR$ are elementary equivalent, our results also hold if the program variables range over $\RR$.
By combining these results with our observations on the complexity of linearization from \cref{subsec:linearizing}, we then analyze the complexity of deciding termination for arbitrary \twn-loops in \cref{subsec:arbitrary-twn}.
In \cref{subsec:uniform}, we show that there is an important subclass of linear loops where our decision procedure for termination works efficiently, i.e., when the number of eigenvalues of the update matrix is bounded, then termination can be decided in polynomial time.
Here, we again use our transformation $\TrOp$ from \cref{sec:transf}.

For our complexity results, we assume the usual dense encoding of univariate polynomials, i.e., a polynomial of degree $k$ is represented as a list of $k+1$ coefficients.
As discussed in \cite{dblp:conf/issac/roche18}, many problems which are considered to be efficiently solvable become intractable if polynomials are encoded sparsely (e.g., as lists of monomials where each monomial is a pair of its non-zero coefficient and its degree).
With densely encoded polynomials, all common representations of algebraic numbers can be converted into each other in polynomial time \cite{handbookmishra}.

\subsection{Complexity of Deciding Termination for Linear-Update Loops}
\label{subsec:linear-update}
\begin{figure}[t]
	\begin{minipage}[t]{0.4\linewidth}
		\vspace*{.14cm}
		\minialg{\linewidth}{}{
			$\WHILEDO{\cond}{
					{\ASSIGN{\vec{x}}{A \cdot \vec{x} + \vec{b}}}}$}
		\vspace*{-.22cm}
		\captionof{figure}{Linear-Update Loop}
		\label{fig:loop:aff}
	\end{minipage}
	\begin{minipage}[t]{0.6\linewidth}
		\vspace*{0pt}
		\minialg{\linewidth}{}{
			$\WHILEDO{\cond \land x_{\vec{b}} = 1}{
					{\ASSIGN{
								\begin{sbmatrix}
									\vec{x} \\
									x_{\vec{b}}
								\end{sbmatrix}
							}{
								\begin{sbmatrix}
									A & \vec{b} \\
									\vec{0}^T & 1
								\end{sbmatrix}
								\cdot
								\begin{sbmatrix}
									\vec{x} \\
									x_{\vec{b}}
								\end{sbmatrix}
							}}}$}
		\vspace*{-.3cm}
		\captionof{figure}{Homogeneous Linear-Update Loop}
		\label{fig:loop:lin}
	\end{minipage}
	\vspace*{-.3cm}
\end{figure}

When analyzing linear-update loops, w.l.o.g.\ we can assume $\vec{b} = \vec{0}$ since a loop of the form in \cref{fig:loop:aff} terminates iff the loop in \cref{fig:loop:lin} terminates, where $x_{\vec{b}}$ is a fresh variable (see \cite{dblp:conf/soda/ouakninepw15,dblp:conf/icalp/hosseinio019}).
Moreover, $\witness$ witnesses (eventual) non-termination for the loop in \cref{fig:loop:aff} iff $
	\begin{sbmatrix}
		\witness\\
		1
	\end{sbmatrix}
$ witnesses (eventual) non-termination for the loop in \cref{fig:loop:lin}.
Note that the only eigenvalue of $
	\begin{sbmatrix}
		A & \vec{b} \\
		\vec{0}^T & 1
	\end{sbmatrix}
$ whose multiplicity increases in comparison to $A$ is $1$.
Thus, to decide termination of linear-update loops with rational or real spectrum, respectively, it suffices to decide termination of \emph{homogeneous} loops of the form $(\cond, A \cdot \vec{x})$ where $A$ has only rational or real eigenvalues.

Such loops can \emph{always} be transformed into \twn-form using our transformation $\TrOp$ from \cref{sec:transf}.
To compute the required automorphism $\eta$, we compute the Jordan normal form $Q$ of $A$ together with the corresponding transformation matrix $T$, i.e., $T$ is an invertible real matrix such that $A=T^{-1} \cdot Q \cdot T$.
Then $Q$ is a triangular real matrix whose diagonal consists of the eigenvalues $\lambda \in \RA$ of $A$.
We define $\eta \in \End{\RA}$ by $\eta(\vec{x}) = T \cdot \vec{x}$.
Then $\eta \in \Aut{\RA}$ has the inverse $\eta^{-1}(\vec{x}) = T^{-1} \cdot \vec{x}$.
Thus, $\Trnv{\eta}{\cond}{A \cdot \vec{x}}$ is a \twn-loop with the update
\[
	(\eta(\vec{x})) \, (A \cdot \vec{x}) \, (\eta^{-1}(\vec{x})) = T \cdot A \cdot T^{-1} \cdot \vec{x} = Q \cdot \vec{x}.
\]
To analyze termination of the loop on $\ring^d$, we have to consider termination of the transformed loop on $F = \widehat{\eta}(\ring^d) = T \cdot \ring^d$ (see \cref{coro:conclusion_transformations}).

The Jordan normal form $Q$ as well as the matrices $T$ and $T^{-1}$ can be computed in polynomial time \cite{jordan,dblp:journals/siamcomp/giesbrecht95}.
Hence, we can decide whether all eigenvalues are rational or real numbers in polynomial time by checking the diagonal entries of $Q$.
Thus, we obtain the following lemma.
\begin{lemma}
	\label{lem:eigenvalues}
	Let $(\cond, A \cdot \vec{x})$ be a linear-update loop.
	\begin{enumerate}
		\item[(a)] It is decidable in polynomial time whether $A$ has only rational or real eigenvalues.
		\item[(b)] If $A$ has only real eigenvalues, we can compute a linear $\eta \in \Aut{\RA}$ such that $\Trnv{\eta}{\cond}{A\cdot \vec{x}}$ is a linear-update \twn-loop in polynomial time.
		\item[(c)] If $(\cond, A \cdot \vec{x})$ is a linear loop, then so is $\Trnv{\eta}{\cond}{A\cdot \vec{x}}$.
	\end{enumerate}
\end{lemma}
So every linear(-update) loop with real spectrum can be transformed into a linear(-update) \twn-loop, i.e., the transformation $\TrOp$ from \cref{sec:transf} is \emph{complete} for such linear(-update) loops.
Note that \cref{lem:eigenvalues} (a) yields an efficient check whether a given linear(-update) loop has rational or real spectrum.

As chaining (\cref{def:chaining}) can clearly be done in polynomial time, w.l.o.g.\ we may assume that $\Trnv{\eta}{\cond}{A \cdot \vec{x}} = (\cond', Q \cdot \vec{x})$ is \tnn.
Next, to analyze termination of a \tnn-loop, our technique of \cref{sec:deciding} (resp.\ \cref{subsec:adaptions}) uses a closed form for the update.
For \tnn-loops $(\cond', Q \cdot \vec{x})$ where $Q$ is a triangular matrix with non-negative diagonal entries, a suitable (i.e., poly-exponential) closed form can be computed in polynomial time analogously to \cite[Prop.\ 5.2]{kincaid19}.
This closed form is linear in $\vec{x}$ (we will discuss this closed form in \cref{subsec:uniform}, see \cref{coro:closed_uniform_hierar_exp}).

According to our approach in \cref{subsec:adaptions}, we now proceed as in \cref{alg:deciding-F} and construct the formula $\exists \vec{x} \in \RA^d.\ \psi_F \land \lia(\cond(\closednorm)) \in \EFO{\ring,\RA}$ in polynomial time due to \cref{coro:eventual_positiveness-F}.
Hence, we get the following lemma.

\begin{lemma}
	\label{lem:polynomial_time_reduction}
	Let $(\cond, A\cdot \vec{x})$ be a linear-update loop with real spectrum.
	Then we can compute a closed formula $\psi \in \EFO{\ring,\RA}$ in polynomial time such that $\psi$ is valid iff the loop is non-terminating.
	If $\cond$ is linear, then so is $\psi$.
\end{lemma}
Note that $\psi$ is existentially quantified.
Hence, if the loop has rational spectrum and coefficients, $\cond$ (and thus also $\psi$) is linear, and $\ring \in \{ \ZZ, \QQ, \RA, \RR \}$, then invalidity of $\psi$ is in \cc{Co-NP} as validity of such formulas is in \cc{NP}, see \cite{dblp:journals/mp/piadm17}.
Thus, we get the first main result of this section.
Here, we fix an inaccuracy in \cite[Thm.\ 42]{sas}, where we also allowed irrational eigenvalues and coefficients, and thus $\psi$ may contain irrational coefficients.
However, to the best of our knowledge, it is not known whether validity of linear formulas from $\EFO{\ring,\RA}$ with irrational algebraic coefficients is in \textnormal{\cc{NP}}.
\begin{theorem}[\cc{Co-NP}-Completeness]
	\label{thm:co_np_completeness}
	For linear loops $(\cond, A \cdot \vec{x} + \vec{b})$ with rational spectrum where $\cond \in \QFFO{\QQ}, A \in \QQ^{d \times d}$, and $\vec{b} \in \QQ^d$, termination over $\ZZ$, $\QQ$, $\RA$, and $\RR$ is \textnormal{\cc{Co-NP}}-complete.
\end{theorem}

\noindent
\cc{Co-NP}-hardness follows from \cc{Co-NP}-hardness of unsatisfiability of propositional formulas:\ let $\xi$ be a propositional formula over the variables $\vec{x}$.
Then the loop $(\xi[x_i / (x_i > 0) \mid 1 \leq i \leq d], \vec{x})$ ter\-mi\-nates iff $\xi$ is unsatisfiable.

We now consider linear-update loops with real spectrum (and possibly non-linear conditions) on $\RA^d$ and $\RR^d$.
Here, non-termination is $\exists\RR$-complete.%
\begin{definition}[$\exists\RR$ \cite{existsr,forallr}]
	\label{def:existsr}
	Let
        \[
          \EFO{\RR}_\top = \{\psi \in \EFO{\RR} \mid \psi \text{ closed and valid}\}.
        \]
	The complexity class $\exists \RR$ is the closure of $\EFO{\RR}_\top$ under poly-time-reductions.
\end{definition}
We have $\cc{NP} \subseteq \exists \RR \subseteq \cc{PSPACE}$ (see \cite{dblp:conf/stoc/canny88}).
By \cref{lem:polynomial_time_reduction}, non-termination of linear-update loops over $\RA$ with real spectrum is in $\exists\RR$.
It is also $\exists\RR$-hard since $(\cond,\vec{x})$ is non-terminating iff $\exists \vec{x} \in \RA^d.\, \cond$ is valid.
So non-termination is $\exists\RR$-complete, i.e., termination is \cc{Co-}$\exists\RR$-complete (where \cc{Co-}$\exists\RR = \forall\RR$ \cite{forallr}).
\begin{theorem}[$\forall\RR$-Completeness]
	\label{thm:forall_r_completeness}
	Termination of linear-update loops with real spectrum over $\RA$ and $\RR$ is $\forall\RR$-complete.
\end{theorem}

\subsection{Complexity of Deciding Termination for \twn-Loops over $\ring \in \{\RA,\RR\}$}
\label{subsec:arbitrary-twn}

By \cref{coro:decidability_reals} (b) termination is decidable for arbitrary \twn-loops over $\RA$ and $\RR$.
So in this section, we discuss the complexity of this decision problem.
First of all, deciding termination of arbitrary \twn-loops over $\RA$ and $\RR$ is $\forall\RR$-hard since termination of linear-update loops with real spectrum over $\RA$ and $\RR$ is $\forall\RR$-hard by \cref{thm:forall_r_completeness} and any such linear-update loop can be transformed into a \twn-loop in polynomial time by \cref{lem:eigenvalues} (b).
Thus, we have a lower bound on the complexity of deciding termination for arbitrary \twn-loops.

To find an upper asymptotic bound for deciding termination of \twn-loops, first note that we can restrict ourselves to \tnn-loops again, as any \twn-loop can be transformed into a \tnn-loop by chaining (\cref{def:chaining}) in polynomial time.

The next step of our decision procedure in \cref{alg:deciding} is to compute closed forms for the update.
However, the loop $\LL_{\mathit{non-pspace}}$ in \eqref{fig:loop:non-poly-ex}
showed that in general, the computation of closed forms
cannot be done in polynomial space.
On the other hand, as mentioned in \cref{subsec:linear-update}, for \emph{linear-update}
loops, closed forms can be computed in polynomial time.
To benefit from this upper bound, we therefore do not proceed directly as in \cref{alg:deciding}, but instead we first linearize the \tnn-loop.
While linearization cannot be computed in polynomial space either
(see \cref{non-pspace linearization}),
in \cref{subsec:linearizing} we formalized and analyzed the complexity of the linearization technique from \cite{oliveira16}.
Given a \tnn-loop $(\cond,\update)$ we can compute a linear-update \tnn-loop $(\cond',\update\,')$ such that $(\cond,\update)$ terminates if and only if $(\cond',\update\,')$ does (\cref{thm:twn_linearizable}).
Clearly, linear-update \tnn-loops over $\RA$ and $\RR$ always have real spectrum as the eigenvalues of a triangular matrix are its diagonal entries.
So by using this linearization, we can give an upper complexity bound for deciding termination of arbitrary \twn-loops.

As linearization can be computed in double exponential time and thus, also in double
exponential space,
and termination of linear-update loops is in $\forall \RR \subseteq $ \cc{PSPACE} $\subseteq$ \cc{EXPTIME} by \cref{lem:polynomial_time_reduction}
(where the size of the linear-update loop may be at most double exponential), we obtain that deciding termination of \twn-loops is in \cc{3-EXPTIME} (i.e., it is between $\forall\RR$ and \cc{3-EXPTIME}).
Moreover, if the number of variables is bounded, checking validity of a formula in $\EFO{\RA}$ is in \cc{P} (see \cite{complexityexistentialreal}).
In this case, combining linearization which
can be computed in exponential time and thus, also in exponential space
when $d$ is bounded (\cref{lem:linearizing-complexity} (b)), and deciding termination of linear-update loops which is polynomial in this case (where the size of the linear-update loop may be at most exponential), we obtain \cref{thm:three_exptime} (b).

\begin{theorem}[Membership in \cc{3-EXPTIME}]
	\label{thm:three_exptime}
	Let $D\in\NN$ be fixed.
	Termination of \twn-loops over $\RA$ and $\RR$
	\begin{enumerate}
		\item[(a)] is in \textnormal{\cc{3-EXPTIME}}.\label{it:general}
		\item[(b)] is in \textnormal{\cc{EXPTIME}} if the number of variables $d$ is at most $D$.\label{it:bounded_vars}
	\end{enumerate}
\end{theorem}

%% file: ptime.tex
\subsection{Complexity of Deciding Termination for Uniform Loops}
\label{subsec:uniform}
In \cref{subsec:linear-update}, we showed that termination of linear loops with rational spectrum is \cc{Co-NP}-complete.
For proving \cc{Co-NP}-hardness, we used the trivial update $\ASSIGN{\vec{x}}{\vec{x}}$ induced by the identity matrix.
Therefore, the question arises whether imposing suitable restrictions to the update matrix (which exclude the identity matrix) leads to a ``more efficient'' decision procedure for termination (assuming \cc{P} $\neq$ \cc{NP}).
We now analyze a special case of linear loops (so-called \emph{uniform} loops) and show that for these loops deciding termination is \emph{polynomial}, if one fixes the number of eigenvalues of the update matrix.

In \cref{subsubsec:problem_statement}, we introduce uniform loops and parameterized decision problems, and state the main result of \cref{subsec:uniform} (\cref{thm:xp}).
To prove it, we show that for uniform loops, instantiating the variables in the loop guard by $\closednorm$ (as required by our decision procedure from \cref{sec:deciding}) results in formulas of a special structure (so-called \emph{interval conditions}, see \cref{subsubsec:hierarchical_expressions,subsubsec:interval_conditions}).
Validity of these formulas can be checked in polynomial time (\cref{subsubsec:algorithm}) which proves \cref{thm:xp} for uniform loops over $\QQ$, $\RA$, and $\RR$.
In \cref{subsec:integer_uniform_loops} we show that our result holds for uniform loops over $\ZZ$ as well.

\subsubsection{Uniform Loops and the Parameterized Complexity Class \normalfont{\cc{XP}}}
\label{subsubsec:problem_statement}
\begin{definition}[Uniform Loop]
	A linear loop $(\cond,A \cdot \vec{x})$ over $\ring \in \{ \ZZ, \QQ, \RA \}$ is \emph{uniform} if each eigenvalue $\lambda$ of $A$ is a \emph{non-negative} number from $\ring$ whose eigenspace w.r.t.\ $A$ is one-dimensional, i.e., $\lambda$ has \emph{geometric multiplicity} $1$.
\end{definition}

The latter property is equivalent to requiring that there is exactly one Jordan block for each eigenvalue in $A$'s Jordan normal form.
To grasp uniform loops intuitively, consider triangular linear loops with updates $x_i \assign \lambda \cdot x_i + p_i$ for all $1 \leq i \leq d$, where the factor $\lambda \geq 0$ is the same for all $i$.
These loops are uniform iff the relation ${\succ_{\update}}$ is total (or equivalently, iff the variables can be ordered such that the super-diagonal of $A$ does not contain zeros).

\begin{lemma}
	\label{lem:uniform_one_eigenvalue}
	A triangular linear loop $(\cond, A \cdot \vec{x})$ where all diagonal entries are identical and non-negative is uniform iff ${\succ_{\update}}$ is a total ordering.
\end{lemma}
\makeproof{lem:uniform_one_eigenvalue}{
	In the following, let $\update = A \cdot \vec{x}$.

	\medskip

	\noindent
	$\impliedby$: Let ${\succ_{\update}}$ be a total ordering and $\lambda$ the unique eigenvalue of $A$, i.e., the diagonal of $A$ only contains $\lambda$.
	We now prove that the matrix $A - \lambda \cdot I^{d \times d}$ has rank at least $d-1$, i.e., its kernel, which is the eigenspace of $A$ w.r.t.\ $\lambda$, has dimension at most $1$.
	As $\lambda$ is an eigenvalue of $A$, its eigenspace then must be exactly one-dimensional.

	Since $A$ is triangular, so is $A - \lambda \cdot I^{d \times d}$.
	As ${\succ_{\update}}$ is a total ordering, the super-diagonal of $A - \lambda \cdot I^{d \times d}$ contains only non-zero values, whereas its diagonal contains only zeros.
	Thus, by deleting the first row and the last column of this matrix, we obtain a triangular $(d - 1) \times (d - 1)$ submatrix $B$ whose diagonal is the super-diagonal of $A - \lambda \cdot I^{d \times d}$, i.e., it contains only non-zero values.
	Thus, the product of the diagonal entries of $B$ is non-zero, i.e., $\det(B) \neq 0$.
	But $B$ is a submatrix of $A - \lambda \cdot I^{d \times d}$, i.e., $A - \lambda \cdot I^{d \times d}$ has a non-zero $(d -1) \times (d -1)$ minor.
	Hence, $\mathrm{rank}(A - \lambda \cdot I^{d \times d}) \geq d -1$.

	\medskip

	\noindent
	$\implies$: Let us assume that ${\succ_{\update}}$ is not a total ordering.
	In this case, the super-diagonal of $A - \lambda \cdot I^{d \times d}$ contains a zero and its diagonal contains only zeros.
	Due to triangularity, its rank can be at most $d-2$.
	Thus, $A$ has at least two linear independent eigenvectors.
}

Thus, loops like the leading example from \cite{dblp:conf/sas/ben-amramdg19} in \cref{fig:loop:run} which is equivalent to \cref{fig:loop:run_equiv} are uniform.
In contrast, a loop is not uniform if each $x_i$ is updated to $\lambda \cdot x_i + c_i$ for constants $c_i \in \ring$.
The reason is that the $x_i$ do not occur in each other's updates.
Hence, we have $x_i \not\succ_{\update} x_j$ and $x_j \not\succ_{\update} x_i$ for all $1 \leq i,j \leq d$.
\begin{figure}[t]
	\begin{minipage}[t]{0.53\linewidth}
		\vspace*{0pt}
		\hspace*{.6cm}	\minialg{4.2cm}{}{
			\While{$x_1 \geq -x_3$}{
				$\ASSIGN{\mat{x_1\\
							x_2\\
							x_3}}{\mat{x_1+x_2\\
							x_2+x_3\\
							x_3-1}}$
			}
		}
		\vspace*{-.1cm}
		\captionof{figure}{Uniform Loop \cite{dblp:conf/sas/ben-amramdg19} via Polynomials}
		\label{fig:loop:run}
	\end{minipage}\hspace*{-.2cm}
	\begin{minipage}[t]{0.5\linewidth}
		\vspace*{0pt}
		\minialg{\linewidth}{}{
			\While{$x_1 \geq -x_3 \land x_4 = 1$}{
				$\ASSIGN{\mat{x_1\\
							x_2\\
							x_3\\
							x_4}}{\rmat{1&1&0&0\\
							0&1&1&0\\
							0&0&1&-1\\
							0&0&0&1} \cdot \mat{x_1\\
							x_2\\
							x_3\\
							x_4}}$
			}
		}
		\vspace*{-.36cm}
		\captionof{figure}{Uniform Loop \cite{dblp:conf/sas/ben-amramdg19} via Matrix}
		\label{fig:loop:run_equiv}
	\end{minipage}
	\vspace*{-.4cm}
\end{figure}

So in particular, a uniform loop cannot have more than one update of the form $x_i \assign x_i$.
However, the loop condition can still be an arbitrary Boolean formula over linear inequations.
Thus, our complexity result is quite surprising since it shows that for this class of loops, termination is easier to decide than satisfiability of the condition (e.g., unsatisfiability of linear formulas over $\RA$ is \cc{Co-NP}-complete).
Intuitively, the reason is that our class \emph{prohibits} multiple updates like $x_i \assign x_i$ where variables ``stabilize'' and where termination is essentially equivalent to unsatisfiability of the condition.

To give an intuition how hard the restriction to uniform loops is, we analyzed
the \emph{TPDB} \cite{tpdb} used at the \emph{Termination and Complexity Competition} \cite{termcomp}.
In the
category for ``Termination of Integer Transition Systems (ITSs)'' we identified 467 polynomial loops with non-constant guard (i.e., termination is not trivial) and 290 (62~\%) of them are uniform loops over $\ZZ$.
Similarly, in the category for ``Complexity Analysis of ITSs'' we found 1,258 such polynomial loops and 452 (36~\%) are uniform.
In fact, in practice one is often interested in termination of triangular loops where after chaining, all variables belong to the eigenvalues 0 or 1.
The reason is that termination is usually easy to show if there are eigenvalues greater than 1, because they lead to exponential growth.
Thus, if the loop terminates, termination is usually reached after few steps (e.g., consider a loop $(x \leq c, 2 \cdot x)$ for any constant $c$).
Hence, the number of eigenvalues $k$ is usually smaller than the number $d$ of program variables.
In this section we show that the complexity for deciding termination of uniform loops is exponential in $k$ but not in $d$.
More precisely, termination of uniform loops is in the \emph{parameterized complexity class}
\cc{XP}, where the \emph{parameter} is the number $k$ of eigenvalues.

\begin{definition}[Parameterized Decision Problem, \cc{XP} \protect{\cite{df99}}]
	A \emph{parameterized decision problem} is a language $\Lan \subseteq \Sigma^* \times \NN$, where $\Sigma$ is a finite alphabet.
	The second component (from $\NN$) is called the \emph{parameter} of the problem.

	A parameterized problem $\Lan$ is \emph{slicewise polynomial} if the time needed for deciding the question ``$(x, k ) \in \Lan?$'' is in $\OO(\vert{}x\vert{}^{f(k)})$ where $f$ is a computable function depending only on $k$.
	The corresponding complexity class is called \textnormal{\cc{XP}}.
\end{definition}

In the remainder of this section, we prove that for any fixed $k \in \NN$, termination of uniform loops with $k$ eigenvalues is decidable in polynomial time.

\begin{theorem}[Parameterized Complexity of $k$-Termination]
	\label{thm:xp}
	We define the parameterized decision problem \emph{$k$-termination} as follows:
	$((\cond,A \cdot \vec{x}),k) \in \Lan_{k\text{-termination}}$ iff the loop $(\cond,A \cdot \vec{x})$ terminates over $\ring$ and $A$ has $k$ eigenvalues.

	For uniform loops, $k$-termination is in \textnormal{\cc{XP}}.
	Moreover, for such loops, $k$-termination over $\RR$ is in \textnormal{\cc{XP}} as well.
\end{theorem}

\subsubsection{Hierarchical Expressions and Partitions}
\label{subsubsec:hierarchical_expressions}

We now elaborate on the closed forms arising from uniform loops.
To this end, we fix a uniform loop $(\cond, A \cdot \vec{x})$.
Let $\spec(A) = \{\lambda_1,\ldots,\lambda_k\}$ be $A$'s eigenvalues where $0 \leq \lambda_1 < \ldots < \lambda_k$, let $Q$ be $A$'s Jordan normal form where the Jordan blocks are ordered such that the numbers on the diagonal are weakly monotonically increasing, and let $T$ be the corresponding transformation matrix, i.e., $A = T^{-1} \cdot Q \cdot T$.
Moreover, let $\eta$ be the automorphism defined by $\eta(\vec{x}) = T \cdot \vec{x}$ and let $\Trnv{\eta}{\cond}{A \cdot \vec{x}} = (\eta^{-1}(\cond), Q \cdot \vec{x}) = (\cond', Q \cdot \vec{x})$ as in \cref{subsec:linear-update}.

Instead of termination of the original loop on $\ring^d$, we now have to prove termination of the transformed loop on $\widehat{\eta}(\ring^d) = T \cdot \ring^d$.
For $\ring \in \{ \QQ, \RA\}$, if the eigenvalues of $A$ are from $\ring$, then the transformation matrix $T$ is an invertible matrix over $\ring$.
Therefore, we obtain $T \cdot \ring^d = \ring^d$.
Hence, we now have to analyze termination of $(\cond', Q \cdot \vec{x})$ over $\ring$.
In contrast, if $\ring = \ZZ$, then even if the eigenvalues of $A$ are integers, the transformation matrix $T$ or its inverse may contain non-integer rational numbers.
Thus, we first regard uniform loops over $\ring \in \{ \QQ, \RA\}$ and discuss the case $\ring = \ZZ$ in \cref{subsec:integer_uniform_loops}.

\begin{figure}[t]
	\begin{minipage}[t]{0.3\linewidth}
		\vspace*{-.27cm}
		\[
			Q_{\lambda} =
			\begin{bmatrix}
				\lambda & 1       & 0      & \cdots  & 0       \\
				0       & \lambda & 1      & \cdots  & 0       \\
				\vdots  & \vdots  & \vdots & \ddots  & \vdots  \\
				0       & 0       & \cdots & \lambda & 1       \\
				0       & 0       & \cdots & 0       & \lambda
			\end{bmatrix}
		\]
		\vspace*{-.1cm}
		\captionof{figure}{Jordan Block}
		\label{fig:jordanBlock}
	\end{minipage}
	\begin{minipage}[t]{0.77\linewidth}
		\[
			Q_\lambda^n =
			\begin{bmatrix}
				\binom{n}{0} \cdot \lambda^n & \binom{n}{1} \cdot \lambda^{n-1} & \binom{n}{2} \cdot \lambda^{n-2} & \cdots & \binom{n}{\nu-1} \cdot \lambda^{n-\nu+1} \\
				0                            & \binom{n}{0} \cdot \lambda^n     & \binom{n}{1} \cdot \lambda^{n-1} & \cdots & \binom{n}{\nu-2} \cdot \lambda^{n-\nu+2} \\
				\vdots                       & \vdots                           & \vdots                           & \ddots & \vdots                                   \\
				0                            & 0                                & 0                                & \cdots & \binom{n}{1} \cdot \lambda^{n-1}         \\
				0                            & 0                                & 0                                & \cdots & \binom{n}{0} \cdot \lambda^{n}
			\end{bmatrix}
		\]
		\vspace*{-.3cm}
		\captionof{figure}{Multiplication of Jordan Block}
		\label{fig:multiplicationJordanBlock}
	\end{minipage}
	\vspace*{-.4cm}
\end{figure}
By uniformity, $Q = \diag(Q_{\lambda_1},\ldots,Q_{\lambda_k})$ has $k$ Jordan blocks $Q_{\lambda_1}$, \ldots, $Q_{\lambda_k}$ where $Q_{\lambda}$ is as in \cref{fig:jordanBlock}.
For each eigenvalue $\lambda$, let $\nu(\lambda)$ be the dimension of $Q_{\lambda}$.
Since each eigenvalue has geometric multiplicity $1$, $\nu(\lambda)$ is the \emph{algebraic multiplicity} of $\lambda$, i.e., the multiplicity as a root of the characteristic polynomial of $A$.
For $\nu = \nu(\lambda)$, \cref{fig:multiplicationJordanBlock}
shows the form of $Q_\lambda^n$ as in, e.g., \cite{dblp:conf/soda/ouakninepw15,kincaid19}, where $\binom{n}{s} = 0$ if $n < s$.
This directly yields a closed form $\closed$ for the $n$-fold application of the update $Q
\cdot \vec{x}$ of the transformed loop.
Since our approach from \cref{sec:deciding}
works by analyzing \emph{eventual} non-termination, we are only interested in validity of formulas for large enough $n$.
Thus, we may assume that $n$ is larger than the algebraic multiplicities  $\nu(\lambda)$
of all eigenvalues  $\lambda \in \spec(A)$.
Then one obtains a resulting normalized closed form  $\closednorm$ which consists of
normalized poly-exponential expressions of a special form, so-called
\emph{hierarchical expressions}.
Here, for any $\alpha \in \Q_\ring[\vec{x}]$, $\deg_{x_i}(\alpha)$ is the highest power of $x_i$ occurring in a monomial of $\alpha$, i.e., it is the degree of $\alpha$ when interpreting all variables besides $x_i$ as constants.

\begin{definition}[Hierarchical Expression]
	\label{def:hierarchical_expression}
	Let $\qlinpoly{\vec{x}}$ denote the set of \emph{linear polynomials} from $\Q_\ring[\vec{x}]$, i.e., of degree at most $1$.
	An expression $h \in \PEN{\ring}[\vec{x}]$ for some ring $\ZZ \leq \ring \leq \RA$ is a \emph{hierarchical expression} over the indices $1\leq i_1<\ldots <i_{\nu} \leq d$ if there exist $1 \leq r \leq \nu$ and $\lambda \in \ring_{> 0}$ such that
	\[
		\mbox{$h = \sum\nolimits_{s=r}^{\nu} \alpha_s \cdot n^{s-r} \cdot \lambda^{n},$}
	\]
	where $\alpha_s \in \qlinpoly{x_{i_s},\ldots,x_{i_{\nu}}}$, $\alpha_s(0,\ldots,0) = 0$, and $\deg_{x_{i_s}}(\alpha_s) = 1$ for $r \leq s \leq \nu$.
	Here, $\alpha_s(v_s,\ldots,v_\nu)$ abbreviates $\alpha_s [x_{i_s}/v_s,\ldots,x_{i_\nu}/v_\nu]$ for $v_s,\ldots,v_\nu \in \RA$.
	We call $\offset(h) = r$ the \emph{offset}, $\nu$ the \emph{order}, and $\base{h} = \lambda$ the \emph{base}
	of $h$.
\end{definition}

\Cref{coro:closed_uniform_hierar_exp} states this observation on $\closednorm$ formally, 
where the \emph{index} $\idx(\lambda)$ is the sum of the algebraic multiplicities of all smaller eigenvalues than $\lambda$, i.e.,
\[
	\mbox{$\idx(\lambda) = \sum\nolimits_{\lambda' \in \spec(A), \lambda' < \lambda} \nu(\lambda')$.}
        \]
If  $A$'s smallest eigenvalue $\lambda_1$ is $0$, then
all entries $q_1, \ldots, q_{\idx(\lambda_2)}$ of $\closednorm$ are 0, i.e., when inserting $\closednorm$ into the loop condition, the variables $x_1,\ldots,x_{\idx(\lambda_2)}$ vanish.
So from now on we assume that 0 is not an eigenvalue of $A$.
(Note that since we just ignore the variables which belong to the eigenvalue zero, we can also permit uniform loops where the eigenvalue 0 may have a higher geometric multiplicity.)

\begin{lemma}
	\label{coro:closed_uniform_hierar_exp}
	For all $\lambda \in \spec(A)$ and $1 \leq r \leq \nu(\lambda)$, the $(\idx(\lambda)+r)$-th element of $\closednorm$ is a hierarchical expression over the indices $\idx(\lambda) + 1,\idx(\lambda) + 2, \ldots, \idx(\lambda) + \nu(\lambda)$ with offset $r$, order $\nu(\lambda)$, and base $\lambda$.
\end{lemma}
\makeproof{coro:closed_uniform_hierar_exp}{Let $Q$ be the Jordan normal form of $A$. From
  the form of $Q_\lambda^n$ in  \cref{fig:multiplicationJordanBlock}, we directly obtain
  the following observation:
  \begin{equation}
    	\label{lem:closed_uniform}
\parbox{10.5cm}{For all $\lambda \in \spec(A)$ and $1 \leq r \leq \nu(\lambda)$, the $(\idx(\lambda)+r)$-th element of $\closed = Q^n \cdot \vec{x}$ is $\sum\nolimits_{s=r}^{\nu(\lambda)} \binom{n}{s-r} \cdot \lambda^{n-s+r} \cdot x_{\idx(\lambda)+s}$.}
\end{equation}

For $s-r \in \NN$, in general $\binom{n}{s-r}$ is not a polynomial in the variable $n$, as one has to distinguish the cases $n < s-r$ and $n \geq s-r$.
But since our approach from \cref{sec:deciding}
analyzes \emph{eventual} non-termination, we are only interested in validity of formulas for large enough $n$.
Thus, we may assume $n\geq s-r$.
Then, $\binom{n}{s-r}$ is indeed a polynomial from $\QQ[n]$ of degree $s-r$, i.e., there are coefficients $c_{s-r,j} \in \QQ$ such that for all $n \geq s-r$ we have $\binom{n}{s-r} = \sum\nolimits_{j=0}^{s-r} c_{s-r,j} \cdot n^j$.

In fact, $c_{s-r,j} = \tfrac{\stir(s-r,j)}{(s-r)!}$ where $\stir$ is the \emph{signed Stirling number of the first kind} (see, e.g., \cite[Ch.\ 6]{dblp:books/aw/gkp1994}).
While $\stir$'s formal definition is not of interest for us, we use $\stir(s-r,s-r) = 1 \neq 0$.
We obtain the following from \eqref{lem:closed_uniform}.
\begin{equation}
	\label{lem:closed_uniform_poly_exp}
\parbox{10cm}{For all $\lambda \in \spec(A)$ and $1 \leq r \leq \nu(\lambda)$, the $(\idx(\lambda)+r)$-th element of $\closednorm$ is $\sum\nolimits_{s=r}^{\nu(\lambda)} (\sum\nolimits_{j=0}^{s-r} c_{s-r,j} \cdot x_{\idx(\lambda) + s} \cdot \lambda^{r-s} \cdot n^j \cdot \lambda^{n}) \in \PEN{\ring}[\vec{x}]$.}
\end{equation}
Re-arranging the order of summation of the closed form in \eqref{lem:closed_uniform_poly_exp} yields
\begin{equation}
	\begin{array}{cl}
		  & \sum\nolimits_{s=r}^{\nu(\lambda)} \left(\sum\nolimits_{j=0}^{s-r} c_{s-r,j} \cdot x_{\idx(\lambda) + s} \cdot \lambda^{r-s} \cdot n^j \cdot \lambda^{n}\right) \\
		= & \sum\nolimits_{s=r}^{\nu(\lambda)}\underbrace{\mbox{$(\sum\nolimits_{j=0}^{\nu(\lambda)-s}
					c_{s-r+j,s-r}\cdot x_{\idx(\lambda)+s+j} \cdot \lambda^{r-s-j})$}}_{{}={}\alpha_s}\cdot n^{s-r} \cdot \lambda^n.
	\end{array}
	\label{eq:hierarchical}
\end{equation}
Here, $\alpha_s$ is a linear polynomial in the variables $x_{\idx(\lambda)+s},\ldots,x_{\idx(\lambda)+\nu(\lambda)}$.
Note that the coefficient of $x_{\idx(\lambda)+s}$ in $\alpha_s$ is $c_{s-r,s-r} \cdot \lambda^{r-s}= \tfrac{\stir(s-r,s-r)}{(s-r)!} \cdot \lambda^{r-s} = \tfrac{1}{(s-r)!} \cdot \lambda^{r-s}\neq 0$.

By \eqref{lem:closed_uniform_poly_exp} and \eqref{eq:hierarchical}, the elements of $\closednorm$ are hierarchical expressions.
Note that here we indeed need $\stir(s -r,s-r) = 1 \neq 0$, i.e., $\deg_{x_{\idx(\lambda)} + s}(\alpha_s) = 1$.
}

\begin{figure}[t]
	\hspace*{-.7cm}
	\minialg{3cm}{}{
		\While{$\cond$}{
			$\ASSIGN{\vec{x}}{A \cdot \vec{x}}$
		}
	}
	\hspace*{-.2cm}
	$
		\begin{array}{rcl}
			A & = & \mat{
			1 & 1 & 0     & 0 & 0 \\
			0 & 1 & 0     & 0 & 0 \\
			0 & 0 & 2     & 1 & 0 \\
			0 & 0 & 0     & 2 & 1 \\
			0 & 0 & 0     & 0 & 2 \\
			}
		\end{array}
	$ \hspace*{.2cm}
	$\closednorm = \mat{
			x_1 + n \cdot x_2 \\
			x_2 \\
			x_3 \cdot 2^n + \left(\tfrac{x_4}{2} - \tfrac{x_5}{8}\right) \cdot n \cdot 2^n + \tfrac{x_5}{8} \cdot n^2 \cdot 2^n \\
			x_4 \cdot 2^n + \frac{x_5}{2} \cdot n \cdot 2^n \\
			x_5 \cdot 2^n
		}$ \vspace*{-.2cm}
	\caption{Uniform Loop and Normalized Closed Form}\label{fig:uniformLeading}
	\vspace*{-.4cm}
\end{figure}

\begin{example}
  \label{ex:hierarchical_expression}
	Consider the uniform loop and its normalized closed form $\closednorm$ in \cref{fig:uniformLeading}, where the update matrix $A$ has $k = 2$ eigenvalues $\lambda_1 = 1$ of algebraic multiplicity $\nu_1 = 2$ and $\lambda_2 = 2$ of algebraic multiplicity $\nu_2 = 3$, both of which have geometric multiplicity one.
	Moreover, $\idx(\lambda_1) = 0$ and $\idx(\lambda_2) = 0+ \nu_1= 2$.
	Here, we have $d = 5$.

	For $\closednorm$ in \cref{fig:uniformLeading}, $h=q_4=x_4 \cdot 2^n + \tfrac{x_5}{2} \cdot n \cdot 2^n$ is a hier\-ar\-chi\-cal expression over the indices $i_1=3$, $i_2=4$, $i_3=5$ with offset $2$, order $3$, and base $2$, as $h = \sum\nolimits_{s=2}^3 \alpha_s \cdot n^{s-2} \cdot 2^n$ for $\alpha_2 = x_4 \in \RA[x_4,x_5]$ and $\alpha_3 = \tfrac{x_5}{2} \in \RA[x_5]$.
\end{example}

To describe the form of the whole vector $\closednorm$, we now introduce \emph{hierarchical partitions}.
To this end,
similar to the concept of solvable loops in \cref{def:solvable}, we consider a partitioning of $\{1,\ldots,d\}$.

\begin{definition}[Hierarchical Partition]\label{def:hierarchical_partition}
	For $k\geq 1$, let $\nu_1,\ldots,\nu_k \in \NN$ form a \emph{$k$-partition} of $d$, i.e., $\nu_1 + \ldots + \nu_k = d$ and $\nu_i > 0$ for all $1 \leq i \leq k$.
	The \emph{blocks} associated to the partition $\nu_1,\ldots,\nu_k$ are $B_1 = \{1,\ldots, \nu_1\}$, $B_2 = \{\nu_1 + 1, \ldots, \nu_1 + \nu_2\}$, \ldots, and $B_k = \{\nu_1 + \ldots + \nu_{k-1} + 1 , \ldots, d\}$.

The hierarchical expressions $h_1,\ldots,h_d$ are a \emph{hierarchical $k$-partition} via $\nu_1,\ldots,\nu_k$ with bases $0 < \lambda_1 < \ldots < \lambda_k$ from $\ring$ if for all $1 \leq i \leq k$:
	\begin{enumerate}
		\item[(a)] $h_j$ is a hierarchical expression over the indices $B_i$ for all $j \in B_i$,
		\item[(b)] $\base{h_j} = \lambda_i$ for all $j \in B_i$,
		\item[(c)] $h_j$ has order $\nu_i$ for all $j \in B_i$,
		\item[(d)] $\offset(h_{\min(B_i)})=1$, and
		\item[(e)] if $j,j+1 \in B_i$, then $\offset(h_j) + 1 = \offset(h_{j+1})$.
	\end{enumerate}
\end{definition}
\noindent

Indeed, when transforming the update of a uniform loop to Jordan normal form, then the normalized closed form is always a hierarchical partition.

\begin{corollary}[$\closednorm$ is Hierarchical Partition]
	\label{coro:uniformity_partition}
	Let $A$ be the update matrix of a uniform loop with eigenvalues $ 0 < \lambda_1 < \ldots < \lambda_k$ and algebraic multiplicities $\nu_1,\ldots,\nu_k$, and let $Q$ be its Jordan normal form where the numbers on the diagonal are weakly monotonically increasing.
	Then the normalized closed form $\closednorm$ of the update $Q$ is a \emph{hierarchical $k$-partition} via $\nu_1,\ldots,\nu_k$ and bases $\lambda_1 < \ldots < \lambda_k$.
	Here the $i$-th block is $B_i = \{\idx(\lambda_i) + 1, \ldots, \idx(\lambda_i) + \nu_i\}$.
\end{corollary}

\begin{example}
	\label{ex:hierarchical_partition}
	In \cref{fig:uniformLeading}, $\vec{h} = \closednorm$ is a hierarchical $2$-partition via $\nu_1 = 2$ and $\nu_2 = 3$, blocks $B_1 = \{1,2\}$, $B_2 = \{3,4,5\}$, and $\lambda_1 = 1 < 2 = \lambda_2$.
	Moreover, $\offset(h_1) = 1$ and $\offset(h_2) = 2$, while $\offset(h_3) = 1$, $\offset(h_4) = 2$, and $\offset(h_5) = 3$.
\end{example}

\subsubsection{Interval Conditions}
\label{subsubsec:interval_conditions}

Our decision procedure in \cref{sec:deciding} instantiates the variables in the polynomials $f$ of the loop guard by $\closednorm = (h_1,\ldots,h_d)$, resulting in poly-exponential expressions $p = f(h_1,\ldots,h_d)$.
We now prove that in the setting of a hierarchical $k$-partition $\closednorm$, the atoms in $\lia(p \triangleright 0)$ (see \cref{lem:eventual_positiveness} and \eqref{lia-def}) are equivalent to so-called \emph{interval conditions}, whose satisfiability is particularly easy to check
(\cref{subsubsec:algorithm}).
In \cref{coro:blockz_correctness}, \cref{lem:valid_equivalent}, and \cref{coro:valid_equivalent}, we introduce the different subformulas occurring in  $\lia(p \triangleright 0)$.
Here, it is convenient to define the \emph{active variables} of polynomials.

\begin{definition}[Active Variables]
	Let $f = c_0 + \sum_{i=1}^d c_i \cdot x_i\in \linpoly{\vec{x}}$.
	We define $\act(f) = \setcomp{x_i}{c_i \neq 0}$.
	If $x_i \in \act(f)$, then $\coeff(f,x_i) = c_i$.
\end{definition}

\begin{example}
	\label{ex:active_vars}
	Consider $f = -x_1 + 3 \cdot x_3 + 4 \in \linpoly{x_1,\ldots,x_5}$.
	Then $\act(f) = \{x_1,x_3\}$, $\coeff(f,x_1) = -1$, and $\coeff(f,x_3) = 3$.
\end{example}

From the addends $\alpha_j \cdot n^{a_j} \cdot b_j^n$ of $p = f(h_1,\ldots,h_d)$ with $\alpha_j \in \qlinpoly{\vec{x}}$, $a_j \in \NN$, and $b_j \in \ring_{>0}$, we again compute the set $\coeffsop(p)$ of marked coefficients as in \cref{def:marked}, which have the form $\alpha_j^{(b_j,a_j)}$.
For any $b \in \ring_{>0}$ and $a \in \NN$, we now define a formula $\blockz(b,a)$ which is equivalent to requiring that all addends $\alpha_j \cdot n^{a_j} \cdot b_j^n$ vanish where $b_j = b$ and where $a_j \geq a$.

\begin{lemma}[Formulas for Vanishing of Addends]
	\label{coro:blockz_correctness}
	Let $h_1,\ldots,h_d$ be a hierarchical $k$-partition via $\nu_1,\ldots,\nu_k$ with bases $0 < \lambda_1 < \ldots < \lambda_k$, and let $f \in \linpoly{\vec{x}}$.
	For any $1 \leq i \leq k$, let $F(i) = \{j \in B_i \mid x_j \in \act(f)\}$.
	Let $M =\linebreak
		\frac{-f(0,\ldots,0)}{\coeff(f,x_{\min F(i)}) \cdot c_{\min F(i)}}$ if $\lambda_{i} = 1$ and $F(i) \neq \emptyset$, where $c_{\min F(i)}$ is the coefficient of $x_{\min F(i)}$ in $h_{\min F(i)}$.\footnote{By \cref{def:hierarchical_partition,def:hierarchical_expression}, $x_{\min F(i)}$ occurs only in a unique (linear) monomial of $h_{\min F(i)}$, whose coefficient is not 0.}
	For any $b \in \ring_{>0}$ and $a \in \NN$, we define $\blockz(b,a)$:
	\begin{itemize}
		\item[(a)] If $\lambda_i \neq 1$ for all $1 \leq i \leq k$ or $\lambda_{i} = 1$ for some $1 \leq i \leq k$ and $F(i) = \emptyset$, then $\blockz(1,0)$ is the formula $f(0,\ldots,0) = 0$.
		\item[(b)] If $\lambda_{i} = 1$ for some $1 \leq i \leq k$ and $F(i) \neq \emptyset$, then $\blockz(1,0)$ is the formula $x_{\min F(i)} = M \land \bigwedge\nolimits_{j \in B_{i}, \min F(i) < j} \, (x_j = 0)$.
		\item[(c)] If $\lambda_{i} = b$ for some $1 \leq i \leq k$, $F(i) \neq \emptyset$, and $(b,a) \neq (1,0)$, then $\blockz(b,a) = \bigwedge\nolimits_{j \in B_{i}, a + \min F(i) \leq j} \, (x_j = 0)$.
		\item[(d)] Otherwise, we define $\blockz(b,a) = \TRUE$.
	\end{itemize}
	Let $p = f(h_1,\ldots,h_d) \in \PEN{\ring}[\vec{x}]$.
	As in \cref{def:marked}, let
        $\coeffsop(p)=\{\alpha_1^{(b_1,a_1)},\ldots,\linebreak \alpha_{\ell}^{(b_\ell,a_\ell)}\}$ where $\alpha^{(b_i,a_i)}_i \preccc \alpha^{(b_{j},a_{j})}_{j}$ for all $1 \leq i < j \leq \ell$.
	Then $\blockz(b,a)$ is equivalent to the requirement that $\alpha_s = 0$ holds for all $\alpha_s^{(b_s,a_s)} \in \coeffsop(p)$ with $b_s = b$ and $a_s \geq a$.
\end{lemma}
\makeproof{coro:blockz_correctness}{We have
	\begin{equation}
		\label{form of p}
		\mbox{$p \; = \; f(0,\ldots,0) + \sum\nolimits_{1 \leq i \leq k, \, r \in F(i)} \, \coeff(f, x_r) \cdot h_r$.}
	\end{equation}
	Moreover, for every $1 \leq i \leq k$ and every $r \in B_i$, the hierarchical expression $h_r$ has the form
	\begin{equation}
		\label{form of h}
		\mbox{$h_r = \sum_{s = r}^{\nu_1+\ldots+\nu_i} \beta_{r,s} \cdot n^{s-r} \cdot \lambda_i^n, \; \text{where} \; \beta_{r,s} \in \qlinpoly{x_s,\ldots, x_{\nu_1+\ldots+\nu_i}}$.}
	\end{equation}
	\begin{itemize}
		\item[(d)] Here, $\lambda_i \neq b$ for all $1 \leq i \leq k$ or $\lambda_i = b$ for some $1 \leq i \leq k$ but $F(i) = \emptyset$.
			Moreover, $b \neq 1$ or $a \geq 1$.
			Hence, \eqref{form of p} and \eqref{form of h}
			imply that there is no $\alpha_s$ with $b_s = b$ and $a_s \geq a$.
		\item[(a)] If $\lambda_{i} \neq 1$ for all $1\leq i \leq k$ or $\lambda_{i} = 1$ for some $1 \leq i \leq k$ and $F(i) = \emptyset$, then in (d) we already showed that there is no $\alpha_s$ with $b_s = 1$ and $a_s \geq 1$.
			However, there is an $\alpha_s$ with $b_s = 1$ and $a_s = 0$, viz.\ $\alpha_s = f(0,\ldots,0)$.
			Thus, we have $\alpha_s = 0$ iff $f(0,\ldots,0) = 0$.
		\item[(c)] The largest number in $B_i$ is $\nu_1 + \ldots + \nu_i$.
			So if $a > \nu_1 + \ldots + \nu_i - \min F(i)$ then we have $\blockz(b,a) = \TRUE$.
			This is sound because then there is no $\alpha_s$ with $b_s = b$ and $a_s = a$.

			Hence, we now consider the case $a \leq \nu_1 + \ldots + \nu_i - \min F(i)$.
			Since $\lambda_i = b$ and $F(i) \neq 0$,\linebreak
			there is an $\alpha_m$ with $b_m = b$.
			Let $1 \leq s_0 \leq \ell$ be the largest number with $b_{s_0} = b = \lambda_{i}$.
			By \eqref{form of p} and \eqref{form of h}, the corresponding addend $\alpha_{s_0} \cdot n^{a_{s_0}} \cdot b_{s_0}^n$ of $p$ has the form $\coeff(f,x_r) \cdot \beta_{r,\nu_1 + \ldots + \nu_i} \cdot n^{\nu_1 + \ldots + \nu_{i} -r} \cdot \lambda_{i}^n$ for the smallest possible $r \in F(i)$.
			So we get
			\[
				\alpha_{s_0} \cdot n^{a_{s_0}} \cdot b_{s_0}^n = \coeff(f,x_{\min F(i)}) \cdot \beta_{\min F(i),\nu_1 + \ldots + \nu_{i}}
				\cdot n^{\nu_1 + \ldots + \nu_{i} - \min F(i)}
				\cdot \lambda_{i}^n.
			\]
			Here, $\beta_{\min F(i),\nu_1 + \ldots + \nu_{i}} \in \qlinpoly{x_{\nu_1 + \ldots + \nu_{i}}}$ and in fact, $\beta_{\min F(i),\nu_1 + \ldots + \nu_{i}} = c \cdot x_{\nu_1 + \ldots + \nu_{i}}$ for some $c \neq 0$.
			Hence, $\alpha_{s_0} = 0$ is equivalent to $x_{\nu_1 + \ldots + \nu_{i}} =0$.

			Now consider the second largest number $s_1 = s_0-1 \leq \ell$ such that $b_{s_1} = b = \lambda_{i}$.
			By \eqref{form of p} and \eqref{form of h}, for $a_{s_1} = \nu_1 + \ldots + \nu_{i} - \min F(i) -1$ we obtain
			\[
				\mbox{$\alpha_{s_1} \cdot n^{a_{s_1}} \cdot b_{s_1}^n = \sum_{r \in F(i), \,	\min F(i) \leq r \leq \min F(i) + 1} \coeff(f,x_r) \cdot \beta_{r,r+a_{s_1}} \cdot n^{a_{s_1}} \cdot \lambda_{i}^n,$}
			\]
			where $\beta_{r,r+a_{s_1}} \in \qlinpoly{x_{r+a_{s_1}}, \ldots, x_{\nu_1 + \ldots + \nu_{i}}}$.
			By taking into account that $x_{\nu_1 + \ldots + \nu_{i}} = 0$, this simplifies to
			\[
				\alpha_{s_1} \cdot n^{a_{s_1}} \cdot b_{s_1}^n = \coeff(f,x_{\min F(i)}) \cdot \beta_{\min F(i),\nu_1 + \ldots + \nu_{i}-1} \cdot n^{\nu_1 + \ldots + \nu_{i}- \min F(i)-1}.
			\]
			Here, $\beta_{\min F(i),\nu_1 + \ldots + \nu_{i} -1} \in \qlinpoly{x_{\nu_1 + \ldots + \nu_{i}-1}, x_{\nu_1 + \ldots + \nu_{i}}}$.
			But when again taking into account that $x_{\nu_1 + \ldots + \nu_{i}} = 0$, in fact we have $\beta_{\min F(i),\nu_1 + \ldots + \nu_{i}-1} = c' \cdot x_{\nu_1 + \ldots + \nu_{i}-1}$ for some $c' \neq 0$.
			Hence, $\alpha_{s_1} = 0$ is equivalent to $x_{\nu_1 + \ldots + \nu_{i}-1} = 0$, or in other words (as $\nu_1 + \ldots + \nu_{i}-1 = \min F(i)+ a_{s_1}$) to $x_{\min F(i)+a_{s_1}} = 0$.

			We repeat this reasoning until we reach an $s' \leq \ell$ with $a_{s'} = a$.
			Thus, $\alpha_s = 0$ for \emph{all} $s \leq \ell$ with $b_s = b = \lambda_{i}$ and $a_s \geq a$ is equivalent to $x_j = 0$ for all $\min F(i) + a \leq j \leq \nu_1 + \ldots + \nu_i$, i.e., to $\bigwedge\nolimits_{j \in B_{i}, \, a + \min F(i) \leq j}\, (x_j = 0)$.
		\item[(b)] To ensure that $\alpha_s = 0$ for all $1 \leq s \leq \ell$ where $b_s = 1$ and $a_s \geq 0$, we have to show that this holds if $a_s \geq 1$ and if $a_s = 0$.
			The former case is equivalent to $\blockz(1,1) = \bigwedge\nolimits_{j \in B_{i}, \, \min F(i) < j}\, (x_j = 0)$ according to (c).

			If $b_s = 1$ and $a_s = 0$, then \eqref{form of p} and \eqref{form of h} imply
			\[
				\mbox{$\alpha_s = f(0,\ldots,0) + \sum\nolimits_{r \in F(i)} \coeff(f,x_r) \cdot \beta_{r,r},$}
			\]
			where $\beta_{r,r} \in \qlinpoly{x_r,\ldots, x_{\nu_1+\ldots+\nu_{i}}}$.
			Clearly $r \in F(i)$ implies $r \geq \min F(i)$.
			Taking into account that $x_j = 0$ for all $\min F(i) < j \leq \nu_1+\ldots+\nu_{i}$, we therefore obtain
			\[
				\alpha_s = f(0,\ldots,0) + \coeff(f,x_{\min F(i)}) \cdot \beta_{\min F(i),\min F(i)},
			\]
			where $\beta_{\min F(i),\min F(i)} = c_{\min F(i)} \cdot x_{\min F(i)}$.
			Therefore,
			\[
				\begin{array}
					{cl}
					     & \alpha_s = 0                                                                       \\[-.1cm]
					\iff & f(0,\ldots,0) +\coeff(f,x_{\min F(i)}) \cdot c_{\min F(i)} \cdot x_{\min F(i)} = 0 \\[-.1cm]
					\iff & x_{\min F(i)} = M.
				\end{array}
			\]
	\end{itemize}

	\vspace*{-.5cm}

}

\begin{example}
	\label{ex:blockzero}
	Consider $f = -x_1 + 3 \cdot x_3 + 4$ from \cref{ex:active_vars} and $\vec{h} = \closednorm$ from \cref{fig:uniformLeading}.
	We have	$p = f(h_1,\ldots,h_5) = -h_1 + 3 \cdot h_3 + 4 =$
	\[
		\mbox{$(-x_1 + 4) -x_2 \cdot n + 3 \cdot x_3 \cdot 2^n + \left(\tfrac{3 \cdot x_4}{2}-\tfrac{3 \cdot x_5}{8}\right)\cdot n \cdot 2^n + \tfrac{3 \cdot x_5}{8} \cdot n^2 \cdot 2^n$.}
	\]
	In the notation of \cref{coro:blockz_correctness}, since $B_1 = \{ 1,2 \}$, $B_2 = \{3,4,5\}$, and $\act(f) = \{ x_1, x_3\}$, we have $F(1) = \{ j \in \{1,2\} \mid x_j \in \{ x_1, x_3\} \}= \{1\}$ and $F(2) = \{ j \in \{3,4,5\} \mid x_j \in \{ x_1, x_3\} \}= \{3\}$.
	Thus, $\min F(1) = 1$, $c_{\min F(1)} = 1$ is the coefficient of $x_1$ in $h_1$, $\coeff(f,x_{\min F(1)}) = -1$ and thus $M = \tfrac{-4}{-1} = 4$.
	Hence,
	\[
		\blockz(1,0) = (x_1 = 4) \land (x_2 = 0) \text{ and }
		\blockz(2,0) = (x_3 = 0) \land (x_4 = 0) \land (x_5 = 0).
	\]
	So $\vec{v} \in \RA^5$ satisfies $\blockz(1,0)$ or $\blockz(2,0)$, respectively, iff all terms with base $1$ or $2$ vanish in $f(h_1,\ldots,h_5)[\vec{x}/\vec{v}]$.
\end{example}

After introducing $\blockz(b,a)$, we are now ready to show that the formulas in \cref{lem:eventual_positiveness}
and \cref{thm:eventual_positiveness} have a special form when considering uniform loops. In the following lemmas, corollaries, and definitions, let $h_1,\ldots,h_d$, $\nu_1, \ldots, \nu_k$, $f$, $F(i)$, $p$, $\coeffs{p}$, and $\ell$ always be as in \cref{coro:blockz_correctness}.

\begin{lemma}
	\label{lem:valid_equivalent}
	For any $1 \leq s_0 \leq \ell$, the formula $\alpha_{s_0} > 0 \land \bigwedge_{s = s_0+1}^{\ell} \left(\alpha_s = 0\right)$ is equivalent to the following formula $\rho_{f,s_0}$, which can be computed in polynomial time from $h_1,\ldots,h_d$ and $f$:
	\begin{enumerate}
		\item[(a)] If $(b_{s_0},a_{s_0}) = (1,0)$ and either $\lambda_{i} \neq 1$ for all $1\leq i \leq k$ or $\lambda_{i_0} = 1$ for some $1\leq i_0 \leq k$ and $F(i_0) = \emptyset$, then $\rho_{f,s_0}$ is
			\[
				\mbox{$f(0,\ldots,0) > 0 \land \bigwedge\nolimits_{i \in \{1,\ldots,k\}, \lambda_i > 1}
						\blockz(\lambda_i,0)$.}
			\]
		\item[(b)] If $(b_{s_0},a_{s_0}) = (1,0)$, $\lambda_{i_0} = 1$ for some $1\leq i_0 \leq k$, and $F(i_0) \neq \emptyset$, then there is a\footnote{$C = \coeff(f,x_{\min F(i_0)}) \cdot c_{\min F(i_0)}$ for the coefficient $c_{\min F(i_0)}$ of $x_{\min F(i_0)}$ in $h_{\min F(i_0)}$.}
			$C \in \Q_\ring$ with $C \neq 0$ such that $\rho_{f,s_0}$ is
			\[
				\mbox{$\sign{C} \cdot x_{\min F(i_0)} + \tfrac{f(0,\ldots,0)}{\card{C}} > 0 \; \land \blockz(1,1) \land \bigwedge\nolimits_{i = i_0 + 1}^{k} \blockz(\lambda_i,0)$.}
			\]
		\item[(c)] If $b_{s_0} < 1$, $f(0,\ldots,0) \neq 0$, and either $\lambda_{i} \neq 1$ for all $1\leq i \leq k$ or $\lambda_{i_0} = 1$ for some $1\leq i_0 \leq k$ and $F(i_0) = \emptyset$, then $\rho_{f,s_0}$ is $\FALSE$.
		\item[(d)] Otherwise, we have $\lambda_{i_0} = b_{s_0}$ for some $1\leq i_0 \leq k$, $F(i_0) \neq \emptyset$, and there is a number\footnote{More precisely, $sg = \signop(\coeff(f,x_{\min F(i_0)}) \cdot c_{\min F(i_0)+ a_{s_0}})$, where $c_{\min F(i_0)+ a_{s_0}}$ is the unique coefficient of $x_{\min F(i_0)+ a_{s_0}}$ in $h_{\min F(i_0)}$'s addend of the form $\beta \cdot n^{a_{s_0}} \cdot b_{s_0}^n$ with $\beta \in \qlinpoly{x_{\min F(i_0) + a_{s_0}}, \ldots, x_{\nu_1 + \ldots + \nu_{i_0}}}$.}
			$sg \in \{1,-1\}$ such that $\rho_{f,s_0}$ is
			\[
				\mbox{$sg \cdot x_{\min F(i_0)+ a_{s_0}} > 0 \land \blockz(\lambda_{i_0},a_{s_0}+1) \land \bigwedge\nolimits_{i = i_0+1}^k \blockz(\lambda_i,0)$.}
			\]
	\end{enumerate}
\end{lemma}
\makeproof{lem:valid_equivalent}{
	\begin{itemize}
		\item[(a)] As in the proof of Case (a) of \cref{coro:blockz_correctness}, we have $\alpha_{s_0} = f(0,\ldots,0)$ and thus $\alpha_{s_0} > 0 \iff f(0,\ldots,0) > 0$.
			In this case, for all ${s_0} < s \leq \ell$ we must have $b_s > 1$.
			Hence, $\bigwedge_{s = s_0+1}^{\ell} \left(\alpha_s = 0\right) \iff \bigwedge\nolimits_{i \in \{1,\ldots,k\}, \lambda_i > 1}
				\blockz(\lambda_i,0)$ by \cref{coro:blockz_correctness}.
		\item[(b)] Now $(b_{s_0},a_{s_0}) = (1,0)$, $\lambda_{i_0} = 1$ for some $1\leq i_0 \leq k$, and $F(i_0) \neq \emptyset$.
			By \cref{coro:blockz_correctness}
			we obtain that $\alpha_s = 0$ for all ${s_0} < s \leq \ell$ with $b_s = 1$ is equivalent to $\blockz(1,1)$, and $\alpha_s = 0$ for all ${s_0} < s \leq \ell$ with $b_s > 1$ is equivalent to $\bigwedge\nolimits_{i = i_0+1}^k \blockz(\lambda_i,0)$.

			Finally, as in the proof of Case (b) of \cref{coro:blockz_correctness}, we get
			\[
				\alpha_{s_0} = f(0,\ldots,0) + \coeff(f,x_{\min F(i_0)}) \cdot \beta_{\min F(i_0),\min F(i_0)},
			\]
			with $\beta_{\min F(i_0),\min F(i_0)} = c_{\min F(i_0)} \cdot x_{\min F(i_0)}$.
			For $C = \coeff(f,x_{\min F(i_0)}) \cdot c_{\min F(i_0)}$,
			\vspace*{-.5cm}

			\[
				\begin{array}
					{cl}
					     & \alpha_{s_0} > 0                                                               \\[-.1cm]
					\iff & f(0,\ldots,0) + C \cdot x_{\min F(i_0)} > 0                                    \\[-.1cm]
					\iff & \tfrac{C}{\card{C}} \cdot x_{\min F(i_0)} + \tfrac{f(0,\ldots,0)}{\card{C}}> 0 \\[-.1cm]
					\iff & \sign{C} \cdot x_{\min F(i_0)} + \tfrac{f(0,\ldots,0)}{\card{C}} > 0.
				\end{array}
			\]
		\item[(c)] If $b_{s_0} < 1$, $f(0,\ldots,0) \neq 0$, and either $\lambda_{i} \neq 1$ for all $1\leq i \leq k$ or $\lambda_{i_0} = 1$ for some $1\leq i_0 \leq k$ and $F(i_0) = \emptyset$, then $\bigwedge_{s = {s_0}+1}^{\ell} \left(\alpha_s = 0\right)$ is $\FALSE$.
			The reason is that $\coeffsop(p)$ contains $\alpha_{s}^{(1,0)}$ with ${s_0} < s$ and since there is either no $\lambda_i = 1$, or $\lambda_{i_0} = 1$ for some $1\leq i_0 \leq k$ but $F(i_0) = \emptyset$, we have $\alpha_{s} = f(0,\ldots,0) \neq 0$.
		\item[(d)] In the remaining case,
			since $b_{s_0} \neq 1$, there must be an $1\leq i_0 \leq k$ with $\lambda_{i_0} = b_{s_0}$ and $F(i_0) \neq \emptyset$.
			By \cref{coro:blockz_correctness}, $\alpha_s = 0$ for all ${s_0} < s \leq \ell$ with $b_s = b_{s_0} = \lambda_{i_0}$ is equivalent to $\blockz(\lambda_{i_0}, a_{s_0} + 1)$, and
			$\alpha_s = 0$ for all ${s_0} < s \leq \ell$ with $b_s \neq b_{s_0}$ is equivalent to $\bigwedge\nolimits_{i = i_0+1}^k \blockz(\lambda_i,0)$.
			Finally, \eqref{form of p} and \eqref{form of h} from the proof of \cref{coro:blockz_correctness}
			imply
			\[
				\mbox{$\alpha_{s_0} = \sum\nolimits_{r \in F(i_0), r \leq \nu_1 + \ldots + \nu_{i_0} - a_{s_0}}
						\coeff(f,x_r) \cdot \beta_{r,r + a_{s_0}} \cdot n^{a_{s_0}} \cdot \lambda_{i_0}^n,$}
			\]
			for $\beta_{r,r + a_{s_0}} \in \qlinpoly{x_{r+a_{s_0}}, \ldots, x_{\nu_1 + \ldots + \nu_{i_0}}}$.
			Clearly $r \in F(i_0)$ implies $r \geq \min F(i_0)$.
			Considering that $x_j = 0$ for all $\min F(i_0) + a_{s_0} < j \leq \nu_1+\ldots+\nu_{i_0}$, we therefore obtain
			\[
				\alpha_{s_0} = \coeff(f,x_{\min F(i_0)}) \cdot \beta_{\min F(i_0),\min F(i_0)+a_{s_0}} \cdot n^{a_{s_0}} \cdot \lambda_{i_0}^n,
			\]
			where $\beta_{\min F(i_0),\min F(i_0)+ a_{s_0}}$ is $c_{\min F(i_0)+a_{s_0}} \cdot\, x_{\min F(i_0)+ a_{s_0}}$.
			Let $sg = \signop(\coeff(f,\linebreak
				x_{\min F(i_0)}) \cdot c_{\min F(i_0)+a_{s_0}})$.
			Thus, $\alpha_{s_0} > 0$ is equivalent to $sg \cdot x_{\min F(i_0)+ a_{s_0}} > 0$.
	\end{itemize}

	\vspace*{-.4cm}
}

\begin{example}
	\label{ex:valid_equivalent_1}
	For $f, \vec{h}, p$ of \cref{ex:blockzero}, $\coeffsop(p)$ is
	\begin{align*}
          &\{\alpha_1^{(1,0)}, \alpha_2^{(1,1)}, \alpha_3^{(2,0)}, \alpha_4^{(2,1)}, \alpha_5^{(2,2)} \} \hspace{7em} \text{where} \\
          &\alpha_1 = -x_1 + 4, \quad \alpha_2 = -x_2, \quad \alpha_3 = 3 \cdot x_3, \quad \alpha_4 = \tfrac{3 \cdot x_4}{2}-\tfrac{3 \cdot x_5}{8}, \quad \alpha_5 = \tfrac{3 \cdot x_5}{8}.
	\end{align*}

	\vspace*{-.3cm}

	Let us compute $\rho_{f,4}$.
	For $\alpha_4^{(2,1)}$, we have $i_0 = 2$, $\min F(i_0) = 3$, and $a_4 = 1$.
	Moreover, we have $sg = \sign{\coeff(f,x_3) \cdot c_4} = \sign{3 \cdot \frac{1}{2}} = 1$ and $\blockz(2,2) = \bigwedge_{j \in \{3,4,5\}, \, 2 + 3 \leq j} \, (x_j = 0) = (x_5 = 0)$ and hence by \cref{lem:valid_equivalent}:
	\[
		\mbox{$
				\begin{array}{rcl}
					(\alpha_4 > 0) \land (\alpha_5 = 0) & \iff & (sg \cdot x_{a_4 + \min F(i_0)} > 0) \wedge \blockz(2,a_4 + 1) \\[-.1cm]
					                                    & \iff & (x_4 > 0) \land (x_5 = 0)
				\end{array}
			$}
	\]

	For $\alpha_1^{(1,0)}\!$, as $\lambda_1 = 1$ and $F(1) = \{1\} \neq \emptyset$, by \cref{lem:valid_equivalent} we use $C = \coeff(f,x_1) \cdot 1 = -1$, $f(0,\ldots,0) = 4$, and $\blockz(1,1) =\bigwedge_{j \in \{1,2\}, 1 + 1 \leq j}
		(x_j = 0) = (x_2 = 0)$ to obtain $\rho_{f,1}$, 	where $\blockz(2,0)= \bigwedge_{j=3}^5 (x_j = 0)$ by \cref{ex:blockzero}:
	\[
		\mbox{$
				\begin{array}{rcl}
					(\alpha_1 > 0) \land \bigwedge\nolimits_{s = 2}^5 (\alpha_s = 0) & \iff & (-x_1 + 4 > 0) \land (x_2 = 0) \land \blockz(2,0)
				\end{array}
			$}
	\]
\end{example}

In addition to the formulas $\rho_{f,s}$ for $1 \leq s \leq \ell$, we also introduce a formula $\rho_{f,0}$ which expresses that all coefficients of $p$ vanish.

\begin{corollary}
	\label{coro:valid_equivalent}
	$\bigwedge\limits_{s = 1}^{\ell} \left(\alpha_s = 0\right)$ is equivalent to $\rho_{f,0}$: $\blockz(1,0) \land \hspace*{-.4cm} \bigwedge\limits_{\lambda \in \{1, \ldots, k\}, \lambda \neq 1} \hspace*{-.5cm} \blockz(\lambda,0)$
\end{corollary}

\begin{example}
	\label{ex:valid_equivalent_2}
	Reconsider \cref{ex:valid_equivalent_1}.
	By \cref{coro:valid_equivalent}, $\bigwedge\nolimits_{s = 1}^5 (\alpha_s = 0)$ is equivalent to $\rho_{f,0} = \blockz(1,0) \land \blockz(2,0)$, where $\blockz(1,0) = (x_1 = 4) \land (x_2 = 0)$ and $\blockz(2,0)=(x_3 = 0) \land (x_4 = 0) \land (x_5 = 0)$ by \cref{ex:blockzero}.
\end{example}

We can now combine \cref{lem:valid_equivalent,coro:valid_equivalent} with \cref{lem:eventual_positiveness} to obtain the following result.
Here, ``$\sic$'' stands for \underline{i}nterval \underline{c}onditions.

\begin{corollary}
	\label{coro:equivalent}
	For ${\triangleright}\in {\{\geq,} {>\}}$, $\lia(p \triangleright 0)$ is equivalent to the formula $\sic(p \triangleright 0)$, where $\sic(p > 0) = \bigvee\nolimits_{s = 1}^{\ell} \rho_{f,s}$ and $\sic(p \geq 0) = \sic(p > 0) \lor \rho_{f,0}$.
\end{corollary}

\begin{example}
	\label{ex:equivalent}
	Reconsider \cref{ex:valid_equivalent_1,ex:valid_equivalent_2}.
	By \cref{coro:equivalent}, $\lia(p > 0)$ is equivalent to $\sic(p>0) = \rho_{f,1}
		\lor \rho_{f,2} \lor \rho_{f,3} \lor \rho_{f,4} \lor \rho_{f,5} =$
	\[
		\mbox{$
				\begin{array}{l@{\,}l@{\,}l@{\,}l@{\,}l@{\,}l@{\,}l}
					     & \left(-x_1 + 4 > 0 \land x_2 = 0 \land \blockz(2,0)\right) & \lor & \left(-x_2 > 0 \land \blockz(2,0)\right) &      &            &            \\[-.1cm]
					\lor & \left(x_3 > 0 \land x_4 = 0 \land x_5 = 0\right)           & \lor & (x_4 > 0 \land x_5 = 0)                  & \lor & (x_5 > 0), & \text{and}
				\end{array}
			$}
	\]
	$\lia(p \geq 0)$ is equivalent to $\sic(p \geq 0) = \sic(p > 0) \lor \left(\blockz(1,0) \land \blockz(2,0)\right)$.
\end{example}
The formulas $\rho_{f,s}$ in \cref{lem:valid_equivalent,coro:valid_equivalent}
are so-called \emph{interval conditions}.

\begin{definition}[Interval Condition]
	\label{Interval Conditions Def}
	For $1\leq i,i' \leq d$, $i \neq i'$, $I \subseteq \{1,\ldots,d\}$, $\mathit{sg}
		\in \{-1,1\}$, and $0 \neq c \in \Q_\ring$, an \emph{interval condition} has the following forms:
	\[
		\mbox{$
				\begin{array}{l@{\qquad\qquad}ll@{\;\;}l@{\; \;}ll}
					\text{(a)} &                  &                &                       & \bigwedge\nolimits_{j \in I}                     & (x_j = 0) \\[-.1cm]
					\text{(b)} & sg \cdot x_i     & > 0 \;\;\land  &                       & \bigwedge\nolimits_{j \in I\setminus \{ i \}}    & (x_j = 0) \\[-.1cm]
					\text{(c)} &                  &                & x_{i'} = c \;\; \land & \bigwedge\nolimits_{j \in I\setminus \{ i' \}}   & (x_j = 0) \\[-.1cm]
					\text{(d)} & sg \cdot x_i     & > 0 \;\; \land & x_{i'} = c \;\; \land & \bigwedge\nolimits_{j \in I\setminus \{ i,i' \}} & (x_j = 0) \\[-.1cm]
					\text{(e)} & sg \cdot x_i + c & > 0 \;\; \land &                       & \bigwedge\nolimits_{j \in I\setminus \{ i \}}    & (x_j = 0)
				\end{array}
			$}
	\]
\end{definition}

\begin{example}
	\label{ex:interval_interval}
	The formulas $\rho_{f,4} = (x_4 > 0) \land (x_5 = 0)$ and $\rho_{f,1} = (-x_1 + 4 > 0)
		\land \bigwedge_{j=2}^5 (x_j = 0)$ from \cref{ex:valid_equivalent_1} are interval conditions as in \cref{Interval Conditions Def} (b) and (e).
\end{example}

\subsubsection{Checking Satisfiability of Interval Conditions}
\label{subsubsec:algorithm}

We now show that to decide satisfiability of the formulas $\sic(p \triangleright 0)$, we only have to regard instantiations of the variables with values from $\{0,\positive,\negative,\nonzero\}$, where $\nonzero$ stands for one additional non-zero value.
There are only polynomially many such instantiations and the particular value for $\nonzero$ is later determined by SMT solving.
This SMT solving only takes polynomial time, because the resulting SMT problem only contains a single variable.
\cref{def:evaluation} instantiates variables accordingly and performs Boolean simplifications as much as possible.

\begin{definition}[Evaluation]
	\label{def:evaluation}
	Let $\rho$ be a propositional formula built from the connectives $\land$ and $\lor$ over atoms of the form $sg \cdot x + c > 0$ and $x = c$ for $sg \in \{1,-1\}$, $c \in \Q_\ring$, and $x \in \{x_1,\ldots,x_d\}$.
	Moreover, let $\vec{v} \in \{0,\positive,\negative,\nonzero\}^d$.
	The \emph{evaluation of $\rho$ w.r.t.\ $\vec{v}$} (written $\rho(\vec{v}){\downarrow}$) results from $\rho(\vec{v}) = \rho[\vec{x}/\vec{v}]$ by simplifying (in)equations without $\nonzero$ to $\TRUE$ or $\FALSE$, and by simplifying conjunctions and disjunctions with $\TRUE$ resp.\ $\FALSE$.
	We write $\vec{v} \maybe \rho$ if $\rho(\vec{v}){\downarrow} \neq \FALSE$.
\end{definition}

For example, if $\rho$ is the formula $(x_1 - \tfrac{5}{2} > 0) \land (x_2 = 0)$ and $\vec{v} = (\nonzero, 0)$, then $\rho(\vec{v}){\downarrow}$ is $\nonzero - \tfrac{5}{2} > 0$.
Hence, $\vec{v} \maybe \rho$.
So in general, $\vec{v} \maybe \rho$ means that $\rho(\vec{v}){\downarrow} = \TRUE$ or that there \emph{could} be a value $w$ for $\nonzero$ such that $\rho[\vec{x}/\vec{v}, \nonzero / w]{\downarrow} = \TRUE$.

Now we define candidate assignments $\satis(\rho_{f,s})$ for the formulas $\rho_{f,s}$ in \cref{lem:valid_equivalent,coro:valid_equivalent} which contain all $\vec{v} \in \{0,\positive,\negative,\nonzero\}^d$ that \emph{may} satisfy $\rho_{f,s}$ (if a suitable value for $\nonzero$ is found).
However, for each Block $B_i$, \emph{at most one} variable $x_j$ with $j \in B_i$ may be assigned a non-zero value (i.e., $\positive$, $\negative$, or $\nonzero$).
Moreover, the value $\nonzero$ may only be used in the block for the eigenvalue $\lambda_i = 1$.

\begin{definition}[Sets of Candidate Assignments]
	\label{def:valAssign}
	For all $0 \leq s \leq \ell$, we define:
	\[
		\begin{array}{ll}
			\satis(\rho_{f,s})\; = \;	\{\vec{v} \in & \{0,\positive,\negative,\nonzero\}^d \, \mid \, \vec{v} \maybe \rho_{f,s},                \\
			[-.1cm]                                & \forall \ 1 \leq i \leq k. \ \text{there is at most one $j \in B_i$ with $v_{j} \neq 0$}, \\
			[-.1cm]                                & v_j = \nonzero \implies j \in B_{i_0} \text{
				where } \lambda_{i_0} = 1 \}
		\end{array}
	\]
\end{definition}

\begin{example}
	\label{ex:valAssign}
	In \cref{ex:valid_equivalent_1,ex:valid_equivalent_2}, for $\rho_{f,4} = (x_4 > 0) \land (x_5 = 0)$, $\vec{v} \maybe \rho_{f,4}$ implies $v_4 = \positive$ and $v_5 = 0$.
	Here, $v_4 = \nonzero$ is not possible, because $4$ does not belong to the block $B_1 = \{1,2 \}$ for the eigenvalue $1$.
	Since at most one value for each block may be non-zero, we have $v_3 = 0$.
	In contrast, $v_1$ and $v_2$ can be arbitrary, but at most one of them may be non-zero.
	Hence, we obtain the following for $\rho_{f,4}$ and for $\rho_{f,0} = (x_1 = 4) \land \bigwedge_{j=2}^5 (x_j = 0)$:
	\[
		\satis(\rho_{f,4})=\{\mat{\positive \\
			0 \\
			0 \\
			\positive \\
			0}\!,\mat{\nonzero \\
			0 \\
			0 \\
			\positive \\
			0}\!,\mat{\negative \\
			0 \\
			0 \\
			\positive \\
			0}\!,\mat{0 \\
			\positive \\
			0 \\
			\positive \\
			0}\!,\mat{0 \\
			\nonzero \\
			0 \\
			\positive \\
			0}\!,\mat{0 \\
			\negative \\
			0 \\
			\positive \\
			0}\!,\mat{0 \\
			0 \\
			0 \\
			\positive \\
			0}\}, \; \satis(\rho_{f,0}) = \{\mat{\nonzero \\
			0 \\
			0 \\
			0 \\
			0}\}
	\]
\end{example}
\begin{lemma}
	\label{lem:cardinality_bound}
	$\card{\satis(\rho_{f,s})} \leq \left(3\!\cdot\!\max \{ \nu_i \mid 1 \leq i \leq k\}
		+ 1\right)^{k}\!$ for all $0\!\leq\!s\!\leq\!\ell$.
\end{lemma}
\makeproof{lem:cardinality_bound}{
	For any $0 \leq s \leq \ell$ we have $\satis(\rho_{f,s}) \subseteq \mathrm{cA}$, where
	\[
		\mathrm{cA} = \{ v \in \{0,\positive,\negative,\nonzero\}^d \mid \forall \ 1 \leq i \leq k. \ \text{there is at most one $j \in B_i$ with $v_{j} \neq 0$} \}.
	\]
	Now we over-approximate the cardinality of $\mathrm{cA}$.
	If $\vec{v} \in \mathrm{cA}$, then for every $1 \leq i \leq k$, we have $v_j \neq 0$ for at most one $j \in B_i$.
	So for the values $v_j$ with $j \in B_i$, there are $3 \cdot \card{B_i} + 1$ possibilities: either exactly one of them is $\positive$, $\negative$, or $\nonzero$, or we have $v_j = 0$ for all $j \in B_i$.
	When combining this result for all $1 \leq i \leq k$ by multiplication, we obtain $\card{\satis(\rho_{f,m})} \leq$
	\[
		\mbox{$\card{\mathrm{cA}} = \prod_{i = 1}^k \left(3 \cdot \card{B_i} + 1\right) = \prod_{i = 1}^k \left(3 \cdot \nu_i + 1\right) \leq \left(3 \cdot \max \{ \nu_i \mid 1 \leq i \leq k\} + 1\right)^k$.}
	\]

	\vspace*{-.5cm}

}

\cref{lem:cardinality_bound} is crucial for our algorithm to decide $k$-termination for uniform loops: $\card{\satis(\rho_{f,s})}$ is bounded by a polynomial in $d$ if $k$ is assumed to be a parameter.
This is because the $\nu_i$ form a $k$-partition of $d$, i.e., $\nu_i \leq d$ for $1 \leq i \leq k$.
Hence, computing $\satis(\rho_{f,s})$ can be done in polynomial time for fixed $k$.

\begin{example}
	In \cref{ex:valAssign}, we have $\nu_1=2$ and $\nu_2=3$, and thus, $k = 2$ and $\max \{ \nu_i \mid 1 \leq i \leq k\} = 3$.
	Here, $\card{\satis(\rho_{f,4})} = 7 \leq 100 = 10^2 = (3 \cdot 3 + 1 )^2$.
\end{example}
This example shows that the bound in \cref{lem:cardinality_bound} is coarse, but it suffices for our analysis.
We now combine \cref{coro:equivalent,def:valAssign} to obtain the sets of candidate assignments for the disjunctions $\sic(p > 0)$ and $\sic(p \geq 0)$.

\begin{corollary}
	\label{coro:cardinality_bound}
        We lift $\satis$ to inequations by defining
	\[
          \begin{array}{l@{\qquad}r@{\;\;}c@{\;\;}l}
            & \satis(\sic(p > 0))    & {}={} & \bigcup\nolimits_{s=1}^\ell \satis(\rho_{f,s}) \\[-.1cm]
            \text{and} & \satis(\sic(p \geq 0)) & {}={} & \satis(\sic(p>0)) \cup \satis(\rho_{f,0}).
          \end{array}
	\]
	Then we have
        \[
          \card{\satis(\sic(p \triangleright 0))} \leq (d+2) \cdot \left(3 \cdot \max \{ \nu_i \mid 1 \leq i \leq k\} + 1\right)^{k}\!.
        \]
\end{corollary}
\makeproof{coro:cardinality_bound}{
	We show that $\ell \leq d+1$ for $\card{\coeffsop(p)}=\ell$.
	Then the result follows from \cref{lem:cardinality_bound} and the definition of $\satis$, since $\satis(\sic(p \triangleright 0)) \subseteq \bigcup\nolimits_{s=0}^\ell \satis(\rho_{f,s})$.
	The number of coefficients of $p$ is determined by the number of terms of the form $\alpha \cdot n^a \cdot b^n$ occurring in $p$.
	But these terms are connected to the bases $0 < \lambda_1 < \ldots < \lambda_k$ of the hierarchical $k$-partition: either $b = \lambda_i$ for some $1 \leq i \leq k$ and $0 \leq a \leq \nu_i-1$ or $a=0$ and $b=1$.
	So we have at most $ 1 + \sum_{i=1}^k \nu_i = 1 + d$ such terms in $p$, i.e., $\ell \leq d + 1$.
}

For a uniform loop with condition $\cond$ and normalized closed form $\closednorm = \vec{h}$, let $\cond(\vec{h})$ contain the atoms $f(\vec{h}) \triangleright 0$, where $f \in \linpoly{\vec{x}}$.
To decide termination, our algorithm computes $\satis(\sic(f(\vec{h})\triangleright 0))$ for all these atoms $f(\vec{h}) \triangleright 0$, and then checks for each of the candidate assignments whether it is a witness for eventual non-termination.
We first lift $\sic$ and $\satis$ to linear formulas.

\begin{definition}[$\sic$ and $\satis$ for Linear Formulas]
	Let $\varphi$ be a linear formula over the atoms $\{f \triangleright 0 \mid \mbox{$f \in \linpoly{\vec{x}}$}, \triangleright \in \{>,\geq\}\}$ and let $\vec{h} = (h_1,\ldots,h_d)$ be a hierarchical $k$-partition.
	Then the formula $\sic(\varphi(\vec{h}))$ results from replacing each atom $f(\vec{h}) \triangleright 0$ in $\varphi(\vec{h}) = \varphi[\vec{x}/\vec{h}]$ by $\sic(f(\vec{h}) \triangleright 0)$. By $\satis(\sic(\varphi(\vec{h})))$ we denote the set $\bigcup\nolimits_{f(\vec{h}) \triangleright 0 \text{ atom in }
			\varphi(\vec{h})} \, \satis(\sic(f(\vec{h}) \triangleright 0))$.
\end{definition}

To analyze termination of uniform loops,
we now present an algorithm to decide whether for a hierarchical $k$-partition $\vec{h}$ and a linear formula $\varphi$,
\begin{align*}
	\exists \vec{x} \in \ring^d, n_0 \in \NN.\ \forall n \in \NN_{> n_0}.\ \varphi(\vec{h}) \tag{see \eqref{eq:decidable_full}}
\end{align*}
is valid.
Our algorithm calls a method SMT($\psi$,$\mathcal{V}$,$\ring$) which checks whether the linear formula $\psi$ in the variables $\mathcal{V}$ is satisfiable.
Here, the variables $\mathcal{V}$ range over $\ring\in\{\ZZ, \QQ,\RA\}$ and the coefficients of the polynomials are from $\Q_\ring$.
(So for $\ring = \ZZ$, one can first multiply all inequations in $\psi$ by the main denominator to result in coefficients from $\ZZ$.)
In our case, $\mathcal{V} = \{ \nonzero \}$ and thus, $\card{\mathcal{V}} = 1$.
With these restrictions, the method SMT has \emph{polynomial} runtime (see \cite{complexityexistentialreal,dblp:journals/mor/lenstra83}). More precisely, SMT is called in \cref{alg:sat} to determine whether $\nonzero$ can be assigned a non-zero value such that $\psi(\vec{v}){\downarrow}$ is satisfiable.
Here, we have to assign \emph{all} occurrences of $\nonzero$ in the formula $\psi(\vec{v}){\downarrow}$ the same value.

\begin{algorithm}
	\caption{Checking Interval Conditions}
	\label{alg:sat}
	\SetAlgoLined
	\KwIn{a formula $\varphi$ over the atoms $\{f \triangleright 0 \mid f \in
			\linpoly{\vec{x}}, \triangleright \in \{>,\geq\}\}$,\linebreak a hierarchical $k$-partition
		$\vec{h} = (h_1,\ldots,h_d)$, and a ring $\ring \in \{ \ZZ, \QQ, \RA\}$}
	\KwResult{$\top$ if $\exists \vec{x} \in \ring^d. \ \sic(\varphi(\vec{h}))$ is valid, $\bot$ otherwise}
	$\psi \assign \sic(\varphi(\vec{h}))$\;
	\ForEach{$\vec{v} \in \satis(\psi)$}{
	$\psi' \assign \psi(\vec{v}){\downarrow}$\;
	\lIf{\normalfont{SMT}(($\psi' \land \nonzero \neq 0),\{\nonzero\},\ring$)}{\KwRet{$\top$}
		}
	}
	\KwRet{$\bot$}
\end{algorithm}

Let us discuss the complexity of \cref{alg:sat}.
The formula $\sic(\varphi(\vec{h}))$ and each element of $\satis(\sic(\varphi(\vec{h})))$ can be computed in polynomial time. By \cref{coro:cardinality_bound}, $\satis(\sic(\varphi(\vec{h})))$ has at most $\card{\varphi} \cdot (d + 2)\cdot \left(3 \cdot \max \{ \nu_i \mid 1 \leq i \leq k \}+ 1\right)^k$ elements, where $\card{\varphi}$ is the number of atoms in $\varphi$ and $\nu_i \leq d$ for all $1 \leq i \leq k$.
Thus, when considering $k$ to be a parameter, $\satis(\sic(\varphi(\vec{h})))$ can be computed in polynomial time.
Moreover, evaluating a formula w.r.t.\ $\vec{v}$ according to \cref{def:evaluation} is possible in polynomial time, too.
Finally, SMT has polynomial runtime as discussed before.
So the runtime of the algorithm is polynomial when regarding $k$ as a parameter.
We now prove that \cref{alg:sat} is sound and complete.

\begin{theorem}
	\label{thm:sat_sound_and_complete}
	\cref{alg:sat} returns $\top$ iff $\exists \vec{x} \in \ring^d. \ \sic(\varphi(\vec{h}))$ is valid.
\end{theorem}
\makeproof{thm:sat_sound_and_complete}{
	\emph{Soundness:}\ If \cref{alg:sat} returns $\top$, then there is a $\vec{v} \in \satis(\sic(\varphi(\vec{h})))$ where $\sic(\varphi(\vec{h}))(\vec{v})$ is satisfiable, i.e., a $w \in \ring$ where $\vec{v}[\nonzero/w]$ is a model of $\sic(\varphi(\vec{h}))$.
	So $\exists \vec{x} \in \ring^d. \ \sic(\varphi(\vec{h}))$ is valid.
	%

	\noindent
	\emph{Completeness:}\ Let $\psi = \sic(\varphi(\vec{h}))$ and let $\exists \vec{x} \in \ring^d. \ \psi$ be valid, i.e., there is a $\vec{v}\in \ring^d$ such that $\psi(\vec{v})$ holds.
	We have to prove that there is a $\vec{v}' \in \satis(\sic(\varphi(\vec{h})))$ where $\psi(\vec{v}'){\downarrow}$ is satisfiable.
	Then the claim follows as the algorithm calls SMT on $\psi(\vec{v}'){\downarrow}$ and thus returns $\top$.

	Note that $\varphi$ is a propositional formula only built from the connectives $\land$ and $\lor$, i.e., it does not contain $\neg$.
	By construction, $\sic(\varphi(\vec{h}))$ results from $\varphi$ by replacing each atom $f \triangleright 0$ by $\sic(f(\vec{h}) \triangleright 0)$.
	So similar to the concept of \emph{fundamental sets} in the proof of \cref{thm:eventual_positiveness}, there is a subset $\{ \sic(f_1(\vec{h}) \triangleright_1 0), \ldots, \sic(f_e(\vec{h}) \triangleright_e 0) \}$ of these formulas such that $\vec{v}$ satisfies them all and such that satisfying these formulas is sufficient for satisfying $\psi$.

	Let $\ell_1$, \ldots, $\ell_e$ be the numbers of coefficients in the poly-exponential expressions $f_1(\vec{h}) $, \ldots, $f_e(\vec{h})$.
	By \cref{coro:equivalent}, $\sic(f_r(\vec{h}) \triangleright_r 0)$ has the form $\bigvee_{s = 1}^{\ell_r} \rho_{f_r,s}$ or $\bigvee_{s = 0}^{\ell_r}
		\rho_{f_r,s}$ for each $1 \leq r \leq e$. So for every $r$ there is \emph{at least} one $s$ where $\rho_{f_r,s}(\vec{v})$ is true. But due to the construction of $\rho_{f_r,s}$ in \cref{lem:valid_equivalent,coro:valid_equivalent}, for every $\vec{v}$ there is \emph{at most}
	one $0 \leq s \leq \ell_r$ where $\rho_{f_r,s}(\vec{v})$ is true.
	Thus, for every $1 \leq r \leq e$, there is a \emph{unique} $0 \leq s_r \leq \ell_r$ where $\rho_{f_r,s_r}(\vec{v})$ is true.

	Let $B_1, \ldots, B_k$ again be the blocks from the $k$-partition $h_1,\ldots,h_d$.
	Note that if there is a block $B_i$ and some $\rho_{f_r,s_r}$ requires $x_j$ with $j \in B_i$ to be non-zero, then $\rho_{f_r,s_r}$ requires all $x_{j'}$\linebreak
	with $j' > j$ and $j' \in B_i$ to be zero.
	Thus, since $\vec{v}$ satisfies \emph{all}
	formulas $\rho_{f_r,s_r}$ for $1 \leq r \leq e$, for each $B_i$ there is \emph{at most} one $r \in B_i$ where some $\rho_{f_r,s_r}$ requires $x_j$ to be non-zero.
	Hence, we can assume that for each block $B_i$ there is at most one $j \in B_i$ where $v_j \neq 0$.
	\begin{enumerate}
		\item \underline{Case $\lambda_{i_0} = 1$ for some $1 \leq i_0 \leq k$:}
		      In this case, for the non-zero entries in $\vec{v}$ belonging to indices in $B_i$ with $i \neq i_0$, only their sign is important since the formulas requiring them to be non-zero are interval conditions according to \cref{Interval Conditions Def} (b) resulting from \cref{lem:valid_equivalent} (d).
		      So these entries can be chosen to be $1$ or $-1$.

		      If there is a (unique) non-zero value $v_j$ with $j \in B_{i_0}$, its value is indeed important: the formulas requiring this value to be non-zero have the form of \cref{Interval Conditions Def} (e) (in \cref{lem:valid_equivalent} (b)) or \cref{Interval Conditions Def} (d) (in \cref{lem:valid_equivalent}
		      (d)) or \cref{Interval Conditions Def} (c) (in \cref{coro:valid_equivalent}).
		      Thus, let
		      \[
			      \mbox{$v'_j =
					      \begin{cases}
						      0,         & \text{if } v_j = 0                          \\
						      \nonzero,  & \text{if } v_j \neq 0 \land j \in B_{i_0}   \\
						      \positive, & \text{if } v_j > 0 \land j \not \in B_{i_0} \\
						      \negative, & \text{if } v_j < 0 \land j \not \in B_{i_0}
					      \end{cases}
				      $}
		      \]
		      As discussed before, $\psi(\vec{v}'){\downarrow}$ is satisfiable as $( \rho_{f_1,s_1} \land \ldots \land \rho_{f_e,s_e}
			      )(\vec{v}'){\downarrow}$ is satisfiable by construction: If the formula contains $\nonzero$, then assigning $\nonzero$ the unique non-zero value $v_j \neq 0$ for $j \in B_{i_0}$ is a satisfying assignment.
		      \smallskip

		\item \underline{Case $\lambda_{i_0} \neq 1$ for all $1 \leq i_0 \leq k$:}
		      In this case, $\rho_{f_1,s_1}$, \ldots, $\rho_{f_e,s_e}$ are interval conditions according to \cref{Interval Conditions Def} (a) or (b).
		      Thus, only the sign of the non-zero values in $\vec{v}$ is important to satisfy these formulas.
		      Hence we define
		      \[
			      \mbox{$v'_j =
					      \begin{cases}
						      0,         & \text{if } v_j = 0 \\
						      \positive, & \text{if } v_j > 0 \\
						      \negative, & \text{if } v_j < 0
					      \end{cases}
				      $}
		      \]
		      Thus, $\psi(\vec{v}'){\downarrow}$ is satisfiable since $( \rho_{f_1,s_1} \land \ldots \land \rho_{f_e,s_e} ) (\vec{v}'){\downarrow}$ is $\TRUE$ by construction.
	\end{enumerate}
	So in both cases $\vec{v}' \in \bigcup\nolimits_{1 \leq r \leq e} \satis\left(\rho_{f_r,s_r}\right) \subseteq \satis(\psi),$
	i.e., \cref{alg:sat} returns $\top$.
}

\begin{example}
	Consider the uniform loop in \cref{fig:uniformLeading}
	where $\varphi = f > 0 \land f' > 0$ for $f = -x_1 + 3 \cdot x_3 + 4$ and $f' = 2 \cdot x_1 - 5$.
	Let $\vec{h} = \closednorm$ as in \cref{fig:uniformLeading}
	and let $p = f(\vec{h})$ and $p' = f'(\vec{h}) = (2 \cdot x_1 - 5) + 2 \cdot x_2 \cdot n$.
	Here, $\psi = \sic(\varphi(\vec{h})) = \sic(p>0) \land \sic(p' > 0)$, where $\sic(p>0) = \bigvee_{s=1}^5\rho_{f,s}$ is stated in \cref{ex:equivalent}.
	Note that $\coeffsop(p') = \{ {\alpha'_1}^{(1,0)}, {\alpha'_2}^{(1,1)} \}$ with $\alpha_1' = 2 \cdot x_1 - 5$ and $\alpha_2' = 2 \cdot x_2$.
	Hence, $\sic(p' > 0) = \rho_{f',1} \vee \rho_{f',2}$.
	To compute $\rho_{f',1}$, for $C = \coeff(f',x_1) \cdot c_1 = 2 \cdot 1 = 2$ and $f'(0,\ldots,0) = -5$, \cref{lem:valid_equivalent} (b) results in $\rho_{f',1} = (x_1 - \tfrac{5}{2} > 0) \land (x_2 = 0)$.
	For $\rho_{f',2}$, with $sg = \sign{\coeff(f',x_1) \cdot c_2} = \sign{2 \cdot 1} = 1$, \cref{lem:valid_equivalent} (d) results in $\rho_{f',2} = (x_2 > 0)$.
	Now from $\sic(p > 0)$ let us choose the disjunct $\rho_{f,1} = (-x_1 + 4 > 0) \land \bigwedge_{j=2}^5 (x_j = 0)$ and from $\sic(p' > 0)$ let us choose the disjunct $\rho_{f',1} = (x_1 - \tfrac{5}{2} > 0) \land (x_2 = 0)$.
	We consider $\vec{v} = (\nonzero, 0, 0, 0, 0)$.
	Then
	\[
		(\rho_{f,1} \land \rho_{f',1})(\vec{v}){\downarrow} \;\;= \;\; ( -\nonzero + 4 > 0) \; \land \; (\nonzero - \tfrac{5}{2} > 0)
	\]
	is satisfiable with the model $\nonzero = 3$. Hence, this model also satisfies $\psi(\vec{v}){\downarrow} \land \nonzero \neq 0$.
	Thus, for both $\ring \in \{ \QQ, \RA \}$, \cref{alg:sat} proves validity of $\exists \vec{x} \in \ring^d. \ \sic(\varphi(\vec{h}))$ and therefore, non-termination of the uniform loop over $\ring$.
	\label{ex:sat}
\end{example}

So for a uniform loop over $\ring \in \{ \QQ, \RA\}$, non-termination is equivalent to validity of $\exists \vec{x} \in \ring^d.\ \forall n \in \NN_{> n_0}.\ \cond(\closednorm)$, which in turn is equivalent to a formula\linebreak
only containing interval conditions.
This insight reduces the search space for\linebreak
proving validity drastically.
Thus, we can now prove \cref{thm:xp} for $\ring \in \{ \QQ, \RA, \RR\}$.

\begin{proof}[Proof of \cref{thm:xp}]
	For $\ring \in \{ \QQ, \RA \}$, we first transform the uniform loop such that the update matrix is in Jordan normal form and then compute the norma\-lized closed form as in \cref{coro:closed_uniform_hierar_exp} in polynomial time.
	This closed form is a\linebreak
	hierarchical partition by \cref{coro:uniformity_partition}.
	By combining \cref{coro:equivalent,thm:sat_sound_and_complete}, \cref{alg:sat} can
	decide validity of the formula from \cref{thm:eventual_positiveness}, i.e., termination of the trans\-formed loop (which is equivalent to termination of the original loop by \cref{coro:conclusion_transformations}).

	As the computation of the equivalent interval conditions in \cref{coro:equivalent} clearly works in polynomial time and we have discussed that \cref{alg:sat} runs in polynomial time when $k$ is assumed to be a parameter, this proves the statement.

	\noindent
	Finally,
	these loops terminate over $\RA$ iff they terminate over $\RR$ by \cref{coro:decidability_reals}.
\end{proof}

For \cref{thm:xp}, it was crucial to \emph{transform} the loop such that the update matrix is in Jordan normal form.
Here we relied on a special closed form for the Jordan normal form, while in \cref{subsec:linear-update} we only used the transformation to argue why the closed form is computable in polynomial time.
Thus, the transformation from \cref{sec:transf} does not only generalize our results from \cref{sec:deciding} to a wider class of loops but it also gives rise to novel results like \cref{thm:xp}.

The approach in the proof of \cref{thm:xp} also works for uniform loops over $\ZZ$ if the update matrix is already in Jordan normal form.
But otherwise,
in addition to $\varphi(\closednorm)$ we also have an update-invariant and $\EFO{\ring,\RA}$-definable subset $F$ which stems from the transformation into Jordan normal form (see \cref{sec:transf}).
Thus, to decide termination we have to decide validity of $\exists \vec{x} \in \RA^d.\ \forall n \in \NN_{> n_0}.\ \cond(\closednorm) \land \psi_F$.
We will discuss this in the next section.

%% file: integer.tex
\subsubsection{Termination of Uniform Loops Over the Integers}
\label{subsec:integer_uniform_loops}

Now we show that deciding termination of uniform loops $(\cond,A\cdot \vec{x})$ over the \emph{integers} is also in \cc{XP}.
Let $A \in \ZZ^{d \times d}$ with $k$ integer eigenvalues each of geometric multiplicity one.
Then there is a matrix $T \in \QQ^{d \times d}$ such that $A = T^{-1} \cdot Q \cdot T$ for a matrix $Q$ in Jordan normal form.
However, in general we do not have $T,T^{-1}\in \ZZ^{d \times d}$ (see, e.g., \cite{sidorov2019}).
As before, let $\eta(\vec{x}) = T \cdot \vec{x}$ and $\cond' = \eta^{-1}(\cond)$.
Then termination of $(\cond,A\cdot \vec{x})$ on $\ZZ^d$ is equivalent to termination of $(\cond',Q \cdot \vec{x})$ on $\widehat{\eta}(\ZZ^d) = T \cdot \ZZ^d$ by \cref{coro:conclusion_transformations}.
Here, $\Lan_T = T \cdot \ZZ^d$ is the set of all \emph{integer} linear combinations of $T$'s columns, i.e., their \emph{lattice}.
In general, we have $\Lan_T \neq \ZZ^d$.

Hence, we now call \cref{alg:sat} with the input $(\cond',\vec{h},\QQ)$, where $\vec{h} = \closednorm$ is the normalized closed form for the update $Q \cdot \vec{x}$.
For $\psi = \sic(\varphi'(\vec{h}))$, we want to find out if $\psi(\vec{v})$ holds for some $\vec{v} \in \Lan_T$.
Such a $\vec{v} \in \Lan_T$ would witness eventual non-termination of $(\cond',Q \cdot \vec{x})$, and since $\Lan_T$ is update-invariant under $Q \cdot \vec{x}$ by \cref{lem:action_preserves_invariance}, this is equivalent to non-termination of $(\cond',Q \cdot \vec{x})$ on $\Lan_T$.

We modify \cref{alg:sat} such that it computes \emph{all} $\vec{v} \in \satis(\psi)$ where $\psi(\vec{v}){\downarrow} \land \nonzero \neq 0$ is satisfied by some $\vec{v}'$ that results from $\vec{v}$ by instantiating $\nonzero$ with a suitable number from $\QQ$.
As shown in \cref{coro:cardinality_bound}, for a fixed number of eigenvalues $k$, there are only polynomially many such candidate assignments $\vec{v}$.
Note that the formula $\varphi'$ is only built from the connectives $\land$ and $\lor$, and $\psi$ results from $\cond'$ by replacing each atom $f \triangleright 0$ by $\sic(f(\vec{h}) \triangleright 0)$.
Hence, for every such $\vec{v}$ there is a subset $\{ \sic(f_1(\vec{h}) \triangleright_1 0), \ldots, \sic(f_e(\vec{h}) \triangleright_e 0) \}$ of these formulas such that $\vec{v}'$ satisfies them all and such that satisfying these formulas is sufficient for satisfying $\psi$.
For each $1 \leq r \leq e$, let $\ell_r = |\coeffsop(f_r(\vec{h}))|$.
By \cref{coro:equivalent}, $\sic(f_r(\vec{h}) \triangleright_r 0)$ has the form $\bigvee_{s = 1}^{\ell_r} \rho_{f_r,s}$ or $\bigvee_{s = 0}^{\ell_r}
	\rho_{f_r,s}$. So for every $r$ there is \emph{at least} one $s$ where $\rho_{f_r,s}(\vec{v}')$ is true. But due to the construction of $\rho_{f_r,s}$ in \cref{lem:valid_equivalent,coro:valid_equivalent}, there is \emph{at most}
one $0 \leq s \leq \ell_r$ where $\rho_{f_r,s}(\vec{v}')$ is true.
Thus, for every $1 \leq r \leq e$, there is a \emph{unique} $0 \leq s_r \leq \ell_r$ where $\rho_{f_r,s_r}(\vec{v}')$ is true.

By \cref{lem:valid_equivalent,coro:valid_equivalent}, all $\rho_{f_r,s_r}$ are interval conditions.
Thus, for each entry $v_j$ of $\vec{v}'$ we can find out whether $x_j = v_j$ is required by some $\rho_{f_r,s_r}$, or whether $v_j = 0$ is just due to setting variables to zero by default, i.e., the formula would still hold when assigning an arbitrary value from $\QQ$ to $v_j$.
So every $\vec{v} \in \satis(\psi)$ gives rise to a certain set of formulas $\{
	\rho_{f_1,s_1}, \ldots, \rho_{f_e,s_e} \}$, which in turn results in a certain \emph{abstract assignment} that indicates for each entry of $\vec{v}'$ whether its actual value is necessary to be a model.

\begin{definition}[Abstract Assignment]
	\label{def:abstract_assignment}
	Let $\mathbb{I}$ be the set of all intervals of the forms $[c,c]$, $(-\infty, c)$, $(c, \infty)$, $(-\infty, \infty)$, or $(c,d)$ for $c,d \in \QQ$ with $c \leq d$.
	Then an \emph{abstract assignment} is an element of $\mathbb{I}^d$.
\end{definition}

For each of the obtained abstract assignments, we now have to check whether it is satisfied by some value from $\Lan_T$.
Let $N \subseteq \{1,\ldots,d\}$ be those indices where $j \in N$ iff the $j$-th component of the abstract assignment is $[0,0]$, i.e., iff the $j$-th component must be 0 in order to satisfy all $\rho_{f_s,m_s}$.
Then we compute a basis of the sublattice $\Lan_N = \{\vec{w} \in \Lan_T \mid w_j =0 \text{ for } j \in N\}$.
To this end, we solve the system of linear equations $\bigwedge_{j \in N} (T \cdot \vec{x})_j = 0$ where $\vec{x} \in \ZZ^d$.
Here as usual, $(T \cdot \vec{x})_j$ denotes the $j$-th component of the vector $T \cdot \vec{x}$.
This problem can be solved in polynomial time (see, e.g., \cite{dblp:conf/issac/storjohannl96,diophantine_polynomial_time}).
Since in general this system contains more variables than equations, the solutions yield a certain linear dependence between the variables.
This dependence can then be used to reduce the number of variables in the system, i.e., it gives rise to a basis of $\Lan_N$, where each basis vector is represented by a $\ZZ$-linear combination of the columns of $T$.
Let $d' \leq d$ be the rank of the sublattice $\Lan_N$ (i.e., the number of its basis vectors) and let $P \in \QQ^{d \times d'}$ be the matrix whose columns form the basis of $\Lan_N$.

Let $N' \subseteq \{1,\ldots,d\}$ be those indices where $j \in N'$ iff the $j$-th component of the abstract assignment is neither $[0,0]$ nor $(-\infty,\infty)$.
So the $j$-th component must be from a certain interval in order to satisfy all $\rho_{f_s,m_s}$.
To ease notation, we define $I_j = (-\infty,\infty)$ if $j \notin N'$ and let $K=\prod_{j=1}^d I_j \subseteq \QQ^d$.
Now we have to decide whether there exists an $X \in \ZZ^{d'}$ such that $P \cdot X \in K$.

Let $B_1, \ldots, B_k$ again be the blocks from the $k$-partition $h_1,\ldots,h_d$.
Note that if there is a block $B_i$ where some $\rho_{f_r,s_r}$ requires $x_j$ with $j \in B_i$ to be non-zero (i.e., $j \in N'$), then $\rho_{f_r,s_r}$ requires all $x_{j'}$ with $j' > j$ and $j' \in B_i$ to be zero (i.e., $j' \in N$).
Thus, since $\vec{v}'$ satisfies \emph{all}
formulas $\rho_{f_r,s_r}$ for $1 \leq r \leq e$, for each block $B_i$ there can be \emph{at most} one $r \in B_i$ where some $\rho_{f_r,s_r}$ requires $x_j$ to be non-zero.
Hence, for each block $B_i$ there is at most one $j \in B_i$ where $j \in N'$.
Note that containment in an interval can be described by at most $2$ inequations, where the strict inequations can be turned into weak ones since the variables only range over the integers.
Thus, to describe the required containment in the intervals for all $x_j$ with $j \in N'$, we need at most $2 \cdot k$ inequations.
In other words, the requirement $P \cdot X \in K$ can be described by $2 \cdot k$ linear inequations where the coefficients are from $\QQ$.
Since linear integer programming with rational coefficients and a fixed number of constraints is possible in polynomial time (see \cite{dblp:journals/mor/lenstra83}), this shows that checking whether a candidate assignment $\vec{v}$ gives rise to a solution in $\Lan_T$ can be done in \cc{XP}.
As $\card{\satis(\psi)}$ is also polynomial for fixed $k$, $k$-termination of uniform loops over $\ZZ$ is in \cc{XP} as well.

%% file: conclusion.tex
\section{Conclusion and Related Work}
\label{sec:conclusion}

In this work, we studied termination of \twn-loops, i.e., loops
where the update $\ASSIGN{\vec{x}}{\update}$ is a triangular system of polynomial equations and the use of non-linearity in $\update$ is mildly restricted.
We first presented a reduction from termination of \twn-loops to $\EFO{\ring}$ in \cref{sec:closed,sec:deciding}.
This implies decidability of termination over $\ring \in \{\RA,\RR\}$ and semi-decidability of non-termination over $\ring \in \{\ZZ,\QQ\}$.

In addition, we showed how to transform certain non-\twn-loops into \twn-form in \cref{sec:transf}, and discussed how this generalizes our results to a wider class of loops.
We also showed that \twn-transformability is semi-decidable.

Afterwards, we analyzed the complexity of deciding termination for different subclasses of \twn-loops.
In \cref{subsec:linearizing}, we first showed that linearizing \twn-loops can be done in
double exponential time.
In \cref{sec:complexity}, we used our transformation and decision procedure to prove \cc{Co-NP}-completeness ($\forall \RR$-completeness) of termination of linear (linear-update) loops with rational (real) spectrum, and based on linearization, that deciding termination of arbitrary \twn-loops over $\RA$ or $\RR$ is in \cc{3-EXPTIME}.

Finally, we showed that for the subclass of uniform loops over $\ring\in \{\ZZ,\QQ,\RA,\linebreak
	\RR\}$, termination can be decided in polynomial time, if the number of eigenvalues of the update matrix is fixed.
So here our decision procedure can be used as an efficient technique for termination analysis.

%% file: related_work.tex
\subsubsection*{Related Work}
\label{part_1:sec:related_work}

In contrast to automated termination analysis (see e.g., \cite{rank,dblp:conf/fmcad/larrazorr13,dblp:conf/cav/brockschmidtcf13,aprove,termcomp,dblp:journals/corr/leikeh15,dblp:conf/cav/ben-amramg17,dblp:conf/vmcai/podelskir04,dblp:conf/cav/bradleyms05,dblp:conf/sas/ben-amramdg19}), we investigated \emph{decidability} of termination for certain classes of loops in \cref{sec:deciding}.
As termination is undecidable in general, decidability results can only be obtained for very restricted classes of programs.

Nevertheless, many techniques used in automated tools for termination analysis (e.g., ranking functions \cite{dblp:conf/vmcai/podelskir04,dblp:conf/sas/ben-amramdg19,dblp:conf/cav/bradleyms05,dblp:journals/jacm/ben-amramg14,dblp:conf/cav/ben-amramg17,rank}) focus on similar classes of loops, since such loops occur as sub-programs in (abstractions of) real programs.
Tools based on these techniques have turned out to be very successful, also for larger classes of programs.
Thus, these tools could benefit from integrating our (semi-)de\-cision procedures and applying them instead of incomplete techniques for any sub-program that can be transformed into a \twn-loop.

Related work on decidability of termination also considers similar (and often more restricted) classes of loops.
For linear conjunctive loops, termination over $\RR$ \cite{dblp:conf/cav/tiwari04,li14,li-witnesses,dblp:journals/fac/xiayzz11} and $\QQ$ \cite{dblp:conf/cav/braverman06} is decidable.
Decidability of termination of linear conjunctive loops over $\ZZ$ was conjectured to be decidable in \cite{dblp:conf/cav/tiwari04}.
After several partial results \cite{bozga14,cav19,dblp:conf/soda/ouakninepw15} this conjecture was confirmed recently in \cite{dblp:conf/icalp/hosseinio019}.
However, \cite{dblp:journals/toplas/ben-amramgm12} shows that for slight generalizations of linear conjunctive loops over $\ZZ$, where a non-deterministic update or a single piecewise update of a variable are allowed, termination is undecidable.
Tiwari \cite{dblp:conf/cav/tiwari04} uses the special case of our \twn-transformation from \cref{sec:transf} where the loop and the automorphism are linear.
In contrast to these results, our approach applies to \emph{non-linear} loops with \emph{arbitrary} conditions over \emph{various rings}.

\emph{Linearization} is another attempt to handle non-linearity, see \cref{subsec:linearizing}.
While the \emph{update} of solvable loops can be linearized \cite{oliveira16}, the \emph{condition} cannot.
Other\-wise, one could linearize any loop $(p = 0, \vec{x})$, which terminates over $\ZZ$ iff $p$ has no
integer root.
By \cite{dblp:conf/icalp/hosseinio019}, this would imply decidability of Hilbert's Tenth Problem.

In the non-linear case, \cite{li16} proves decidability of termination for conjunctive loops on $\RR^d$ for the case that the condition defines a compact and connected subset of $\RR^d$.
In \cite{xiaz10}, decidability of termination of conjunctive linear-update loops on $\RR^d$ with the \emph{non-zero minimum property} is shown, which covers conjunctive linear-update loops with real spectrum.
For general conjunctive linear-update loops on $\RR^d$ undecidability is conjectured.
Moreover, \cite{dblp:conf/iccsa/wusbz10} shows that termination of conjunctive linear-update loops where the update matrix has only periodic real eigenvalues is decidable, which also covers conjunctive linear-update loops with real spectrum.
Here, a special case of our transformation from \cref{sec:transf} with linear automorphisms is used.
In combination with \cite{oliveira16}, \cite{dblp:conf/iccsa/wusbz10,xiaz10} both yield a decision procedure for termination of conjunctive \twn-loops over $\RR$.
Furthermore, \cite{dblp:conf/concur/neumanno020} proves that termination of (not necessarily conjunctive) linear-update loops is decidable if the condition describes a compact set.
Finally, \cite{xu13} gives sufficient criteria for (non-)termination of solvable loops and \cite{li17} presents sufficient conditions under which termination of non-deterministic non-linear loops on $\RR^d$ can be reduced to satisfiability of a semi-algebraic system.

For linear-update loops with real spectrum over $\RR$, we prove $\forall \RR$-complete\-ness of termination, whereas \cite{xiaz10,dblp:conf/iccsa/wusbz10} do not give tight complexity results.
The approach from \cite{xu13} is incomplete, whereas we present a complete reduction from termination to the respective existential fragment of the first-order theory.
The work in \cite{li17} is orthogonal to ours as it only applies to loops that satisfy certain non-trivial conditions.
Moreover, we consider loops with arbitrary conditions over various rings, while \cite{li16,xiaz10,li17,dblp:conf/iccsa/wusbz10} only consider conjunctive loops over $\RR$ and \cite{dblp:conf/concur/neumanno020} only considers loops over $\RR$ where the condition defines a compact set.

Regarding complexity, \cite{dblp:conf/soda/ouakninepw15} proves that termination of conjunctive linear loops over $\ZZ$ with update $\vec{x} \assign A \cdot \vec{x} + \vec{b}$ is in \cc{PSPACE} if $\vert{}\vec{x}\vert{} \leq 4$ resp.\ in \cc{EXPSPACE} if $A$ is dia\-gonalizable.
Moreover, in \cite{dblp:journals/jacm/ben-amramg14} it is shown that existence of a linear (lexicographic) ranking function for linear conjunctive loops over $\QQ$ or $\ZZ$ is \cc{Co-NP}-complete.

Our \cc{Co-NP}-completeness result is or\-tho\-go\-nal to those results as we allow disjunctions in the condition.
Moreover, \cc{Co-NP}-completeness also holds for termination over $\ZZ$, while \cite{dblp:conf/cav/braverman06,dblp:conf/cav/tiwari04} only consider termination over $\QQ$ resp.\ $\RR$.
Additionally, we showed that $k$-termination of uniform loops over $\ZZ$, $\QQ$, $\RA$, and $\RR$ is in \cc{XP}, where the parameter $k$ is the number of eigenvalues.
This result is also orthogonal to \cite{dblp:conf/cav/braverman06,dblp:conf/cav/tiwari04} since we again allow disjunctions in the condition.
Furthermore, existence of a linear (lexicographic) ranking function is not necessary for termination of linear loops.
We refer to \cite{thesismarcel} for further discussion on possible extensions of our results to uniform loops over $\ring$, where however the eigenvalues are \emph{not} from $\ring$.

Several works exploit the existence of closed forms for solvable (or similar classes of) loops, e.g., to analyze termination on a \emph{given} input, to infer runtime bounds, or to reason about invariants \cite{lpar20,kincaid19,oliveira16,kovacs08,solvable-maps,dblp:conf/vmcai/humenbergerk21}.
While our approach covers solvable loops with real eigenvalues (by \cref{cor:solvable}), it also applies to loops which are not solvable, see \cref{ex:linear_transf}.
Our transformation of \cref{sec:transf} may also be of interest for other techniques for solvable or other sub-classes of polynomial loops, as it may be used to extend the applicability of such approaches.

\subsubsection*{Acknowledgments}
\vspace*{-.3cm}
We thank Alberto Fiori for help with the example $\LL_{\mathit{non-pspace}}$ from \eqref{fig:loop:non-poly-ex} and Arno van den Essen for useful discussions.

%% file: main.bbl
\providecommand{\noopsort}[1]{}\providecommand*{\hyphen}-

%% file: main.bbl
\begin{thebibliography}{66}
\providecommand{\natexlab}[1]{#1}
\providecommand{\url}[1]{{#1}}
\providecommand{\urlprefix}{URL }
\expandafter\ifx\csname urlstyle\endcsname\relax
  \providecommand{\doi}[1]{DOI~\discretionary{}{}{}#1}\else
  \providecommand{\doi}{DOI~\discretionary{}{}{}\begingroup
  \urlstyle{rm}\Url}\fi
\providecommand{\eprint}[2][]{\url{#2}}

\bibitem[{Alias et~al.(2010)Alias, Darte, Feautrier, and Gonnord}]{rank}
Alias C, Darte A, Feautrier P, Gonnord L (2010) Multi-dimensional rankings,
  program termination, and complexity bounds of flowchart programs. In: Proc.\
  SAS, LNCS 6337, pp 117--133, \doi{10.1007/978-3-642-15769-1_8}

\bibitem[{Basu et~al.(2006)Basu, Pollack, and Roy}]{complexityexistentialreal}
Basu S, Pollack R, Roy MF (2006) Algorithms in Real Algebraic Geometry.
  Algorithms and Comp.\ in Math.\ 10, Springer, \doi{10.1007/3-540-33099-2}

\bibitem[{Basu\noopsort{1} and Mishra(2017)}]{handbookmishra}
Basu\noopsort{1} S, Mishra B (2017) Computational and quantitative real
  algebraic geometry. In: Goodman JE, O'Rourke J, T\'oth CD (eds) Handbook of
  Discrete and Computational Geometry, 3rd Ed., CRC, pp 969--1002

\bibitem[{Ben-Amram et~al.(2012)Ben-Amram, Genaim, and
  Masud}]{dblp:journals/toplas/ben-amramgm12}
Ben-Amram AM, Genaim S, Masud AN (2012) On the termination of integer loops.
  {ACM} Trans\ Prog\ Lang\ Syst 34(4), \doi{10.1145/2400676.2400679}

\bibitem[{Ben-Amram\noopsort{1} and
  Genaim(2014)}]{dblp:journals/jacm/ben-amramg14}
Ben-Amram\noopsort{1} AM, Genaim S (2014) Ranking functions for
  linear-constraint loops. J\ {ACM} 61(4), \doi{10.1145/2629488}

\bibitem[{Ben-Amram\noopsort{1} and Genaim(2017)}]{dblp:conf/cav/ben-amramg17}
Ben-Amram\noopsort{1} AM, Genaim S (2017) On multiphase-linear ranking
  functions. In: Proc.\ CAV, LNCS 10427, pp 601--620,
  \doi{10.1007/978-3-319-63390-9_32}

\bibitem[{Ben-Amram\noopsort{1} et~al.(2019)Ben-Amram\noopsort{1},
  Dom{\'{e}}nech, and Genaim}]{dblp:conf/sas/ben-amramdg19}
Ben-Amram\noopsort{1} AM, Dom{\'{e}}nech JJ, Genaim S (2019) Multiphase-linear
  ranking functions and their relation to recurrent sets. In: Proc.\ SAS, LNCS
  11822, pp 459--480, \doi{10.1007/978-3-030-32304-2_22}

\bibitem[{Bozga et~al.(2014)Bozga, Iosif, and Konecn{\'{y}}}]{bozga14}
Bozga M, Iosif R, Konecn{\'{y}} F (2014) Deciding conditional termination. Log\
  Methods Comput\ Sci 10(3), \doi{10.2168/LMCS-10(3:8)2014}

\bibitem[{Bradley et~al.(2005)Bradley, Manna, and
  Sipma}]{dblp:conf/cav/bradleyms05}
Bradley AR, Manna Z, Sipma HB (2005) Linear ranking with reachability. In:
  Proc.\ CAV, LNCS 3576, pp 491--504, \doi{10.1007/11513988_48}

\bibitem[{Braverman(2006)}]{dblp:conf/cav/braverman06}
Braverman M (2006) Termination of integer linear programs. In: Proc.\ CAV, LNCS
  4144, pp 372--385, \doi{10.1007/11817963_34}

\bibitem[{Brockschmidt et~al.(2013)Brockschmidt, Cook, and
  Fuhs}]{dblp:conf/cav/brockschmidtcf13}
Brockschmidt M, Cook B, Fuhs C (2013) Better termination proving through
  cooperation. In: Proc.\ CAV, LNCS 8044, pp 413--429,
  \doi{10.1007/978-3-642-39799-8_28}

\bibitem[{Canny(1988)}]{dblp:conf/stoc/canny88}
Canny JF (1988) Some algebraic and geometric computations in \cc{PSPACE}. In:
  Proc.\ STOC, pp 460--467, \doi{10.1145/62212.62257}

\bibitem[{Cohen(1969)}]{cohen69}
Cohen PJ (1969) Decision procedures for real and $p$-adic fields. Commun\ Pure
  Appl\ Math 22(2):131--151, \doi{10.1002/cpa.3160220202}

\bibitem[{Dai and Xia(2012)}]{dblp:conf/ictac/daix12}
Dai L, Xia B (2012) Non-termination sets of simple linear loops. In: Proc.\
  ICTAC, LNCS 7521, pp 61--73, \doi{10.1007/978-3-642-32943-2_5}

\bibitem[{Dantzig(1949)}]{dantzig}
Dantzig GB (1949) Programming in a linear structure. Econometrica 17:73--74,
  {R}eport of the September 9, 1948 meeting in Madison

\bibitem[{Downey and Fellows(1999)}]{df99}
Downey RG, Fellows MR (1999) Parameterized Complexity. Monographs in Computer
  Science, Springer, \doi{10.1007/978-1-4612-0515-9}

\bibitem[{van~den Essen and Hubbers(1996)}]{vandenessen1996121}
van~den Essen A, Hubbers E (1996) Polynomial maps with strongly nilpotent
  {J}acobian matrix and the {J}acobian conjecture. Linear Algebra Its Appl
  247:121 -- 132, \doi{10.1016/0024-3795(95)00095-X}

\bibitem[{van~den Essen\noopsort{1}(2000)}]{polyautomorphisms}
van~den Essen\noopsort{1} A (2000) Polynomial Automorphisms and the Jacobian
  Conjecture. Springer, \doi{10.1007/978-3-0348-8440-2}

\bibitem[{Frohn and Giesl(2019)}]{cav19}
Frohn F, Giesl J (2019) Termination of triangular integer loops is decidable.
  In: Proc.\ CAV, LNCS 11562, pp 269--286, \doi{10.1007/978-3-030-25543-5_24}

\bibitem[{Frohn et~al.(2020)Frohn, Hark, and Giesl}]{sas}
Frohn F, Hark M, Giesl J (2020) Termination of polynomial loops. In: Proc.\
  {SAS}, LNCS 12389, pp 89--112, \doi{10.1007/978-3-030-65474-0_5}

\bibitem[{Frohn\noopsort{2}(2020)}]{frohn20}
Frohn\noopsort{2} F (2020) A calculus for modular loop acceleration. In: Proc.\
  TACAS, LNCS 12078, pp 58--76, \doi{10.1007/978-3-030-45190-5_4}

\bibitem[{Frumkin(1977)}]{diophantine_polynomial_time}
Frumkin MA (1977) Polynomial time algorithms in the theory of linear
  diophantine equations. In: Proc.\ FCT, LNCS 56, pp 386--392,
  \doi{10.1007/3-540-08442-8_106}

\bibitem[{Giesbrecht(1995)}]{dblp:journals/siamcomp/giesbrecht95}
Giesbrecht M (1995) Nearly optimal algorithms for canonical matrix forms.
  {SIAM} J\ Comput 24(5):948--969, \doi{10.1137/S0097539793252687}

\bibitem[{Giesl et~al.(2017)Giesl, Aschermann, Brockschmidt, Emmes, Frohn,
  Fuhs, Hensel, Otto, Pl{\"{u}}cker, Schneider-Kamp, Str{\"{o}}der, Swiderski,
  and Thiemann}]{aprove}
Giesl J, Aschermann C, Brockschmidt M, Emmes F, Frohn F, Fuhs C, Hensel J, Otto
  C, Pl{\"{u}}cker M, Schneider-Kamp P, Str{\"{o}}der T, Swiderski S, Thiemann
  R (2017) Analyzing program termination and complexity automatically with
  \textsf{AProVE}. J\ Autom\ Reason 58(1):3--31,
  \doi{10.1007/s10817-016-9388-y}

\bibitem[{Giesl et~al.(2019)Giesl, Rubio, Sternagel, Waldmann, and
  Yamada}]{termcomp}
Giesl J, Rubio A, Sternagel C, Waldmann J, Yamada A (2019) The termination and
  complexity competition. In: Proc.\ TACAS, LNCS 11429, pp 156--166,
  \doi{10.1007/978-3-030-17502-3_10}

\bibitem[{Gomory(1960)}]{gomorymilp}
Gomory R (1960) An algorithm for the mixed integer problem. Tech. Rep. RM-2597,
  The RAND Corporation,
  \urlprefix\url{https://www.rand.org/pubs/research_memoranda/RM2597.html}

\bibitem[{Graham et~al.(1994)Graham, Knuth, and
  Patashnik}]{dblp:books/aw/gkp1994}
Graham RL, Knuth DE, Patashnik O (1994) Concrete Mathematics: {A} Foundation
  for Computer Science, 2nd Ed. Addison-Wesley

\bibitem[{Hark et~al.(2020)Hark, Frohn, and Giesl}]{lpar20}
Hark M, Frohn F, Giesl J (2020) Polynomial loops: Beyond termination. In:
  Proc.\ LPAR, EPiC 73, pp 279--297, \doi{10.29007/nxv1}

\bibitem[{Hark\noopsort{1}(2021)}]{thesismarcel}
Hark\noopsort{1} M (2021) Towards complete methods for automated complexity and
  termination analysis of (probabilistic) programs. PhD thesis, RWTH Aachen
  University, {Germany}, \doi{10.18154/RWTH-2021-06073}

\bibitem[{Hosseini et~al.(2019)Hosseini, Ouaknine, and
  Worrell}]{dblp:conf/icalp/hosseinio019}
Hosseini M, Ouaknine J, Worrell J (2019) Termination of linear loops over the
  integers. In: Proc.\ ICALP, LIPIcs 132, \doi{10.4230/LIPIcs.ICALP.2019.118}

\bibitem[{Humenberger and Kov{\'{a}}cs(2021)}]{dblp:conf/vmcai/humenbergerk21}
Humenberger A, Kov{\'{a}}cs L (2021) Algebra-based synthesis of loops and their
  invariants (invited paper). In: Proc.\ {VMCAI}, LNCS 12597, pp 17--28,
  \doi{10.1007/978-3-030-67067-2_2}

\bibitem[{Kantorovich(1960)}]{kantorovich}
Kantorovich LV (1960) Mathematical methods of organizing and planning
  production. Manage Sci 6(4):366–422, \doi{10.1287/mnsc.6.4.366}

\bibitem[{Kauers and Paule(2011)}]{tetrahedron}
Kauers M, Paule P (2011) The Concrete Tetrahedron -- Symbolic Sums, Recurrence
  Equations, Generating Functions, Asymptotic Estimates. Springer,
  \doi{10.1007/978-3-7091-0445-3}

\bibitem[{Kincaid et~al.(2019)Kincaid, Breck, Cyphert, and Reps}]{kincaid19}
Kincaid Z, Breck J, Cyphert J, Reps TW (2019) Closed forms for numerical loops.
  Proc {ACM} Program Lang 3({POPL}), \doi{10.1145/3290368}

\bibitem[{Kov{\'{a}}cs(2008)}]{kovacs08}
Kov{\'{a}}cs L (2008) Reasoning algebraically about $p$-solvable loops. In:
  Proc.\ TACAS, LNCS 4963, pp 249--264, \doi{10.1007/978-3-540-78800-3_18}

\bibitem[{Land and Doig(1960)}]{branchbound}
Land AH, Doig AG (1960) An automatic method of solving discrete programming
  problems. Econometrica 28(3):497--520, \doi{10.2307/1910129}

\bibitem[{Landau(1909)}]{landau}
Landau E (1909) Handbuch der Lehre von der Verteilung der Primzahlen. Teubner

\bibitem[{Larraz et~al.(2013)Larraz, Oliveras, Rodr\'iguez-Carbonell, and
  Rubio}]{dblp:conf/fmcad/larrazorr13}
Larraz D, Oliveras A, Rodr\'iguez-Carbonell E, Rubio A (2013) Proving
  termination of imperative programs using {Max-SMT}. In: Proc.\ FMCAD, pp
  218--225, \doi{10.1109/FMCAD.2013.6679413}

\bibitem[{Leike and Heizmann(2015)}]{dblp:journals/corr/leikeh15}
Leike J, Heizmann M (2015) Ranking templates for linear loops. Log\ Methods
  Comput\ Sci 11(1), \doi{10.2168/LMCS-11(1:16)2015}

\bibitem[{{Lenstra Jr.}(1983)}]{dblp:journals/mor/lenstra83}
{Lenstra Jr} HW (1983) Integer programming with a fixed number of variables.
  Math\ Oper\ Res 8(4):538--548, \doi{10.1287/moor.8.4.538}

\bibitem[{Li(2014)}]{li14}
Li Y (2014) A recursive decision method for termination of linear programs. In:
  Proc.\ SNC, pp 97--106, \doi{10.1145/2631948.2631966}

\bibitem[{Li(2016)}]{li16}
Li Y (2016) Termination of single-path polynomial loop programs. In: Proc.\
  ICTAC, LNCS 9965, pp 33--50, \doi{10.1007/978-3-319-46750-4_3}

\bibitem[{Li(2017{\natexlab{a}})}]{li17}
Li Y (2017{\natexlab{a}}) Termination of semi-algebraic loop programs. In:
  Proc.\ SETTA, LNCS 10606, pp 131--146, \doi{10.1007/978-3-319-69483-2_8}

\bibitem[{Li(2017{\natexlab{b}})}]{li-witnesses}
Li Y (2017{\natexlab{b}}) Witness to non-termination of linear programs. Theor\
  Comput\ Sci 681:75--100, \doi{10.1016/j.tcs.2017.03.036}

\bibitem[{Matijasevi\v{c}(1970)}]{mr0258744}
Matijasevi\v{c} JV (1970) The {D}iophantineness of enumerable sets. Dokl Akad
  Nauk SSSR 191:279--282

\bibitem[{Neumann et~al.(2020)Neumann, Ouaknine, and
  Worrell}]{dblp:conf/concur/neumanno020}
Neumann E, Ouaknine J, Worrell J (2020) On ranking function synthesis and
  termination for polynomial programs. In: Proc.\ {CONCUR}, LIPIcs 171,
  \doi{10.4230/LIPIcs.CONCUR.2020.15}

\bibitem[{de~Oliveira et~al.(2016)de~Oliveira, Bensalem, and
  Prevosto}]{oliveira16}
de~Oliveira S, Bensalem S, Prevosto V (2016) Polynomial invariants by linear
  algebra. In: Proc.\ ATVA, LNCS 9938, pp 479--494,
  \doi{10.1007/978-3-319-46520-3_30}

\bibitem[{Ouaknine et~al.(2015)Ouaknine, Pinto, and
  Worrell}]{dblp:conf/soda/ouakninepw15}
Ouaknine J, Pinto JS, Worrell J (2015) On termination of integer linear loops.
  In: Proc.\ SODA, pp 957--969, \doi{10.1137/1.9781611973730.65}

\bibitem[{Pia et~al.(2017)Pia, Dey, and Molinaro}]{dblp:journals/mp/piadm17}
Pia AD, Dey SS, Molinaro M (2017) Mixed-integer quadratic programming is in
  \cc{NP}. Math\ Program 162(1-2):225--240, \doi{10.1007/s10107-016-1036-0}

\bibitem[{Podelski and Rybalchenko(2004)}]{dblp:conf/vmcai/podelskir04}
Podelski A, Rybalchenko A (2004) A complete method for the synthesis of linear
  ranking functions. In: Proc.\ VMCAI, LNCS 2937, pp 239--251,
  \doi{10.1007/978-3-540-24622-0_20}

\bibitem[{Renegar(1992)}]{renegar}
Renegar J (1992) On the computational complexity and geometry of the
  first-order theory of the reals, {Part I}: Introduction. {P}reliminaries.
  {T}he geometry of semi-algebraic sets. {T}he decision problem for the
  existential theory of the reals. J\ Symb\ Comput 13(3):255--300,
  \doi{10.1016/S0747-7171(10)80003-3}

\bibitem[{Robinson(1949)}]{robinson49}
Robinson J (1949) Definability and decision problems in arithmetic. J\ Symb\
  Log 14(2):98--114, \doi{10.2307/2266510}

\bibitem[{Roch and Villard(1996)}]{jordan}
Roch JL, Villard G (1996) Fast parallel computation of the {J}ordan normal form
  of matrices. Parallel Process\ Lett 06(02):203--212,
  \doi{10.1142/S0129626496000200}

\bibitem[{Roche(2018)}]{dblp:conf/issac/roche18}
Roche DS (2018) What can (and can't) we do with sparse polynomials? In: Proc.\
  ISSAC, pp 25--30, \doi{10.1145/3208976.3209027}

\bibitem[{Rodr\'iguez-Carbonell and Kapur(2004)}]{solvable-maps}
Rodr\'iguez-Carbonell E, Kapur D (2004) Automatic generation of polynomial loop
  invariants: Algebraic foundation. In: Proc.\ ISSAC, pp 266--273,
  \doi{10.1145/1005285.1005324}

\bibitem[{Schaefer(2009)}]{existsr}
Schaefer M (2009) Complexity of some geometric and topological problems. In:
  Proc.\ GD, LNCS 5849, pp 334--344, \doi{10.1007/978-3-642-11805-0_32}

\bibitem[{Schaefer and Stefankovic(2017)}]{forallr}
Schaefer M, Stefankovic D (2017) Fixed points, {N}ash equilibria, and the
  existential theory of the reals. Theory Comput Syst 60(2):172--193,
  \doi{10.1007/s00224-015-9662-0}

\bibitem[{Sidorov(2019)}]{sidorov2019}
Sidorov SV (2019) On the similarity of certain integer matrices with single
  eigenvalue over the ring of integers. Math\ Notes 105(5):756--762,
  \doi{10.1134/S0001434619050122}

\bibitem[{Storjohann and Labahn(1996)}]{dblp:conf/issac/storjohannl96}
Storjohann A, Labahn G (1996) Asymptotically fast computation of {H}ermite
  normal forms of integer matrices. In: Proc.\ ISSAC, pp 259--266,
  \doi{10.1145/236869.237083}

\bibitem[{Tarski(1998)}]{tarski_decidability}
Tarski A (1998) A decision method for elementary algebra and geometry. In:
  Caviness BF, Johnson JR (eds) Quantifier Elimination and Cylindrical
  Algebraic Decomposition, Springer, pp 24--84,
  \doi{10.1007/978-3-7091-9459-1}, originally appeared in 1951, U California
  Press, Berkeley and Los Angeles

\bibitem[{Tiwari(2004)}]{dblp:conf/cav/tiwari04}
Tiwari A (2004) Termination of linear programs. In: {Proc.\ CAV}, LNCS 3114, pp
  70--82, \doi{10.1007/978-3-540-27813-9_6}

\bibitem[{TPDB(2003 -- 2021)}]{tpdb}
TPDB (2003 -- 2021) {Termination Problems Data Base}.
  \urlprefix\url{http://termination-portal.org/wiki/TPDB}

\bibitem[{Wu et~al.(2010)Wu, Shen, Bi, and Zeng}]{dblp:conf/iccsa/wusbz10}
Wu B, Shen L, Bi Z, Zeng Z (2010) Termination of loop programs with polynomial
  guards. In: Proc.\ ICCSA, LNCS 6019, pp 482--496,
  \doi{10.1007/978-3-642-12189-0_42}

\bibitem[{Xia and Zhang(2010)}]{xiaz10}
Xia B, Zhang Z (2010) Termination of linear programs with nonlinear
  constraints. J\ Symb\ Comput 45(11):1234--1249,
  \doi{10.1016/j.jsc.2010.06.006}

\bibitem[{Xia et~al.(2011)Xia, Yang, Zhan, and
  Zhang}]{dblp:journals/fac/xiayzz11}
Xia B, Yang L, Zhan N, Zhang Z (2011) Symbolic decision procedure for
  termination of linear programs. Formal Aspects Comput 23(2):171--190,
  \doi{10.1007/s00165-009-0144-5}

\bibitem[{Xu and Li(2013)}]{xu13}
Xu M, Li ZB (2013) Symbolic termination analysis of solvable loops. J\ Symb\
  Comput 50:28--49, \doi{10.1016/j.jsc.2012.05.005}

\end{thebibliography}
